\documentclass[twocolumn,pra]{revtex4}
\usepackage{amssymb}
\usepackage{amsfonts}
\usepackage{amsmath}
\usepackage{amsxtra}
\usepackage{amscd}
\usepackage{amsthm}
\usepackage{graphicx}
\usepackage{epsf}
\usepackage{bbold}
\usepackage{xcolor}
\usepackage{mathalfa}
\usepackage{eucal}
\usepackage{bm}
\usepackage{array}

\newcommand{\PreserveBackslash}[1]{\let\temp=\\#1\let\\=\temp}
\newcolumntype{C}[1]{>{\PreserveBackslash\centering}p{#1}}
\newcolumntype{R}[1]{>{\PreserveBackslash\raggedleft}p{#1}}
\newcolumntype{L}[1]{>{\PreserveBackslash\raggedright}p{#1}}

\renewcommand{\arraystretch}{1.2}

\begin{document}

\title{Catalog of noninteracting tight-binding models with two energy bands in one dimension}

\author{Edward McCann}
\email{ed.mccann@lancaster.ac.uk}
\affiliation{Physics Department, Lancaster University, Lancaster, LA1 4YB, United Kingdom}

\begin{abstract}
We classify Hermitian tight-binding models describing noninteracting electrons on a one-dimensional periodic lattice with two energy bands. To do this, we write a generalized Rice-Mele model with two orbitals per unit cell, including all possible complex-valued long-range hoppings consistent with Hermicity. We then apply different forms of time-reversal, charge-conjugation and chiral symmetry in order to constrain the parameters, resulting in an array of possible models in different symmetry classes. For each symmetry class, we define a single, canonical form of the Hamiltonian and identify models that are related to the canonical form by an off-diagonal unitary transformation in the atomic basis. The models have either symmorphic or nonsymmorphic nonspatial symmetries (time $T$, chiral and charge-conjugation).
The nonsymmorphic category separates into two types of state of matter: an insulator with a $\mathbb{Z}_2$ topological index in the absence of nonsymmorphic time-reversal symmetry or, in the presence of nonsymmorphic time-reversal symmetry, a metallic state. The latter is an instance of Kramer's degeneracy with one degeneracy point in the Brillouin zone as opposed to no degeneracy points in symmorphic systems with $T^2 = 1$ and two in symmorphic systems with $T^2 = - 1$.
\end{abstract}

\maketitle


\begin{table}[t]
\begin{center}
\caption{\label{tablesummary}Summary of non-interacting, one-dimensional tight-binding models with two bands. The first column indicates the Cartan label of the symmetry class. Labels $T^2$, $C^2$ and $S^2$ indicate the type of time-reversal, charge-conjugation and chiral symmetries, where `$0$' indicates the absence of symmetry and `NS' indicates a nonsymmorphic symmetry. Column `No.' indicates the number of distinct models in the atomic basis (such models are related by an off-diagonal unitary transformation, not a diagonal change of gauge), column `Canonical' shows the location in the text of the canonical form of the Bloch Hamiltonian $H(k)$, column `State' gives the state of matter of a typical member of the class.}
\begin{tabular}{ L{0.8cm} | C{0.7cm} | C{0.7cm} | C{0.7cm} | C{0.7cm} | C{1.9cm} | C{1.8cm}}
\hline
Class & $T^2$ & $C^2$ & $S^2$ & No. & Canonical & State \\
\hline \hline
A & $0$ & $0$ & $0$ & $1$ & Eqs.(\ref{ad0})-(\ref{adz}) & insulator \\
\hline
AI & $1$ & $0$ & $0$ & $2$ & Table~\ref{tablesns3} &  insulator \\
\hline
AII & $-1$ & $0$ & $0$ & $1$ & Table~\ref{tablesns3} &  gapless \\
\hline
AIII & $0$ & $0$ & $1$ & $3$ & Table~\ref{tablescf} & $\mathbb{Z}$ insulator \\
\hline
BDI & $1$ & $1$ & $1$ & $3$ & Table~\ref{tablescf} & $\mathbb{Z}$ insulator \\
\hline
D & $0$ & $1$ & $0$ & $2$ & Table~\ref{tablesns3} & $\mathbb{Z}_2$ insulator \\
\hline
C & $0$ & $-1$ & $0$ & $1$ & Table~\ref{tablesns3} & insulator \\
\hline
DIII & $-1$ & $1$ & $1$ & $2$ & Table~\ref{tablescf} & gapless \\
\hline
CI & $1$ & $-1$ & $1$ & $2$ & Table~\ref{tablescf} & insulator \\
\hline
A & $0$ & $0$ & NS & $1$ & Table~\ref{tablescf} & $\mathbb{Z}_2$ insulator \\ 
\hline
AI & $1$ & NS & NS & $1$ & Table~\ref{tablescf} & $\mathbb{Z}_2$ insulator \\ 
\hline
D & NS & $1$ & NS & $1$ & Table~\ref{tablescf} & gapless \\
\hline
AIII & NS & NS & $1$ & $1$ & Table~\ref{tablescf} & gapless \\
\hline
A & NS & $0$ & $0$ & $1$ & Table~\ref{tablesns3} & gapless \\
\hline
A & $0$ & NS & $0$ & $1$ & Table~\ref{tablesns3} & $\mathbb{Z}_2$ insulator \\
\hline
\end{tabular}
\end{center}
\end{table}

\begin{figure}[t]
\includegraphics[scale=0.41]{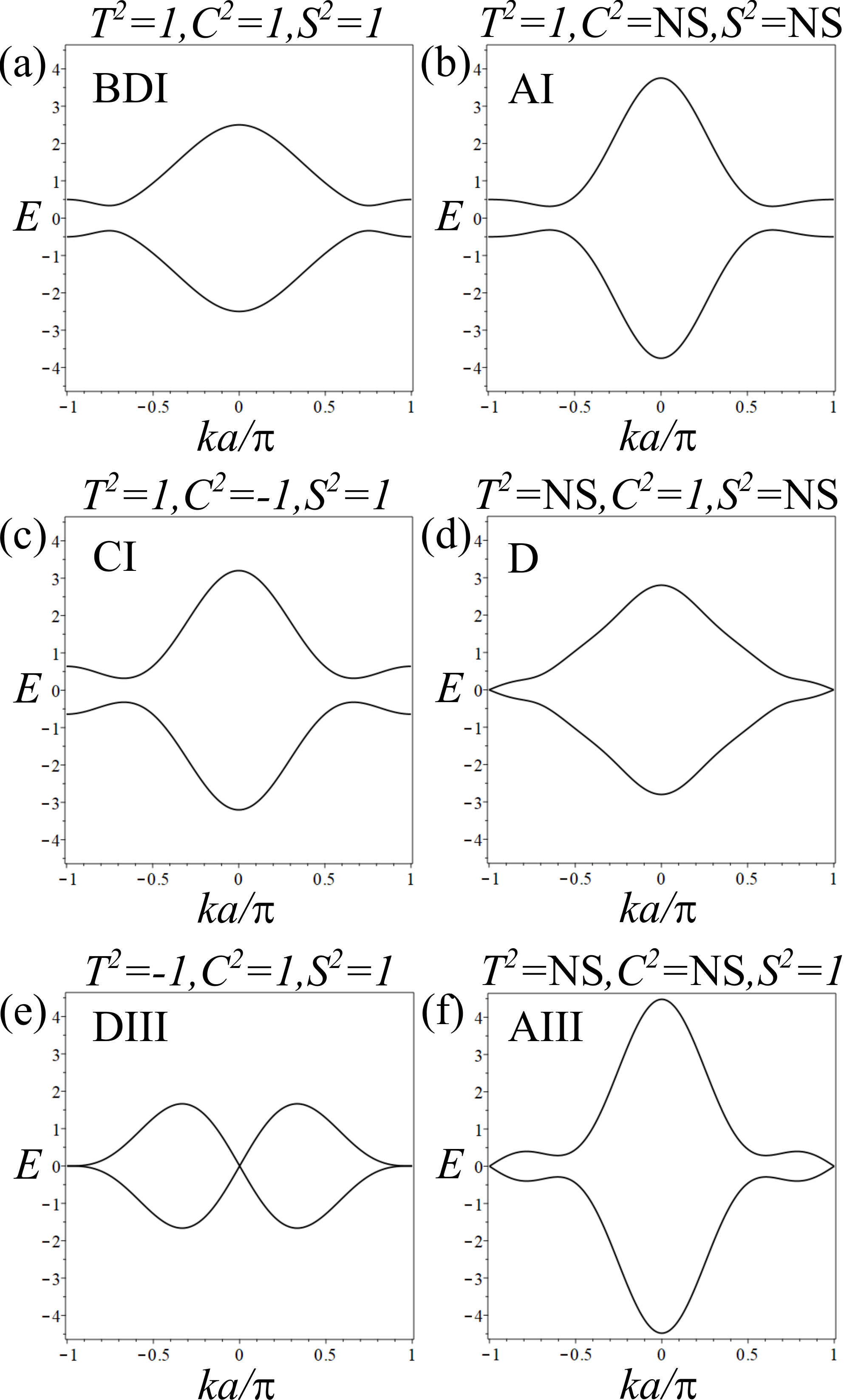}
\caption{Bulk band structure $E(k)$ for classes with time-reversal, charge-conjugation and chiral symmetry using the canonical form for each class (Table~\ref{tablesummary}) with generic parameter values.
(a) Symmorphic class BDI with $x_0 = 2.0$, $x_1 = 1.0$, $x_2 = 0.5$, $\tilde{y}_1 = 0.8$, $\tilde{y}_2 = 0.4$.
(b) Nonsymmorphic class AI with $x_0 = 2.0$, $x_1 = 1.0$, $x_2 = 0.5$, $\beta_0 =1.6$, $\beta_1 = 0.8$, $\beta_2 = 0.4$.
(c) Symmorphic class CI with $x_0 = 2.0$, $x_1 = 1.0$, $x_2 = 0.5$, 
$y_0 = 1.6$, $y_1 = 0.8$, $y_2 = 0.4$.
(d) Nonsymmorphic class D with $\tilde{x}_1 = 1.0$, $\tilde{x}_2 = 0.5$, $\beta_0 =1.6$, $\beta_1 = 0.8$, $\beta_2 = 0.4$.
(e) Symmorphic class DIII with $\tilde{x}_1 = 1.0$, $\tilde{x}_2 = 0.5$, $\tilde{y}_1 = 0.8$, $\tilde{y}_2 = 0.4$.
(f) Nonsymmorphic class AIII with $\alpha_0 = 2.0$, $\alpha_1 = 1.0$, $\alpha_2 = 0.5$, $\beta_0 =1.6$, $\beta_1 = 0.8$, $\beta_2 = 0.4$.
Higher-order parameters are zero.
}\label{figbulk1}
\end{figure}

\section{Introduction}

The study of the topological properties of materials according to the ten-fold way classification of nonspatial symmetries (time-reversal, charge-conjugation and chiral)~\cite{wigner51,wigner58,dyson62a,gade93,verbaarschot93,verbaarschot94,altland97,schnyder08,kitaev09,ryu10,chiu16} has expanded to include materials with nonsymmorphic crystal symmetries~\cite{liu14,shiozaki14,young15,wang16,shiozaki16,varjas17,kruthoff17,herrera22} which combine a nonprimitive lattice translation with either a mirror reflection or a rotation~\cite{lax74}. 
Recently, nonprimitive lattice translations have been incorporated into nonsymmorphic nonspatial symmetries~\cite{mong10,otrokov19,gong19,zhang19,niu20,fang15,shiozaki15,zhao16,arkinstall17,marques19,brzezicki20,allen22,yang22}. A particular example is that of three-dimensional antiferromagnetic topological insulators with nonsymmorphic magnetic symmetry~\cite{mong10,otrokov19,gong19,zhang19,niu20} (i.e., nonsymmorphic time-reversal symmetry).

In one dimension, the Su-Schrieffer-Heeger (SSH) model~\cite{ssh79,ssh80} is probably the simplest example of a topological insulator~\cite{hasan10,asboth16,cayssol21} and, as such, it has been studied extensively~\cite{takayama80,kivelson83,fulga11,chen11,gangadharaiah12,pershoguba12,li14,li15,asboth16,guo16,velasco17,bercioux17,rhim17,rhim18,liu18,munoz18,perezgonzalez19,chen20,scollon20,pletyukhov20,lin20,vanmiert20,han20,hetenyi21,chen21,cayssol21,fuchs21,zurita21} and realized experimentally in different platforms~\cite{kim12,cheon15,shim15,kim17,drost17,lee19,huda20,kiczynski22,atala13,przysiezna15,meier16,meier18,cooper19,kang2020}.
The SSH model has alternating hopping strengths and may be considered as a particular phase of the Rice-Mele model~\cite{ricemele82} which has alternating onsite energies as well as alternating hopping strengths. Surprisingly, it was recently shown that a phase of the Rice-Mele model which has alternating onsite energies but constant hopping strengths, the charge-density-wave (CDW) model~\cite{kivelson83,shiozaki15,brzezicki20,cayssol21,fuchs21,allen22}, is an example of a system exhibiting nonsymmorphic chiral symmetry.

It is natural to wonder how many different symmetry classes may be realized in minimal models (one-dimensional models with only two energy bands), to determine the form of such models in the atomic basis, and, particularly, to ask whether it is possible to find further examples with nonsymmorphic nonspatial symmetries including nonsymmorphic time-reversal symmetry.
In this paper, we answer these questions by writing down a generalized Rice-Mele model with two electronic orbitals per unit cell, including all possible complex-valued long-range hoppings consistent with Hermicity. 
We then apply different forms of nonspatial symmetries in order to constrain the parameters, resulting in an array of possible models in different symmetry classes.
Within random matrix theory~\cite{wigner51,wigner58,dyson62a}, it is common to write generic matrices representing symmetry classes~\cite{altland97,schnyder08,kitaev09,chiu16} and this is also possible with continuous Dirac Hamiltonians~\cite{ryu10}.
There are recent studies of generic tight-binding models with four bands~\cite{zurita21,matveeva22} that have included some representatives of some of the symmorphic symmetry classes. 
Here, we focus on tight-binding models in both position and reciprocal space, and we identify every distinct model within each symetry class according to the ten-fold way of nonspatial symmetries~\cite{schnyder08,kitaev09,ryu10,chiu16} and including nonsymmorphic nonspatial symmetries~\cite{shiozaki16}, albeit with only two energy bands.

Our results are summarized in Table~\ref{tablesummary}. The first column indicates the Cartan label of the symmetry class taken from the ten-fold way classification~\cite{wigner51,wigner58,dyson62a,gade93,verbaarschot93,verbaarschot94,altland97,schnyder08,kitaev09,ryu10,chiu16}.
Columns labeled $T^2$, $C^2$, $S^2$ show the form of time-reversal, charge-conjugation and chiral symmetry, respectively, where a zero indicates an absence of symmetry and `NS' indicates a nonsymmorphic symmetry.
For systems with nonsymmorphic symmetries, we adopt the classification of Ref.~\cite{shiozaki16} whereby the symmetry class is assigned by neglecting the nonsymmorphic symmetries (as if `NS' were replaced by a zero), and, subsequently, the nonsymmorphic symmetries are used to subdivide the symmetry classes. This is why there are three (sub)classes with nonsymmorphic symmetry labeled as class A.

The column labeled `No.' in Table~\ref{tablesummary} indicates the number of distinct models in the symmetry class. We consider distinct models to be related by off-diagonal unitary transformations (from the atomic basis), whereas models related by diagonal gauge transformations are not considered to be distinct. For example, the three distinct models in the BDI class are well known: the SSH model~\cite{ssh79,ssh80}, the Creutz model~\cite{creutz99,guo16} and the Shockley $sp$ orbital model~\cite{shockley39,vanderbilt93,fuchs21}.
The number of models in each class may be deduced simply by considering possible combinations of Pauli spin matrices, as described later.
Although there are distinct models within a typical symmetry class, they may be related by unitary transformations and, thus, it is possible to write them all in a single `canonical' form~\cite{canonicalnote} as indicated in the column `Canonical'. For classes with chiral symmetry, the canonical form is written in the basis in which the chiral symmetry operator is diagonal~\cite{ryu10,chiu16,guo16}, which we refer to as the `chiral' basis.

The final column in Table~\ref{tablesummary} shows the state of matter of members of the symmetry class,
and Fig.~\ref{figbulk1} shows representative band structures for classes with time-reversal, charge-conjugation and chiral symmetry.
For two energy bands $E_{\pm} (k)$, chiral symmetry imposes 
$E_{\pm}(k) = - E_{\mp}(k)$ and time-reversal symmetry imposes
$E_{\pm}(-k) = E_{\pm}(k)$, Fig.~\ref{figbulk1}.
The states of matter agree with predictions of the tenfold way classification for the symmorphic models~\cite{schnyder08,kitaev09,ryu10,chiu16} and with the classification of Ref.~\cite{shiozaki16} for nonsymmorphic models.
Exceptions are the symmorphic DIII and the nonsymmorphic D class which we find are both gapless due to the presence of time-reversal symmetry with Kramer's degeneracy and only two orbitals, as opposed to a four-orbital Bogoliubov de Gennes representation~\cite{budich13,li16}. With two bands, there are also not enough degrees of freedom to realize the CII (chiral symplectic) symmetry class~\cite{gholizadeh18,matveeva22}.

Table~\ref{tablesummary} shows that the nonsymmorphic category separates into only two types of state of matter: an insulator with a $\mathbb{Z}_2$ topological index~\cite{shiozaki15,shiozaki16} in the absence of nonsymmorphic time-reversal symmetry or, in the presence of nonsymmorphic time-reversal symmetry, a metallic state~\cite{zhao16}. The latter is an instance of Kramer's degeneracy with one degeneracy point in the Brillouin zone, Fig.~\ref{figbulk1}(d), (f), as opposed to no degeneracy points in symmorphic systems with $T^2 = 1$ and two in symmorphic systems with $T^2 = - 1$, Fig.~\ref{figbulk1}(e).

In Section~\ref{s:grm} we describe the generalized Rice-Mele model~\cite{ricemele82} which is a periodic tight-binding model with two orbitals per unit cell. We consider a time-independent and Hermitian Hamiltonian describing noninteracting fermions and excluding superconducting pairing. In the main text, we describe nearest-neighbor coupling, but we generalize to include hoppings of all ranges in Appendix~\ref{a:longrange}. Section~\ref{s:sym} describes the symmetries that are applied to the generalized Rice-Mele model in order to constrain the parameter values and identify models within each symmetry class.
The rest of the paper considers each symmetry class in turn, with symmorphic models with chiral symmetry in Section~\ref{s:s} and without chiral symmetry in Section~\ref{s:sns}.
Nonsymmorphic models with chiral symmetry are discussed in Section~\ref{s:ns}, those without chiral symmetry in Section~\ref{s:nsns}.
The paper ends with a brief conclusion, Section~\ref{s:conc}.
Appendix~\ref{a:spatial} discusses the role of spatial-inversion symmetry~\cite{teo08,hughes11,chen11,chiu13,morimoto13,shiozaki14,chiu16,fuchs21}, which is not the focus of the classification in the main text.

\section{Methodology}\label{s:methodology}

\subsection{Generalized Rice-Mele model}\label{s:grm}

\begin{figure}[t]
\includegraphics[scale=0.40]{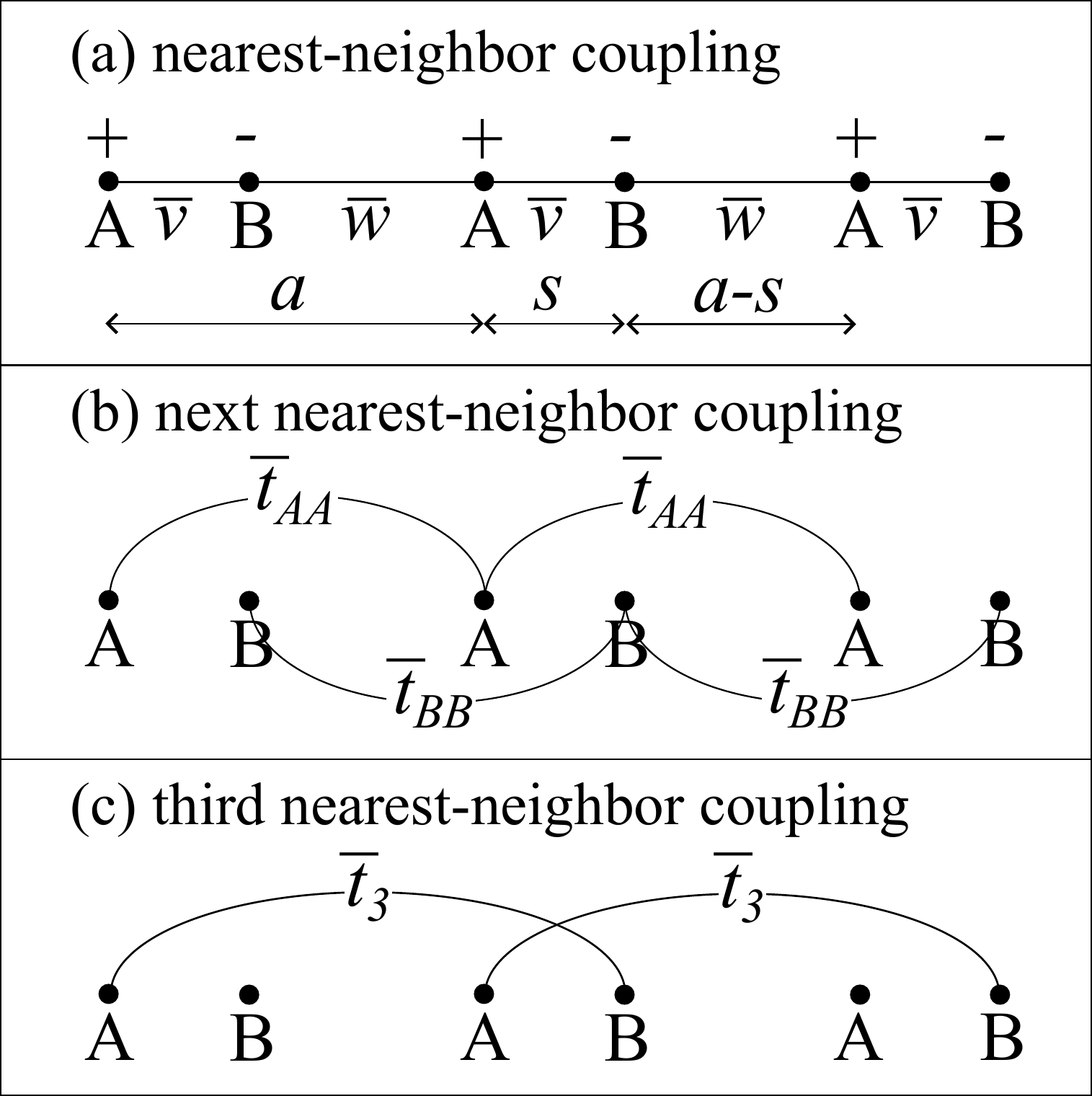}
\caption{Generalized Rice-Mele model~(\ref{cgrm1},\ref{cgrm2}) with up to third nearest-neighbor coupling. (a) shows nearest-neighbor coupling with staggered hoppings $v$ and $w$ and staggered onsite energies, `$+$' represents $\epsilon_0 + u$, `$-$' represents $\epsilon_0 - u$. The lattice constant is $a$, and a B orbital is located at intracell distance $s$ to the right of an A orbital. (b) next nearest-neighbor coupling $t_{AA}$ between A orbitals and $t_{BB}$ between B orbitals. (c) third nearest-neighbor coupling $t_3$ between an A orbital and the second B orbital to its right. Bars over parameters indicate that they are complex numbers, in general.
}\label{figmodels1}
\end{figure}

In position space for a system of $J$ atoms with open boundary conditions, we consider a generalized Rice-Mele model~\cite{ricemele82} Hamiltonian, Fig.~\ref{figmodels1}, as a $J \times J$ Hermitian matrix in a basis of atomic orbitals,
\begin{eqnarray}
H \!=\! \begin{pmatrix}
\epsilon_0 + u & v e^{i \phi_v} & t_{AA} e^{i \phi_{AA}} & t_3 e^{i \phi_3} \\
ve^{-i \phi_v} & \epsilon_0 -u & we^{i \phi_w} & t_{BB}e^{i \phi_{BB}} \\
t_{AA} e^{-i \phi_{AA}} & we^{-i \phi_w} & \epsilon_0 + u & v e^{i \phi_v} \\
t_3e^{-i \phi_3} & t_{BB}e^{-i \phi_{BB}} & ve^{-i \phi_v} & \epsilon_0 -u \\
\end{pmatrix} \!\! ,\label{cgrm1}
\end{eqnarray}
where we write the matrix for just $J = 4$ atoms.
As in the usual Rice-Mele model~\cite{ricemele82}, there are alternating onsite energies parameterized by $u$ and alternating nearest-neighbor hoppings $v$ and $w$. In addition, we include a uniform contribution $\epsilon_0$ to the onsite energy, next-nearest couplings $t_{AA}$ and $t_{BB}$, and third-nearest coupling $t_3$. All parameters, other than the onsite energies $\epsilon_0 \pm u$, are complex, with phases explicitly written in Eq.~(\ref{cgrm1}).

There are two different orbitals per unit cell, labeled A and B. With lattice constant $a$, we consider an A orbital to be separated by distance $s$ from the next B orbital to the right, which is then separated by distance $a-s$ from the next A to the right, Fig.~\ref{figmodels1}.
The Bloch Hamiltonian in $k$ space written in the type II or `canonical' representation~\cite{bena09,cayssol21} with A and B Bloch orbitals is
\begin{eqnarray}
H (k,s) &=& \mbox{\boldmath$\sigma$} \cdot {\bf d}  (k,s) , \label{cgrm2} \\
d_0 (k,s) &=& \epsilon_0 + t_{AA} \cos (ka + \phi_{AA}) + t_{BB} \cos (ka + \phi_{BB}) , \nonumber \\
d_x (k,s) &=& v \cos (ks + \phi_v) + w \cos [k(a-s) + \phi_w] \nonumber \\
&& \qquad + t_3 \cos [k(a+s) + \phi_3] , \nonumber \\
d_y (k,s) &=& - v \sin (ks + \phi_v) + w \sin [k(a-s) + \phi_w] \nonumber \\
&& \qquad - t_3 \sin [k(a+s) + \phi_3] , \nonumber \\
d_z (k,s) &=& u + t_{AA} \cos (ka + \phi_{AA}) - t_{BB} \cos (ka + \phi_{BB}) , \nonumber
\end{eqnarray}
with $\mbox{\boldmath$\sigma$} = (I,\sigma_x,\sigma_y,\sigma_z)$ and ${\bf d} = (d_0,d_x,d_y,d_z)$, where $\sigma_i$ are Pauli matrices and $I$ is the $2 \times 2$ identity matrix.
We can write
\begin{eqnarray}
H (k,s) = U (k,s) H (k,0) U^{\dagger} (k,s) , \label{uks1}
\end{eqnarray}
where
\begin{eqnarray}
U (k,s) = \begin{pmatrix}
e^{iks/2} & 0 \\
0 & e^{-iks/2}
\end{pmatrix} , \label{uks2}
\end{eqnarray}
and $H (k,0)$ coincides with the Hamiltonian written in the type I or `periodic' representation~\cite{bena09,cayssol21}

Our aim is to describe all models with nearest-neighbor coupling. We include next-nearest couplings $t_{AA}$ and $t_{BB}$, and third-nearest coupling $t_3$ because they become nearest-neighbor in the limit of intracell spacing $s=0$. We consider only $0 \leq s \leq a/2$ without loss of generality, so do not include third-nearest coupling between an A orbital and the second B orbital to its left. Note, however, that it is straightforward to include all orders of coupling as described in Appendix~\ref{a:longrange}.

The Hamiltonian~(\ref{cgrm1}) in position space does not explicitly depend on the distances $a$ and $s$ although, in a real physical system, one would expect the values of the tight-binding parameters to depend on the distances $a$ and $s$. However, we are not interested in the numerical values of the parameters, but general constraints on the existence of their real and imaginary parts. Thus, we treat the generalized Rice-Mele model as a toy model with no contraints initially on the values of the tight-binding parameters other than that the Hamiltonian is Hermitian and, in particular, the tight-binding parameters are considered to be independent of $a$ and $s$.
In general, the generalized Rice-Mele model, Eqs.~(\ref{cgrm1},\ref{cgrm2}), does not satisfy spatial inversion, chiral, time-reversal, or charge-conjugation symmetry; we apply chiral, time-reversal, or charge-conjugation symmetry in order to place some further constraints on the tight-binding parameters.
We nominally describe spinless electrons, although the cases with time-reversal symmetry $T^2 = -1$ may be interpreted as describing spinful electrons, Sections~\ref{s:d3} and~\ref{s:a2}.

\subsection{Symmetries}\label{s:sym}

We consider the non-spatial symmetries of chiral symmetry ($S$), time-reversal symmetry ($T$), and charge-conjugation symmetry ($C$).
In $k$ space, they are expressed as
\begin{eqnarray}
\mathrm{chiral}\!\!: \quad U_S^{\dagger} (k,s) H (k,s) U_S (k,s) &=& - H (k,s) , \label{sk} \\
\mathrm{time}\!\!: \quad U_T^{\dagger} (k,s) H^{\ast} (k,s) U_T (k,s) &=& H (-k,s) , \label{tk} \\
\mathrm{charge}\!\!: \quad U_C^{\dagger} (k,s) H^{\ast} (k,s) U_C (k,s) &=& - H (-k,s) , \label{ck} \\
\mathrm{space}\!\!: \quad U_P^{\dagger} (k,s) H (k,s) U_P (k,s) &=& H (-k,s) , \label{pk} 
\end{eqnarray}
where $U_S$, $U_T$, $U_C$, and $U_P$ are unitary matrices with
\begin{eqnarray}
U_S (k,s) U_S (k,s)  &=& I , \label{sone} \\
U_T (k,s) U_T^{\ast} (-k,s)  &=& \pm I , \label{tone} \\
U_C (k,s) U_C^{\ast} (-k,s)  &=& \pm I . \label{cone}
\end{eqnarray}
These three conditions are written in shorthand as $S^2 = 1$, $T^2 = \pm 1$, and $C^2 = \pm 1$.
The operators are related as $U_S (k,s) =U_C^{\ast} (k,s) U_T (-k,s)$~\cite{symmetrynotes}.

In addition to the three non-spatial symmetries, we also include spatial-inversion symmetry ($P$) for completeness. Although it is not our focus, many of the models we discuss will satisfy spatial-inversion symmetry~\cite{fuchs21}, and we describe it in more detail in Appendix~\ref{a:spatial}.
Using the unitary transformation~(\ref{uks1},\ref{uks2}) relating $H(k,s)$ to $H(k,0)$, we can write
\begin{eqnarray}
U_S (k,s) &=& U(k,s) U_S (k,0) U^{\dagger} (k,s) , \label{usks} \\
U_T (k,s) &=& U^{\dagger} (k,s) U_T (k,0) U (k,s) , \label{utks} \\
U_C (k,s) &=& U^{\dagger} (k,s) U_C (k,0) U (k,s) , \label{ucks} \\
U_P (k,s) &=& U (k,s) U_P (k,0) U (k,s) . \label{upks}
\end{eqnarray}
Note that the different forms of these transformations are due to the presence or absence of complex conjugation and inversion of $k$ in the definitions of the symmetries, Eqs.~(\ref{sk})-(\ref{pk}).

Given an energy eigenvalue equation $H(k,s) \psi_{\pm} (k,s) = E_{\pm}(k) \psi_{\pm} (k,s)$ for two energy bands $E_{\pm}(k)$ and eigenstates $\psi_{\pm} (k,s)$, the effect of the symmetries is
\begin{eqnarray}
\mathrm{chiral}\!\!: \qquad E_{\pm}(k) &=& - E_{\mp}(k) , \label{es} \\
\mathrm{time}\!\!: \qquad E_{\pm}(-k) &=& E_{\pm}(k) , \label{et} \\
\mathrm{charge}\!\!: \qquad E_{\pm}(-k) &=& - E_{\mp}(k) , \label{ec} \\
\mathrm{space}\!\!: \qquad E_{\pm}(-k) &=& E_{\pm}(k) . \label{ep} 
\end{eqnarray}

For symmorphic models, symmetry operators $U_S$, $U_T$, $U_C$, $U_P$, when present, are independent of $k$ for intracell distance $s=0$ (or, equivalently, when using the type I or `periodic' representation~\cite{bena09,cayssol21}). For nonsymmorphic symmetries, the symmetry operators are independent of $k$ for intracell distance $s=a/2$ because this implicitly takes into account a spatial translation by half of the unit cell $a$;
when transformed back to $s=0$ using Eqs.~(\ref{usks})-(\ref{upks}), they acquire a $k$ dependence.
For the two band models we consider, the $k$ independent symmetry operators may be represented as either Pauli matrices or the identity matrix, as appropriate~\cite{symmetrynotes}.
Application of chiral symmetry requires $d_0$ and one of the other components, $d_i$ for $i=x$, $y$, or $z$, to be zero whereas time-reversal and charge-conjugation symmetries require each component of the ${\bf d}$ vector to be either even or odd functions of the wave vector $k$.

In position space, the symmetry operations are written as
\begin{eqnarray}
\mathrm{chiral}\!\!: \quad {\mathcal S}^{\dagger} H {\mathcal S} &=& - H , \quad {\mathcal S}{\mathcal S} = I , \label{spos} \\
\mathrm{time}\!\!: \quad {\mathcal T}^{\dagger} H^{\ast} {\mathcal T} &=& H , \quad {\mathcal T}{\mathcal T}^{\ast} = \pm I , \\
\mathrm{charge}\!\!: \quad {\mathcal C}^{\dagger} H^{\ast} {\mathcal C} &=& - H , \quad {\mathcal C}{\mathcal C}^{\ast} = \pm I , \\
\mathrm{space}\!\!: \quad {\mathcal P}^{\dagger} H {\mathcal P} &=& H , \label{ppos}
\end{eqnarray}
where ${\mathcal S}$, ${\mathcal T}$, ${\mathcal C}$, and ${\mathcal P}$ are unitary matrices, and ${\mathcal S} ={\mathcal T}^{\ast} {\mathcal C}$.
For systems with symmorphic symmetries, ${\mathcal S}$, ${\mathcal T}$, ${\mathcal C}$ act locally within a unit cell and may be represented by $J \times J$ matrices $S_x$, $S_y$ or $S_z$, acting in the atomic basis as
\begin{eqnarray}
S_x = \begin{pmatrix}
0 & 1 & 0 & 0 & \cdots & 0 & 0 \\
1 & 0 & 0 & 0 & \cdots & 0 & 0 \\
0 & 0 & 0 & 1 & \cdots & 0 & 0 \\
0 & 0 & 1 & 0 & \cdots & 0 & 0 \\
\vdots & \vdots & \vdots & \vdots & \vdots & \vdots & \vdots \\
0 & 0 & 0 & 0 & \cdots & 0 & 1 \\
0 & 0 & 0 & 0 & \cdots & 1 & 0 \\
\end{pmatrix} . \label{sx}
\end{eqnarray}
\begin{eqnarray}
S_{y} = \begin{pmatrix}
0 & -i & 0 & 0 & \cdots & 0 & 0 \\
i & 0 & 0 & 0 & \cdots & 0 & 0 \\
0 & 0 & 0 & -i & \cdots & 0 & 0 \\
0 & 0 & i & 0 & \cdots & 0 & 0 \\
\vdots & \vdots & \vdots & \vdots & \vdots & \vdots & \vdots \\
0 & 0 & 0 & 0 & \cdots & 0 & -i \\
0 & 0 & 0 & 0 & \cdots & i & 0 \\
\end{pmatrix} , \label{sy}
\end{eqnarray}
\begin{eqnarray}
S_z = \begin{pmatrix}
1 & 0 & 0 & \cdots & 0 & 0 \\
0 & -1 & 0 & \cdots & 0 & 0 \\
0 & 0 & 1 & \cdots & 0 & 0 \\
\vdots & \vdots & \vdots & \vdots & \vdots & \vdots \\
0 & 0 & 0 & \cdots & 1 & 0 \\
0 & 0 & 0 & \cdots & 0 & -1 \\
\end{pmatrix} . \label{sz}
\end{eqnarray}
Operations ${\mathcal T}$ and ${\mathcal C}$ may also be represented by a $J \times J$ identity matrix, $I$.
When representing chiral symmetry ${\mathcal S}$, operator $S_z$ (or $\sigma_z$ in $k$ space) is usually referred to as sublattice symmetry.
Note that we refer to chiral symmetry as any unitary operator ${\mathcal S}$ satisfying Eq.~(\ref{spos}), and, in the atomic basis, this may include $S_x$ or $S_y$ as well as $S_z$. In these cases (when symmorphic), it is possible to perform a unitary transformation from the atomic basis to the chiral basis in which ${\mathcal S}$ is represented by $S_z$.

Spatial inversion symmetry (parity) ${\mathcal P}$ is non-local and may be represented by $J \times J$ matrices $P_x$, $P_y$, $P_z$, or $P_I$, where
\begin{eqnarray}
P_x = \begin{pmatrix}
0 & 0 & 0 & \cdots & 0 & 1 \\
0 & 0 & 0 & \cdots & 1 & 0 \\
\vdots & \vdots & \vdots & \vdots & \vdots & \vdots \\
0 & 0 & 1 & \cdots & 0 & 0 \\
0 & 1 & 0 & \cdots & 0 & 0 \\
1 & 0 & 0 & \cdots & 0 & 0 
\end{pmatrix} \!\! , \label{px}
\end{eqnarray}
\begin{eqnarray}
P_y = \begin{pmatrix}
0 & 0 & 0 & \cdots & 0 & 1 \\
0 & 0 & 0 & \cdots & -1 & 0 \\
\vdots & \vdots & \vdots & \vdots & \vdots & \vdots \\
0 & 0 & -1 & \cdots & 0 & 0 \\
0 & 1 & 0 & \cdots & 0 & 0 \\
-1 & 0 & 0 & \cdots & 0 & 0 
\end{pmatrix} \!\! , \label{py}
\end{eqnarray}
\begin{eqnarray}
P_z = \begin{pmatrix}
0 & 0 & 0 & 0 & \cdots & 1 & 0 \\
0 & 0 & 0 & 0 & \cdots & 0 & -1 \\
\vdots & \vdots & \vdots & \vdots & \vdots & \vdots & \vdots \\
0 & 0 & 1 & 0 & \cdots & 0 & 0 \\
0 & 0 & 0 & -1 & \cdots & 0 & 0 \\
1 & 0 & 0 & 0 & \cdots & 0 & 0 \\
0 & -1 & 0 & 0 & \cdots & 0 & 0 
\end{pmatrix} \!\! , \label{pz}
\end{eqnarray}
\begin{eqnarray}
P_I = \begin{pmatrix}
0 & 0 & 0 & 0 & \cdots & 1 & 0 \\
0 & 0 & 0 & 0 & \cdots & 0 & 1 \\
\vdots & \vdots & \vdots & \vdots & \vdots & \vdots & \vdots \\
0 & 0 & 1 & 0 & \cdots & 0 & 0 \\
0 & 0 & 0 & 1 & \cdots & 0 & 0 \\
1 & 0 & 0 & 0 & \cdots & 0 & 0 \\
0 & 1 & 0 & 0 & \cdots & 0 & 0 
\end{pmatrix} \!\! . \label{pi}
\end{eqnarray}

Nonsymmorphic symmetry involves a translation $T_{a/2}$~\cite{allen22} by the atomic spacing $a/2$,
\begin{eqnarray}
T_{a/2} = \begin{pmatrix}
0 & 1 & 0 & 0 & \cdots & 0 & 0 \\
0 & 0 & 1 & 0 & \cdots & 0 & 0 \\
0 & 0 & 0 & 1 & \cdots & 0 & 0 \\
0 & 0 & 0 & 0 & \cdots & 0 & 0 \\
\vdots & \vdots & \vdots & \vdots & \vdots & \vdots & \vdots \\
0 & 0 & 0 & 0 & \cdots & 0 & 1 \\
1 & 0 & 0 & 0 & \cdots & 0 & 0 \\
\end{pmatrix} . \label{ta2}
\end{eqnarray}
It could also be expressed as a matrix product $T_{a/2} S_z$ of sublattice symmetry $S_z$~(\ref{sz}) with $T_{a/2}$,
\begin{eqnarray}
T_{a/2} S_z = \begin{pmatrix}
0 & -1 & 0 & 0 & \cdots & 0 & 0 \\
0 & 0 & 1 & 0 & \cdots & 0 & 0 \\
0 & 0 & 0 & -1 & \cdots & 0 & 0 \\
0 & 0 & 0 & 0 & \cdots & 0 & 0 \\
\vdots & \vdots & \vdots & \vdots & \vdots & \vdots & \vdots \\
0 & 0 & 0 & 0 & \cdots & 0 & 1 \\
1 & 0 & 0 & 0 & \cdots & 0 & 0 \\
\end{pmatrix} , \label{SCDW}
\end{eqnarray}
where $T_{a/2} S_z$ is written for an odd number of atoms, $J$.
These two possibilities correspond to $\sigma_x$ or $\sigma_y$, respectively, in $k$ space.

\begin{table*}[t]
\begin{center}
\caption{\label{tabless}Symmorphic models with chiral symmetry, showing the values of the tight-binding parameters of the generalized Rice-Mele model with $d_0 = 0$, Eq.~(\ref{cgrm4}), $\epsilon_0 = 0$, $t_{BB} = - t_{AA}$ and $\phi_{BB} = \phi_{AA}$. `arb.' indicates that the parameter can take any arbitrary real value. The second column `$\mathcal{P}$' shows the form of spatial-inversion symmetry in position space for an even number of orbitals, where `$0$' indicates it is absent. The same models with long-range hoppings are shown in Table~\ref{tableAs}.
}
\begin{tabular}{ L{1.2cm} | C{0.6cm} || C{1.0cm} | C{1.0cm} | C{1.0cm} | C{1.0cm} | C{1.0cm} | C{1.0cm} | C{1.0cm} | C{1.0cm} | C{1.0cm}}
\hline
name & $\mathcal{P}$ & $u$ & $t_{AA}$ & $\phi_{AA}$ & $v$ & $\phi_v$ & $w$ & $\phi_w$ & $t_3$ & $\phi_3$  \\ [1pt]
\hline \hline
$H_{\mathrm{S,AIII}}^{(0,0,z)}$ & $0$ & $0$ & $0$ & n/a & arb. & arb. & arb. & arb. & arb. & arb. \\ [1pt]
\hline
$H_{\mathrm{S,AIII}}^{(0,0,y)}$ & $0$  & arb. & arb. & arb. & arb. & $0$ & arb. & arb. & $w$ & $\phi_w$ \\ [1pt]
\hline
$H_{\mathrm{S,AIII}}^{(0,0,x)}$ & $0$  & arb. & arb. & arb. & arb. & $\pi/2$ & arb. & arb. & $-w$ & $\phi_w$ \\ [1pt]
\hline \hline
$H_{\mathrm{S,BDI}}^{(I,z,z)}$ & $P_x$  & $0$ & $0$ & n/a & arb. & $0$ & arb. & $0$ & arb. & $0$ \\ [1pt]
\hline
$H_{\mathrm{S,BDI}}^{(x,z,y)}$ & $P_x$  & $0$ & arb. & $\pi/2$ & arb. & $0$ & arb. & $0$ & $w$ & $0$ \\ [1pt]
\hline
$H_{\mathrm{S,BDI}}^{(I,x,x)}$ & $P_z$  & arb. & arb. & $0$ & $0$ & n/a & arb. & $0$ & $-w$ & $0$ \\ [1pt]
\hline
$H_{\mathrm{S,BDI}}^{(z,I,z)}$ & $P_y$  & $0$ & $0$ & n/a & arb. & $\pi/2$ & arb. & $\pi/2$ & arb. & $\pi/2$ \\ [1pt]
\hline
$H_{\mathrm{S,BDI}}^{(x,I,x)}$ & $P_y$  & $0$ & arb. & $\pi/2$ & arb. & $\pi/2$ & arb. & $\pi/2$ & $-w$ & $\pi/2$ \\ [1pt]
\hline
$H_{\mathrm{S,BDI}}^{(z,x,y)}$ & $P_z$  & arb. & arb. & $0$ & $0$ & n/a & arb. & $\pi/2$ & $w$ & $\pi/2$ \\ [1pt]
\hline \hline
$H_{\mathrm{S,DIII}}^{(y,x,z)}$ & $P_z$  & $0$ & $0$ & n/a & $0$ & n/a & arb. & arb. & $-w$ & $-\phi_w$ \\ [1pt]
\hline
$H_{\mathrm{S,DIII}}^{(y,z,x)}$ & $P_x$  & $0$ & arb. & $\pi/2$ & $0$ & n/a & arb. & $0$ & $-w$ & $0$ \\ [1pt]
\hline
$H_{\mathrm{S,DIII}}^{(y,I,y)}$ & $P_y$  & $0$ & arb. & $\pi/2$ & $0$ & n/a & arb. & $\pi/2$ & $w$ & $\pi/2$ \\ [1pt]
\hline \hline
$H_{\mathrm{S,CI}}^{(x,y,z)}$ & $P_I$  & $0$ & $0$ & n/a & arb. & arb. & arb. & arb. & $w$ & $-\phi_w$ \\ [1pt]
\hline
$H_{\mathrm{S,CI}}^{(I,y,y)}$ & $P_I$  & arb. & arb. & $0$ & arb. & $0$ & arb. & $0$ & $w$ & $0$  \\ [1pt]
\hline
$H_{\mathrm{S,CI}}^{(z,y,x)}$ & $P_I$  & arb. & arb. & $0$ & arb. & $\pi/2$ & arb. & $\pi/2$ & $-w$ & $\pi/2$ \\ [1pt]
\hline
\end{tabular}
\end{center}
\end{table*}

For symmorphic models in position space, symmetries generally only hold for an even number of atoms, $J$, unless the matrix representing the symmetry is diagonal. An example is the SSH model~\cite{ssh79,ssh80} for which time-reversal ($I$), charge-conjugation ($S_z$), and chiral ($S_z$) symmetry are all diagonal, so they hold for an odd number of atoms; spatial inversion symmetry ($P_x$) does not hold in the SSH model for an odd number of atoms, however.
For nonsymmorphic models in position space, nonspatial symmetries are represented either by nonsymmorphic operations, $T_{a/2}$ or $T_{a/2} S_z$, or by diagonal ones, $I$ or $S_z$ (e.g. it is possible to combine symmorphic time with nonsymmorphic charge-conjugation and chiral symmetries). Thus, they nominally apply to systems with either even or odd numbers of atoms, except that, in both instances, the ends of a system with open boundary conditions break the nonsymmorphic symmetry~\cite{shiozaki15,allen22}.
Spatial inversion symmetry for nonsymmorphic models in position space is represented by $P_x$ or $P_y$, and its presence, for a given symmetry class, depends on whether the number of atoms is even or odd as indicated in Table~\ref{tablenss}.

Symmorphic models are labeled $H_{\mathrm{S,n}}^{(i,j,\ell)}$ where $n$ indicates the Cartan label of the symmetry class, Table~\ref{tablesummary}. Index $i = I,x,y,z$ indicates the form of time-reversal symmetry as either the identity matrix ($I$) in both $k$ and position space, or a Pauli spin matrix $\sigma_i$ in $k$ space and matrix $S_i$ in position space; likewise index $j = I,x,y,z$ for charge-conjugation symmetry
and $\ell = x,y,z$ for chiral symmetry (for which the unit matrix is not possible).
Nonsymmorphic models are labeled $H_{\mathrm{NS,n}}^{(i,j,\ell)}$ where $n$ indicates the Cartan label and index $i = I,x,y,z$ indicates the form of time-reversal symmetry as either the identity matrix ($I$) in both $k$ and position space, Pauli spin matrix $\sigma_x$ in $k$ space and matrix $T_{a/2}$ in position space, Pauli spin matrix $\sigma_y$ in $k$ space and matrix $T_{a/2}S_z$ in position space, or Pauli spin matrix $\sigma_z$ in $k$ space and matrix $S_z$ in position space; likewise index $j = I,x,y,z$ for charge-conjugation symmetry, and index $\ell = x,y,z$ for chiral symmetry (for which the unit matrix is not possible).

\begin{figure}[t]
\includegraphics[scale=0.45]{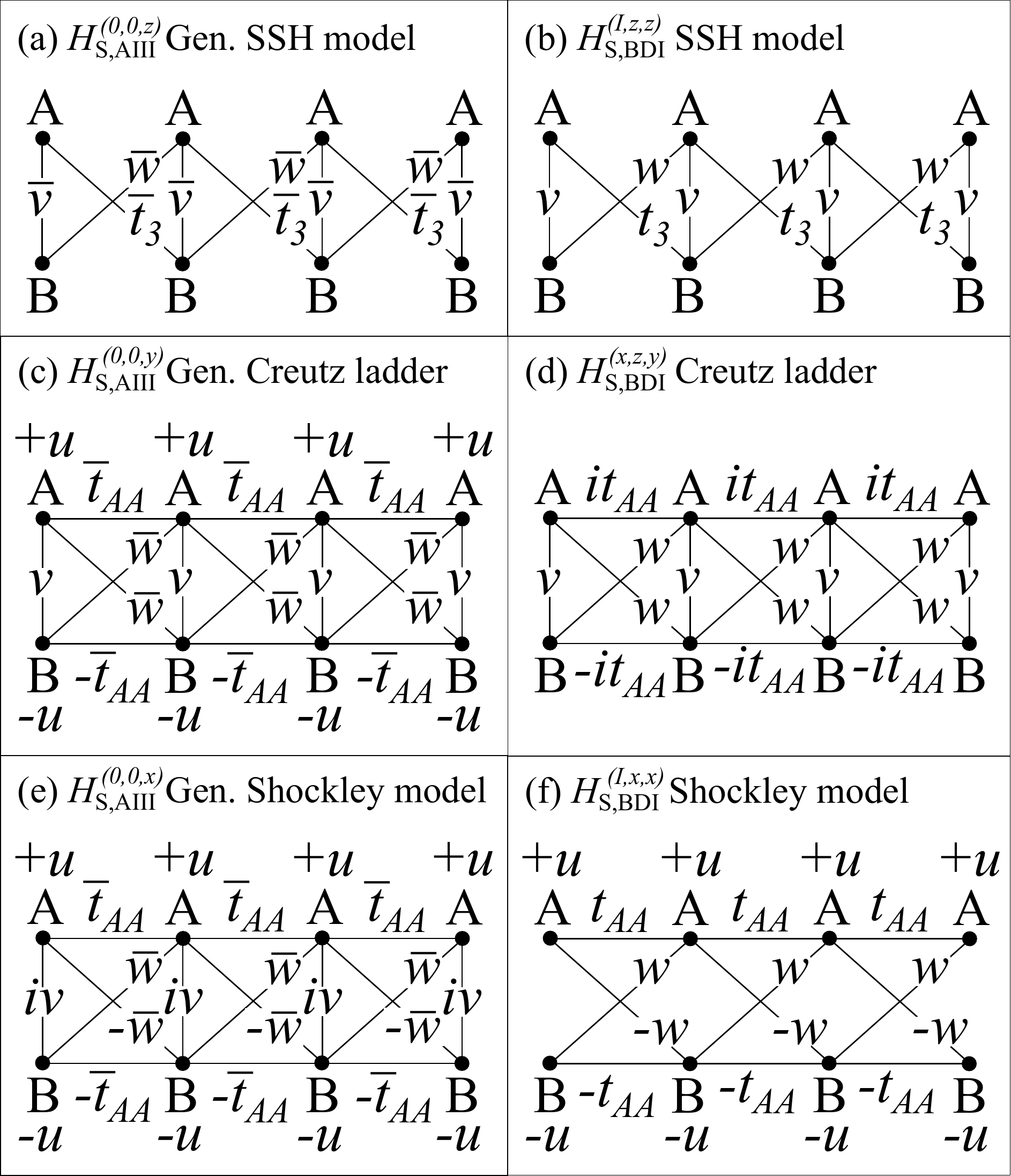}
\caption{Tight-binding models in the AIII symmetry class ($T^2 = 0$, $C^2 = 0$, $S^2 = 1$) on the left, in the BDI symmetry class ($T^2 = 1$, $C^2 = 1$, $S^2 = 1$) on the right, shown with intracell spacing $s=0$. 
(a) Shows $H_{\mathrm{S,AIII}}^{(0,0,z)}$, the generalized SSH model with staggered couplings $v$ and $w$ plus third-nearest neighbor hopping $t_3$.
(b) Shows the SSH model, $H_{\mathrm{S,BDI}}^{(I,z,z)}$, with staggered couplings $v$ and $w$ (as usually considered) plus third-nearest neighbor hopping $t_3$.
(c) $H_{\mathrm{S,AIII}}^{(0,0,y)}$, the generalized Creutz ladder with onsite energies $\pm u$, hopping $\pm t_{AA}e^{i \phi_{AA}}$ along each chain, intracell hopping $v$, and intercell hoppings $w e^{i \phi_w}$ between each chain.
(d) The Creutz ladder, $H_{\mathrm{S,BDI}}^{(x,z,y)}$, with onsite energies $\pm u$ and imaginary hopping $\pm i t_{AA}$ along each chain, intracell hopping $v$, and intercell hoppings $\pm w$ between each chain.
(e) $H_{\mathrm{S,AIII}}^{(0,0,x)}$, the generalized Shockley $s$-$p$ orbital model with onsite energies $\pm u$ and hopping $\pm t_{AA}e^{i \phi_{AA}}$ along each chain, intracell hopping $iv$, and intercell hoppings $\pm w e^{i \phi_w}$ between each chain.
(f) The Shockley $sp$ orbital model, $H_{\mathrm{S,BDI}}^{(I,x,x)}$, with onsite energies $\pm u$ and hopping $\pm t_{AA}$ along each chain, and intercell hoppings $\pm w$ between each chain.
Bars over parameters indicate that they are complex numbers, in general.
A finite lateral spacing between A and B orbitals is shown for clarity, but this separation can be zero without loss of generality.
}\label{figchiral1}
\end{figure}

\renewcommand{\arraystretch}{1.3}
\begin{table*}[t]
\begin{center}
\caption{\label{tablescf}Canonical forms of the Bloch Hamiltonian $H(k)$ for non-interacting, one-dimensional tight-binding models with two bands and chiral symmetry. The first column indicates the Cartan label of the symmetry class where `NS' indicates a nonsymmorphic subclass. Labels $U_T$, $U_C$, $U_S$ and $U_P$ indicate the form of time-reversal, charge-conjugation, chiral and spatial-inversion symmetries in $k$ space, where `$I$' is the identity matrix and `$0$' indicates the absence of symmetry. 
Columns `$\tilde{d}_x$' and `$\tilde{d}_y$' indicate components of the canonical form $\tilde{H} (k,s) = \mbox{\boldmath$\sigma$} \cdot {\bf \tilde{d}}$ where intracell spacing $s=0$ for symmorphic systems, $s=a/2$ for nonsymmorphic systems.}
\begin{tabular}{ L{1.4cm} | C{0.7cm} | C{0.7cm} | C{0.7cm} | C{0.7cm} | C{5.5cm} | C{6.0cm}}
\hline
Class & $U_T$ & $U_C$ & $U_S$ & $U_P$ & $\tilde{d}_x$ & $\tilde{d}_y$ \\ 
\hline \hline
AIII & $0$ & $0$ & $\sigma_z$ & $0$ & $\frac{x_{0}}{2} + \sum_{n=1}^{\infty} [ x_{n} \cos (kna) + \tilde{x}_{n} \sin (kna) ]$ & $\frac{y_{0}}{2} + \sum_{n=1}^{\infty} [ y_{n} \cos (kna) + \tilde{y}_{n} \sin (kna) ]$ \\ 
\hline
BDI & $I$ & $\sigma_z$ & $\sigma_z$ & $\sigma_x$ & $\frac{x_{0}}{2} + \sum_{n=1}^{\infty} x_{n} \cos (kna)$ & $\sum_{n=1}^{\infty} \tilde{y}_{n} \sin (kna)$ \\ 
\hline
DIII & $\sigma_y$ & $\sigma_x$ & $\sigma_z$ & $\sigma_z$ & $\sum_{n=1}^{\infty} \tilde{x}_{n} \sin (kna)$ & $\sum_{n=1}^{\infty} \tilde{y}_{n} \sin (kna)$ \\ 
\hline
CI & $\sigma_x$ & $\sigma_y$ & $\sigma_z$ & $I$ & $\frac{x_{0}}{2} + \sum_{n=1}^{\infty} x_{n} \cos (kna)$ & $\frac{y_{0}}{2} + \sum_{n=1}^{\infty} y_{n} \cos (kna)$ \\
\hline
A (NS) & $0$ & $0$ & $\sigma_z$ & $0$ & $\frac{x_{0}}{2} + \sum_{n=1}^{\infty} [ x_{n} \cos (kna) + \tilde{x}_{n} \sin (kna) ]$ & $\sum_{n=0}^{\infty} [ \beta_n \cos [ ka (n+\tfrac{1}{2} ) ] + \tilde{\beta}_n \sin [ ka (n+\tfrac{1}{2} ) ]]$ \\
\hline
AI (NS) & $\sigma_x$ & $\sigma_y$  & $\sigma_z$ & $I$ & $\frac{x_{0}}{2} + \sum_{n=1}^{\infty} x_{n} \cos (kna)$ & $\sum_{n=0}^{\infty} \beta_n \cos [ ka (n+\tfrac{1}{2} ) ]$ \\
\hline
D (NS) & $\sigma_z$ & $I$ & $\sigma_z$ & $\sigma_y$ & $\sum_{n=1}^{\infty} \tilde{x}_{n} \sin (kna)$ & $\sum_{n=0}^{\infty} \beta_n \cos [ ka (n+\tfrac{1}{2} ) ]$ \\
\hline
AIII (NS) & $\sigma_x$ & $\sigma_y$  & $\sigma_z$ & $I$ & $\sum_{n=0}^{\infty} \alpha_n \cos [ ka (n+\tfrac{1}{2} ) ]$ & $\sum_{n=0}^{\infty} \beta_n \cos [ ka (n+\tfrac{1}{2} ) ]$ \\
\hline
\end{tabular}
\end{center}
\end{table*}
\renewcommand{\arraystretch}{1.2}

\begin{table*}[t]
\begin{center}
\caption{\label{tabless2}Symmorphic models with chiral symmetry showing how their parameters, Table~\ref{tabless}, correspond to the canonical forms in Table~\ref{tablescf}.}
\begin{tabular}{ L{1.2cm} | C{1.5cm} | C{3.0cm} | C{3.0cm} | C{1.5cm} | C{3.0cm} | C{3.0cm}}
\hline
model & $x_0$ & $x_1$ & $\tilde{x}_1$ & $y_0$ & $y_1$ & $\tilde{y}_1$ \\
\hline \hline
$H_{\mathrm{S,AIII}}^{(0,0,z)}$ & $2v\cos \phi_v$  & $w\cos \phi_w+t_3\cos \phi_3$  & $-w \sin \phi_w-t_3\sin \phi_3$ & $-2v\sin \phi_v$ & $w\sin \phi_w-t_3\sin \phi_3$  & $w \cos \phi_w-t_3\cos \phi_3$  \\ [1pt]
\hline
$H_{\mathrm{S,AIII}}^{(0,0,y)}$ & $2v$ & $2w \cos \phi_w$ & $-2w \sin \phi_w$ & $-2u$ & $-2t_{AA} \cos \phi_{AA}$ & $2t_{AA} \sin \phi_{AA}$  \\ [1pt]
\hline
$H_{\mathrm{S,AIII}}^{(0,0,x)}$ & $2u$ & $2t_{AA} \cos \phi_{AA}$ & $-2t_{AA} \sin \phi_{AA}$ & $2v$ & $-2w \sin \phi_w$ & $-2w \cos \phi_w$  \\ [1pt]
\hline \hline
$H_{\mathrm{S,BDI}}^{(I,z,z)}$ & $2v$ & $w+t_3$ & $0$ & $0$ & $0$ & $w-t_3$  \\ [1pt]
\hline
$H_{\mathrm{S,BDI}}^{(x,z,y)}$ & $2v$ & $2w$ & $0$ & $0$ & $0$ & $2t_{AA}$  \\ [1pt]
\hline
$H_{\mathrm{S,BDI}}^{(I,x,x)}$ & $2u$ & $2t_{AA}$ & $0$ & $0$ & $0$ & $-2w$  \\ [1pt]
\hline
$H_{\mathrm{S,BDI}}^{(z,I,z)}$ & $2v$ & $w-t_3$ & $0$ & $0$ & $0$ & $w+t_3$  \\ [1pt]
\hline
$H_{\mathrm{S,BDI}}^{(x,I,x)}$ & $2v$ & $2w$ & $0$ & $0$ & $0$ & $-2t_{AA}$  \\ [1pt]
\hline
$H_{\mathrm{S,BDI}}^{(z,x,y)}$ & $2u$ & $-2t_{AA}$ & $0$ & $0$ & $0$ & $-2w$  \\ [1pt]
\hline \hline
$H_{\mathrm{S,DIII}}^{(y,x,z)}$ & $0$ & $0$ & $-2w\sin\phi_w$ & $0$ & $0$ & $2w\cos\phi_w$  \\ [1pt]
\hline
$H_{\mathrm{S,DIII}}^{(y,z,x)}$ & $0$ & $0$ & $-2t_{AA}$ & $0$ & $0$ & $-2w$  \\ [1pt]
\hline
$H_{\mathrm{S,DIII}}^{(y,I,y)}$ & $0$ & $0$ & $-2w$ & $0$ & $0$ & $2t_{AA}$  \\ [1pt]
\hline \hline
$H_{\mathrm{S,CI}}^{(x,y,z)}$ & $2v\cos \phi_v$ & $2w\cos \phi_w$ & $0$ & $-2v\sin \phi_v$ & $2w\sin \phi_w$ & $0$  \\ [1pt]
\hline
$H_{\mathrm{S,CI}}^{(I,y,y)}$ & $2v$ & $2w$ & $0$ & $-2u$ & $-2t_{AA}$ & $0$  \\ [1pt]
\hline
$H_{\mathrm{S,CI}}^{(z,y,x)}$ & $2u$ & $2t_{AA}$ & $0$ & $2v$ & $-2w$ & $0$  \\ [1pt]
\hline
\end{tabular}
\end{center}
\end{table*}

\section{Symmorphic models with chiral symmetry}\label{s:s}

\subsection{Symmetry class AIII with $T^2 = 0$, $C^2 = 0$, $S^2 = 1$}\label{s:a3}

All versions of chiral symmetry require $d_0 = 0$. From Eq.~(\ref{cgrm2}), this may be satisfied with $\epsilon_0 = 0$, $t_{BB} = - t_{AA}$ and $\phi_{BB} = \phi_{AA}$. Hence, with $d_0 = 0$, we consider
\begin{eqnarray}
H (k,s) &=& \mbox{\boldmath$\sigma$} \cdot {\bf d}  (k,s) , \label{cgrm4} \\ 
d_x (k,s) &=& v \cos (ks + \phi_v) + w \cos [k(a-s) + \phi_w] \nonumber \\
&& \qquad + t_3 \cos [k(a+s) + \phi_3] , \nonumber \\
d_y (k,s) &=& - v \sin (ks + \phi_v) + w \sin [k(a-s) + \phi_w] \nonumber \\
&& \qquad - t_3 \sin [k(a+s) + \phi_3] , \nonumber \\
d_z (k,s) &=& u + 2t_{AA} \cos (ka + \phi_{AA})  , \nonumber
\end{eqnarray}
with ${\bf d} = (0,d_x,d_y,d_z)$.

We begin by considering models with symmorphic chiral symmetry only (i.e., generally no time-reversal or charge-conjugation symmetry) which belong to the AIII (chiral unitary) symmetry class.
To identify such models, we determine the cases for which $H(k,0)$ satisfies chiral symmetry~(\ref{sk}) with $U_S (k,0)$ being independent of $k$. We find there are three such models, corresponding to $U_S (k,0) = \sigma_z$, $\sigma_y$, or $\sigma_x$, and denoted $H_{\mathrm{S,AIII}}^{(0,0,z)}$, $H_{\mathrm{S,AIII}}^{(0,0,y)}$, $H_{\mathrm{S,AIII}}^{(0,0,x)}$, respectively, Fig.~\ref{figchiral1}.
The values of the tight-binding parameters for each model are summarized in Table~\ref{tabless}.
Note that we could consider a more general form of chiral symmetry with $U_S (k,s=0)$ consisting of a linear combination of Pauli matrices, but this would lead to a more constrained version of the Hamiltonians that we already consider, so we exclude this possibility

The three models, $H_{\mathrm{S,AIII}}^{(0,0,i)}$, $i = x,y,z$, are distinct in the atomic basis, but they are all equivalent after a similarity transformation into the chiral basis, i.e., the basis in which the chiral operator is $\tilde{U}_S (k,0) = \sigma_z$ in $k$ space. For each model we identify the unitary operator $R$ for which the similarity transformation may be expressed as
\begin{eqnarray}
\tilde{H}_{\mathrm{S,AIII}}^{(0,0,i)} (k,0) = R^{\dagger} H_{\mathrm{S,AIII}}^{(0,0,i)}(k,0) R \equiv \mbox{\boldmath$\sigma$} \cdot {\bf \tilde{d}} , \label{sim1}
\end{eqnarray}
so that the Hamiltonian $\tilde{H}_{\mathrm{S,AIII}}^{(0,0,i)} (k,0)$ is written in the canonical form where ${\tilde d}_x$ and ${\tilde d}_y$ are represented by the Fourier series of arbitrary, real $2\pi$ periodic functions (with ${\tilde d}_z = 0$), as shown in Table~\ref{tablescf}.
This canonical form has chiral symmetry $\tilde{U}_S (k,0) = \sigma_z$ in $k$ space ($S_z$ in position space).
Table~\ref{tabless2} shows how the tight-binding parameters of the three models correspond to $x_n$, $\tilde{x}_n$, $y_n$, $\tilde{y}_n$ of the canonical form,~Table~\ref{tablescf}.
Energies eigenvalues are $E_{\pm}(k) = \pm \sqrt{{\tilde d}_x^2 + {\tilde d}_y^2}$, and the system is generally an insulator.
Since spatial-inversion symmetry is broken, the spectrum generally isn't an even function of $k$.
As ${\tilde d}_x$ and ${\tilde d}_y$ are arbitrary, real $2\pi$ periodic functions,
the path of the ${\bf \tilde{d}}$ vector in the ${\tilde d}_x$-${\tilde d}_y$ plane defines an integer $\mathbb{Z}$ winding number, in agreement with expectations from the tenfold way classification~\cite{schnyder08,kitaev09,ryu10,chiu16}. We describe the winding number in more detail in Section~\ref{s:bd1}.

In the following, we briefly describe each model, $H_{\mathrm{S,AIII}}^{(0,0,i)}$, $i = x,y,z$, in more detail.
Chiral symmetry $U_S (k,0) = \sigma_z$ demands that $d_z (k,0) = 0$. For arbitrary $k$, this requires $u=0$ and $t_{AA} = 0$, yielding $H_{\mathrm{S,AIII}}^{(0,0,z)} (k,s) = \mbox{\boldmath$\sigma$} \cdot {\bf d}^{(z)} (k,s)$ where
\begin{eqnarray*}
d_x^{(z)} (k,s) &=& v \cos (ks + \phi_v) + w \cos [k(a-s) + \phi_w] \\
&& \qquad + t_3 \cos [k(a+s) + \phi_3] , \\
d_y^{(z)} (k,s) &=& - v \sin (ks + \phi_v) + w \sin [k(a-s) + \phi_w] \\
&& \qquad  - t_3 \sin [k(a+s) + \phi_3] , \\
d_z^{(z)} (k,s) &=& 0 .
\end{eqnarray*}
This can be viewed as a generalized version of the SSH model~\cite{ssh79,ssh80,velasco17}: arbitrary, complex A-B hoppings are allowed as long as all A-A or B-B hoppings are absent, as shown schematically in Fig.~\ref{figchiral1}(a) for intracell spacing $s=0$.
It is an unusual case because $U_S (k,s) = \sigma_z$ is independent of $k$ for all intracell spacing $s$, including $s \neq 0$.
In position space, chiral symmetry is given by $S_z$, Eq.~(\ref{sz}).
Note that we include hopping $t_3$ because, for $s=0$, it is a nearest-neighbor hopping, Fig.~\ref{figchiral1}(a) (and the generalization to longer-range hopping is straightforward, Appendix~\ref{a:longrange}). In the absence of $t_3$ or longer range hoppings, i.e., with only ${\bar v} = ve^{i\phi_v}$ and ${\bar w} = we^{i\phi_w}$, then it is possible to gauge away the phases $\phi_v$ and $\phi_w$~\cite{matveeva22}, giving the conventional SSH model in symmetry class BDI, Fig.~\ref{figchiral1}(b).

Chiral symmetry $U_S (k,0) = \sigma_y$ demands that $d_y = 0$.
This can be achieved with $\phi_v = 0$, $t_3 = w$ and $\phi_3 = \phi_w$,
yielding $H_{\mathrm{S,AIII}}^{(0,0,y)} (k,0) = \mbox{\boldmath$\sigma$} \cdot {\bf d}^{(y)} (k,0)$ where
\begin{eqnarray*}
d_x^{(y)} (k,0) &=& v + 2 w \cos (ka + \phi_w) , \\
d_y^{(y)} (k,0) &=& 0 , \\
d_z^{(y)} (k,0) &=& u + 2t_{AA} \cos (ka + \phi_{AA}) .
\end{eqnarray*}
This can be viewed as a generalized version of the Creutz ladder~\cite{creutz99,guo16,junemann17}, Fig.~\ref{figchiral1}(c). Note that it is possible to find $H_{\mathrm{S,AIII}}^{(0,0,y)} (k,s)$ for arbitrary $s$ using the unitary transformation~(\ref{uks1}).
This can be transformed to the canonical form, Table~\ref{tablescf}, with parameters given in Table~\ref{tabless2} using $R=R_y R_r$, where
\begin{eqnarray}
R_y = \frac{1}{\sqrt{2}}
\begin{pmatrix}
1 & 1 \\
i & -i 
\end{pmatrix} , \label{ry}
\end{eqnarray}
and $R_r = \mathrm{diag} (1,i)$. Here, $R_y$ transforms to the chiral basis and $R_r$ rotates the $d_x$-$d_y$ plane around the $d_z$ axis.
In the atomic basis in position space, a transformation to the chiral basis is
$\tilde{H}_{\mathrm{S,AIII}}^{(0,0,y)} = {\mathcal R}_r^{\dagger} {\mathcal R}_y^{\dagger} H_{\mathrm{S,AIII}}^{(0,0,y)} {\mathcal R}_y {\mathcal R}_r$
where
\begin{eqnarray*}
{\mathcal R}_y \!=\! \frac{1}{\sqrt{2}} \begin{pmatrix}
1 & 1 & 0 & 0 & 0 & \cdots  \\
i & -i & 0  & 0 & 0 & \cdots \\
0 & 0  & 1 & 1 & 0 & \cdots \\
0 & 0 & i & -i & 0 & \cdots \\
0 & 0 & 0 & 0 & 1 & \cdots \\
\vdots & \vdots & \vdots & \vdots & \vdots & \ddots
\end{pmatrix} \!\! ,
\end{eqnarray*}
and ${\mathcal R}_r = \mathrm{diag} (1,i,1,i,1,i,\ldots)$.
With this transformation, $\tilde{H}_{\mathrm{S,AIII}}^{(0,0,y)}$ is of the same general (canonical) form as $H_{\mathrm{S,AIII}}^{(0,0,z)}$.

Chiral symmetry $U_S (k,0) = \sigma_x$ demands that $d_x = 0$ which can be achieved with $\phi_v = \pi/2$, $t_3 = - w$ and $\phi_3 = \phi_w$,
yielding $H_{\mathrm{S,AIII}}^{(0,0,x)} (k,0) = \mbox{\boldmath$\sigma$} \cdot {\bf d}^{(x)} (k,0)$ where
\begin{eqnarray*}
d_x^{(x)} (k,0) &=& 0 , \\
d_y^{(x)} (k,0) &=& - v + 2 w \sin (ka + \phi_w) , \\
d_z^{(x)} (k,0) &=& u + 2t_{AA} \cos (ka + \phi_{AA}) .
\end{eqnarray*}
This can be viewed as a generalized version of the Shockley $sp$ orbital model~\cite{shockley39,vanderbilt93,fuchs21}, Fig.~\ref{figchiral1}(e).
It can be transformed to the canonical form, Table~\ref{tablescf}, with parameters given in Table~\ref{tabless2} using $R=R_x$, where
\begin{eqnarray}
R_x = \frac{1}{\sqrt{2}}
\begin{pmatrix}
1 & 1 \\
1 & -1 
\end{pmatrix} . \label{rx}
\end{eqnarray}
In the atomic basis in position space, a transformation to the chiral basis is
$\tilde{H}_{\mathrm{S,AIII}}^{(0,0,x)} = {\mathcal R}_x^{\dagger} H_{\mathrm{S,AIII}}^{(0,0,x)} {\mathcal R}_x$
where
\begin{eqnarray*}
{\mathcal R}_x \!=\! \frac{1}{\sqrt{2}} \begin{pmatrix}
1 & 1 & 0 & 0 & 0 & \cdots  \\
1 & -1 & 0  & 0 & 0 & \cdots \\
0 & 0  & 1 & 1 & 0 & \cdots \\
0 & 0 & 1 & -1 & 0 & \cdots \\
0 & 0 & 0 & 0 & 1 & \cdots \\
\vdots & \vdots & \vdots & \vdots & \vdots & \ddots
\end{pmatrix} \!\! .
\end{eqnarray*}
With this transformation, $\tilde{H}_{\mathrm{S,AIII}}^{(0,0,x)}$ is of the same general (canonical) form as $H_{\mathrm{S,AIII}}^{(0,0,z)}$.

\subsection{Symmetry class BDI with $T^2=1$, $C^2=1$, $S^2=1$}\label{s:bd1}

We apply time-reversal symmetry~(\ref{tk}) to the three models with chiral symmetry described in Section~\ref{s:a3}.
We consider $U_T (k,0)$ to be independent of $k$, and this means that 
$U_T (k,0) = I$, $\sigma_x$ or $\sigma_z$ (for which $T^2 = +1$), or $U_T (k,0) = \sigma_y$ (for which $T^2 = -1$).
Then, with definite chiral symmetry and time-reversal symmetry, we determine the form of charge-conjugation symmetry $U_C$ using $U_S (k,0) =U_C^{\ast} (k,0) U_T (-k,0)$, where $C^2 = \pm 1$.
We then group models in the same symmetry class together, beginning with the BDI (chiral orthogonal) class, $T^2 = 1$, $C^2 = 1$,  $S^2 = 1$.

In this class, the canonical form of the Bloch Hamiltonian $\tilde{H}_{\mathrm{S,BDI}} (k,0) = \mbox{\boldmath$\sigma$} \cdot {\bf \tilde{d}}$
is given in Table~\ref{tablescf} where ${\tilde d}_x$ is represented as an even-in-$k$ $2\pi$ periodic function, 
${\tilde d}_y$ is an odd $2\pi$ periodic function,
and ${\tilde d}_z = 0$,
where parameters $x_n$ and $\tilde{y}_n$ are real.
Note that a form with ${\tilde d}_x$ odd and ${\tilde d}_y$ even is equivalent~\cite{canonicalnote}, because it is related by a rotation of the ${\tilde d}_x$-${\tilde d}_y$ axes about ${\tilde d}_z$.
Ref.~\cite{chen20} expressed the SSH model in an equivalent form with ${\tilde d}_x + i {\tilde d}_y$ written as a complex Fourier series.
The canonical form has time-reversal symmetry $U_T (k,0) = I$, charge-conjugation symmetry $U_C (k,0) = \sigma_z$, and chiral symmetry $U_S (k,0) = \sigma_z$.

\begin{figure}[t]
\includegraphics[scale=0.39]{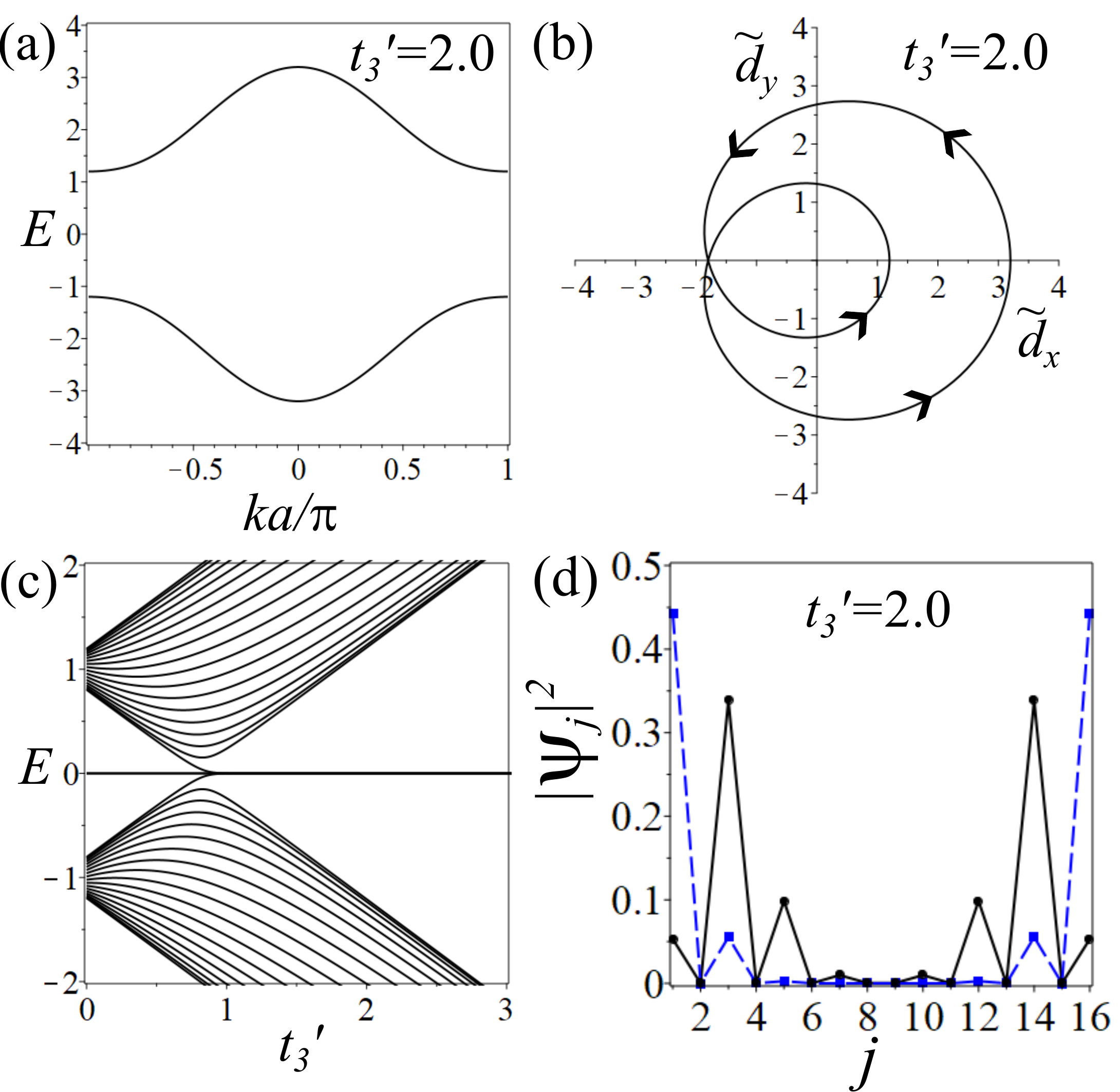}
\caption{Edge states and winding number for the SSH model, $H_{\mathrm{S,BDI}}^{(I,z,z)}$, in the BDI symmorphic symmetry class ($T^2=1$, $C^2=1$, $S^2= 1$).
(a) Bulk band structure $E(k)$ dominated by second order terms in the canonical form, Table~\ref{tablescf}.
(b) Is the corresponding trajectory of the ${\bf \tilde{d}}$ vector in the ${\tilde d}_x$-${\tilde d}_y$ plane defining a winding number of $W = 2$.
(c) Energy levels $E$ in position space as a function of parameter $t_3^{\prime}$ with $J=40$ atoms.
(d) The probability density $|\psi_j|^2$ per site $j=1,2,\ldots,16$ for the four zero energy states in the SSH model for $t_3^{\prime} = 2.0$ and $J=16$ atoms. Note that the levels come in pairs with equal probability densities, so only two of the four are distinguishable.
Parameter values are $t_3 = t_5 = 0$, $v = 0.2$, and $w=1.0$. In (a), (b), (d), $t_3^{\prime}$ is fixed, $t_3^{\prime} = 2.0$. In terms of the canonical form, Table~\ref{tablescf}, then $x_0 = 2v = 0.4$, $x_1 = \tilde{y}_1 = w = 1.0$, and $x_2 = \tilde{y}_2 = t_3^{\prime}$ with $x_n = \tilde{y}_n = 0$ for $n \geq 3$.}\label{figwind1}
\end{figure}

Energy eigenvalues are $E_{\pm} (k) = \pm \sqrt{{\tilde d}_x^2 + {\tilde d}_y^2}$, and the system is generally an insulator, Fig.~\ref{figbulk1}(a).
The path of the ${\bf \tilde{d}}$ vector in the ${\tilde d}_x$-${\tilde d}_y$ plane defines an integer $\mathbb{Z}$ winding number $W$, in agreement with expectations from the tenfold way classification~\cite{schnyder08,kitaev09,ryu10,chiu16}.
As a hypothetical example, Fig.~\ref{figwind1}(a) shows the band structure when the system is dominated by second order terms $x_2$, $\tilde{y}_2$ in the canonical form, Table~\ref{tablescf}, and Fig.~\ref{figwind1}(b) shows the corresponding path of the ${\bf \tilde{d}}$ vector with a winding number of $W = 2$.

In position space, one can set ${\cal S} = S_x$, $S_y$, or $S_z$ with ${\cal T} = I$, $S_x$, or $S_z$ and ${\cal C} = I$, $S_x$, or $S_z$, and search for combinations of them; there are only six possibilities because ${\cal T} \neq S_y$, ${\cal C} \neq S_y$, Table~\ref{tabless}.
Of these, three are related by diagonal gauge transformations as seen by multiplying two of ${\cal S}$, ${\cal T}$, or ${\cal C}$ by $S_z$ (which is diagonal).
The three remaining distinct models have all been discussed widely in previous literature: $H_{\mathrm{S,BDI}}^{(I,z,z)}$ is the SSH model~\cite{ssh79,ssh80}, $H_{\mathrm{S,BDI}}^{(x,z,y)}$ is the Creutz ladder~\cite{creutz99,li15,guo16,zurita21}, and $H_{\mathrm{S,BDI}}^{(I,x,x)}$ is the Shockley $sp$ orbital model~\cite{shockley39,vanderbilt93,fuchs21}, as shown schematically in Fig.~\ref{figchiral1}. They are distinct in the sense that the unitary transformation from the atomic basis to the canonical form, Table~\ref{tablescf}, is off-diagonal~(\ref{ry},\ref{rx}).

The three models related by diagonal gauge transformations correspond to an additional version of each of the distinct three such that their A-B hopping parameters are imaginary rather than real, Table~\ref{tabless}.
Formally, the diagonal transformation in the atomic basis in position space has a period of $2a$,
\begin{eqnarray}
{\mathcal R}_i \!=\! \begin{pmatrix}
1 & 0 & 0 & 0 & 0 & \cdots \\
0 & -i & 0  & 0 & 0 & \cdots \\
0 & 0  & -1 & 0 & 0 & \cdots \\
0 & 0 & 0 & i & 0 & \cdots \\
0 & 0 & 0 & 0 & 1 & \cdots \\
\vdots & \vdots & \vdots & \vdots & \vdots & \ddots
\end{pmatrix} \!\! , \label{ri}
\end{eqnarray}
so that $H_{\mathrm{S,BDI}}^{(z,I,z)} = {\mathcal R}_i^{\dagger} H_{\mathrm{S,BDI}}^{(I,z,z)} {\mathcal R}_i$, with a flip of sign of $t_3$, etc.

Table~\ref{tabless2} shows how the tight-binding parameters of the models correspond to $x_n$, $\tilde{y}_n$ of the canonical form, Table~\ref{tablescf}. Since each model consists of nearest-neighbor only, $x_n = \tilde{y}_n = 0$ for $n \geq 2$, the winding number $W$ can only take two values ($0$ or $1$)~\cite{asboth16,matveeva22}.
In terms of the canonical form, Table~\ref{tablescf}, it is given by
\begin{eqnarray}
W = \begin{cases}
0 &\mbox{if } |x_0| > 2 |x_1| , \\
1 & \mbox{if } |x_0| < 2 |x_1| ,
\end{cases}
\end{eqnarray}
and the system is gapless if $|x_0| = 2 |x_1|$.
Expressions for $x_0$, $x_1$ in terms of tight-binding parameters for different models may be read off Table~\ref{tabless2}. For example, for the SSH model, $H_{\mathrm{S,BDI}}^{(I,z,z)}$, then $W = 1$ if $|v| < |w+t_3|$.
The SSH model is usually considered with $t_3 = 0$~\cite{ssh79,ssh80,asboth16,cayssol21}, but we include it because it is a nearest-neighbor hopping for intracell spacing $s=0$, and because its inclusion does not affect the symmetry classification of the model.
It is straightforward to generalize the models to longer range hoppings, as described in Appendix~\ref{a:longrange}, and terms beyond nearest-neighbor hopping in the SSH model have been described previously~\cite{li14,perezgonzalez19,chen20,hetenyi21}.
Note that the spinless Kitaev chain~\cite{kitaev01,kitaev09,guo16,matveeva22} also belongs to the BDI symmetry class, but we do not discuss it here because superconducting pairing is not included in the generalized Rice-Mele model~(\ref{cgrm1},\ref{cgrm2}).

In the atomic basis in position space, edge states will be located on different atomic sites in the three models. Given an eigenstate $\psi_{\mathrm{SSH}}$ of the SSH model, $H_{\mathrm{S,BDI}}^{(I,z,z)}$, a corresponding state in the Creutz ladder, $H_{\mathrm{S,BDI}}^{(x,z,y)}$, is ${\mathcal R}_y {\mathcal R}_r \psi_{\mathrm{SSH}}$ and in the Shockley $sp$ orbital model, $H_{\mathrm{S,BDI}}^{(I,x,x)}$, it is ${\mathcal R}_x \psi_{\mathrm{SSH}}$.
The SSH model is usually considered for $t_3 =0$, which, according to Table~\ref{tabless2}, is equivalent to $t_{AA} = w$ in the Creutz ladder and $t_{AA} = -w$ in the Shockley $sp$ orbital model. Then, in the fully dimerized limit of $v=0$, one can see from Fig.~\ref{figchiral1}(b) that there will be an edge state fully localized on the left most A atom in the SSH model~\cite{asboth16}. Owing to the off-diagonal transformations to the bases of the other models, such a state will be shared equally between the left-most A and left-most B atom in the same limit of the Creutz ladder ($v=0$) and the Shockley $sp$ orbital model ($u=0$), Fig.~\ref{figchiral1}(d) and (f).

As an illustration, Fig.~\ref{figwind1}(c) shows the energy levels in position space of the SSH model, $H_{\mathrm{S,BDI}}^{(I,z,z)}$, as a function of parameter $t_3^{\prime}$ with $t_3 = t_5 = 0$. With these choices, parameters of the canonical form are $x_0 = 2v$, $x_1 = \tilde{y}_1 = w$ and $x_2 = \tilde{y}_2 = t_3^{\prime}$. With $w=1$ and $v=0.2$, then there is a phase transition from winding number $W=2$ for $t_3^{\prime} >1$ to $W=1$ for $t_3^{\prime} < 1$. As shown in Fig.~\ref{figwind1}(c) and (d), these phases support four zero-energy edge states for $t_3^{\prime} >1$ and two zero-energy edge states for $t_3^{\prime} < 1$.

\begin{figure}[t]
\includegraphics[scale=0.45]{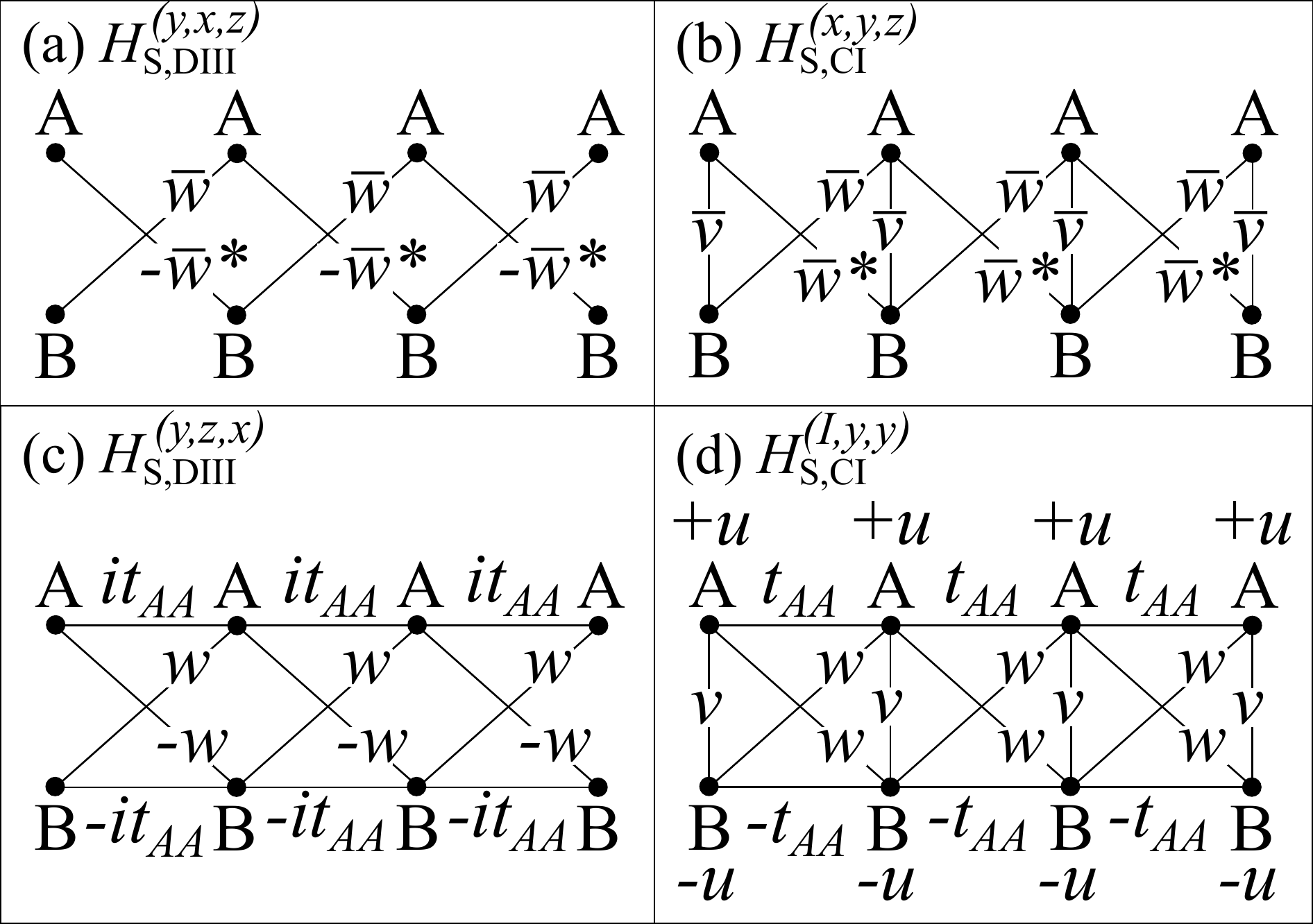}
\caption{Tight-binding models in the DIII symmetry class ($T^2 = -1$, $C^2 = 1$, $S^2 = 1$) on the left, in the CI symmetry class ($T^2 = 1$, $C^2 = -1$, $S^2 = 1$) on the right, shown with intracell spacing $s=0$. 
(a) Shows $H_{\mathrm{S,DIII}}^{(y,x,z)}$ with complex intercell hoppings $\pm we^{\pm i\phi_w}$. As shown here, with only nearest neighbor hopping, $H_{\mathrm{S,DIII}}^{(y,x,z)}$ is simply two disconnected linear chains; higher-order hopping is required to couple the chains, Table~\ref{tableAs}.
(b) Shows $H_{\mathrm{S,CI}}^{(x,y,z)}$ with complex intracell hopping $ve^{i\phi_v}$, and complex intercell hoppings $we^{\pm i\phi_w}$.
(c) $H_{\mathrm{S,DIII}}^{(y,z,x)}$ with imaginary, staggered second-neighbor hopping $\pm it_{AA}$, and intercell hoppings $\pm w$.
(d) $H_{\mathrm{S,CI}}^{(I,y,y)}$ with staggered onsite energies $\pm u$, staggered second-neighbor hopping $\pm t_{AA}$, intracell hopping $v$ and intercell hopping $w$.
A finite lateral spacing between A and B orbitals is shown for clarity, but this separation can be zero without loss of generality.
}\label{figchiral2}
\end{figure}

\subsection{Symmetry class DIII with $T^2=-1$, $C^2=1$, $S^2=1$}\label{s:d3}

Models with chiral symmetry $S^2=1$, time $T^2=-1$, and charge conjugation $C^2=1$ are in the DIII symmetry class which is usually associated with a Bogoliubov de Gennes Hamiltonian.
This symmetry class must have time ${\cal T} = S_y$. There are three possibilities for chiral symmetry, ${\cal S}= S_x$, $S_y$, or $S_z$, giving three models, Table~\ref{tabless}. However, multiplying ${\cal S}$ and ${\cal C}$ by $S_z$ shows that one of them is trivially related to another by a diagonal gauge transformation.
Thus, there are two distinct models ($H_{\mathrm{S,DIII}}^{(y,x,z)}$ and $H_{\mathrm{S,DIII}}^{(y,z,x)}$) as shown schematically in Fig.~\ref{figchiral2}. They are distinct in the sense that the unitary transformation relating them is off-diagonal.
Model $H_{\mathrm{S,DIII}}^{(y,z,x)}$ has an additional version $H_{\mathrm{S,DIII}}^{(y,I,y)}$ that is the one related by a trivial, diagonal transformation such that its A-B hopping parameters are imaginary rather than real, Table~\ref{tabless}.

In this class, the canonical form of the Bloch Hamiltonian $\tilde{H}_{\mathrm{sym}} (k,0) = \mbox{\boldmath$\sigma$} \cdot {\bf \tilde{d}}$
is where both ${\tilde d}_x$ and ${\tilde d}_y$ are represented by odd-in-$k$ $2\pi$ periodic functions, Table~\ref{tablescf},
parameters $\tilde{x}_n$ and $\tilde{y}_n$ are real, and ${\tilde d}_z (k,0) = 0$.
Table~\ref{tabless2} shows how the tight-binding parameters of the three models correspond to $\tilde{x}_n$, $\tilde{y}_n$ of the canonical form, Table~\ref{tablescf}. Since each model consists of nearest-neighbor only, $\tilde{x}_n = \tilde{y}_n = 0$ for $n \geq 2$, although it is straightforward to generalize to longer range hoppings, as described in Appendix~\ref{a:longrange}.

Energy eigenvalues are $E_{\pm} (k) = \pm \sqrt{{\tilde d}_x^2 + {\tilde d}_y^2}$. Owing to Kramer's degeneracy (with $T^2=-1$), states at $k=0$ and $k= \pm \pi/a$ are degenerate because these $k$ values are time-reversal invariant. As there are only two bands and the bands are degenerate, the system is gapless, Fig.~\ref{figbulk1}(e).

As a representative of this symmetry class, let us briefly discuss $H_{\mathrm{S,DIII}}^{(y,x,z)}$, Fig.~\ref{figchiral2}(a).
With only nearest-neighbor coupling, $\pm we^{\pm i\phi_w}$, the model is simply two disconnected linear chains; higher-order hopping is required to couple the chains, Table~\ref{tableAs}.
As time $T^2 = -1$, the A atoms can be considered as spin-up electrons and the B atoms as spin-down electrons, with the coupling ($\pm we^{\pm i\phi_w}$) describing spin-orbit coupling~\cite{bychkovrashba84,mii14} between them,
\begin{eqnarray*}
H_{\mathrm{S,DIII}}^{(y,x,z)} (k,0) = \begin{pmatrix}
0 & -2iwe^{i\phi_w} \sin (ka) \\
2iwe^{-i\phi_w} \sin (ka) & 0
\end{pmatrix} .
\end{eqnarray*}
Owing to the presence of chiral symmetry, there is no coupling between spins of the same orientation (A-A or B-B) here, however it has been taken into account previously~\cite{mii14} as $d_0 (k,0) = 2t \cos (ka)$ for nearest-neighbor A-A and B-B hopping with parameter $t = t_{AA} = t_{BB}$.

\begin{figure}[t]
\includegraphics[scale=0.43]{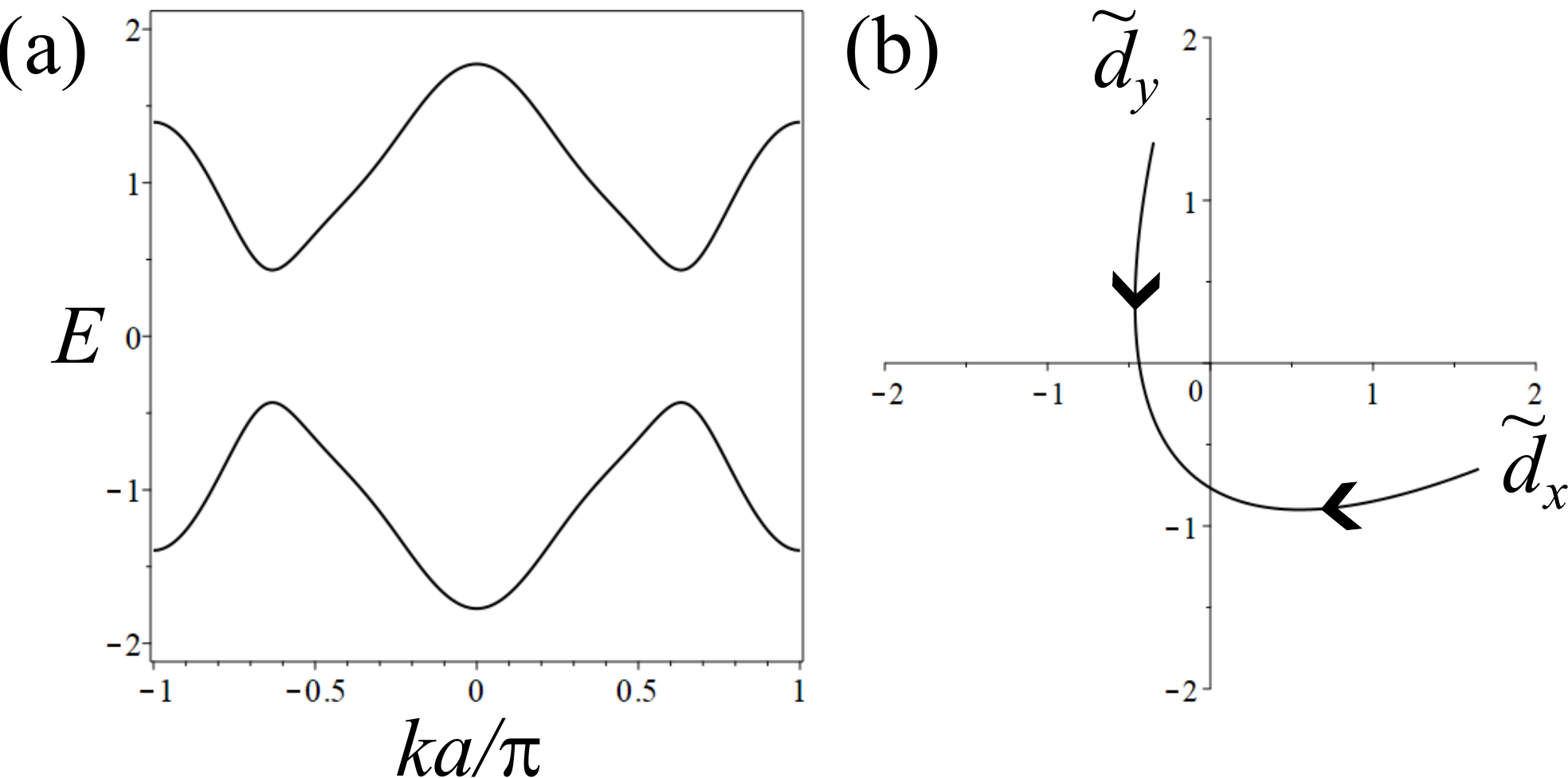}
\caption{(a) Typical band structure $E(k)$ in the CI symmetry class ($T^2 = 1$, $C^2 = -1$, $S^2 = 1$) using the canonical form, Table~\ref{tablescf}. Parameter values are $x_0 = 0.5$, $x_1 = 1.0$, $x_2 = 0.4$, $y_0 = - 0.3$, $y_1 = - 1.0$, $y_2 = 0.5$, with $x_n = y_n = 0$ for $n \geq 3$.
(b) is the corresponding trajectory of the ${\bf \tilde{d}}$ vector in the ${\tilde d}_x$-${\tilde d}_y$ plane. It traces an open path for $-\pi < ka \leq 0$ and then reverses along the same path for $0 < ka \leq \pi$ to end at the original starting point.
}\label{figwind2}
\end{figure}

\subsection{Symmetry class CI with $T^2=1$, $C^2=-1$, $S^2=1$}

Models with chiral symmetry $S^2=1$, time $T^2=1$, and charge conjugation $C^2=-1$ are in the CI symmetry class which is usually associated with a Bogoliubov de Gennes Hamiltonian.
In this class, the canonical form of the Bloch Hamiltonian $\tilde{H}_{\mathrm{sym}} (k,0) = \mbox{\boldmath$\sigma$} \cdot {\bf \tilde{d}}$
is where both ${\tilde d}_x$ and ${\tilde d}_y$ are represented by even-in-$k$ $2\pi$ periodic functions, Table~\ref{tablescf},
parameters $x_n$ and $y_n$ are real, and ${\tilde d}_z (k,0) = 0$.

Energy eigenvalues are $E_{\pm} (k) = \pm \sqrt{{\tilde d}_x^2 + {\tilde d}_y^2}$, and the system is generally an insulator, Fig.~\ref{figbulk1}(c).
In agreement with expectations from the tenfold way classification~\cite{schnyder08,kitaev09,ryu10,chiu16}, it is topologically trivial.
As a hypothetical example, Fig.~\ref{figwind2}(a) shows the band structure for typical parameter values using the canonical form, Table~\ref{tablescf}, and Fig.~\ref{figwind2}(b) shows the corresponding path of the ${\bf \tilde{d}}$ vector in the ${\tilde d}_x$-${\tilde d}_y$ plane. It traces an open path for $-\pi < ka \leq 0$ and then reverses along the same path for $0 < ka \leq \pi$ to end at the original starting point.
Note the dips in the band structure at $ka \approx \pm \pi/2$ arising from higher-order terms in the canonical form, Table~\ref{tablescf}, which would result in van Hove singularities in the density of states.

This symmetry class must have charge conjugation ${\cal C} = S_y$. There are three possibilities for chiral symmetry, ${\cal S}= S_x$, $S_y$, or $S_z$, giving three models, Table~\ref{tabless}. However, multiplying ${\cal S}$ and ${\cal T}$ by $S_z$ shows that one of them is trivially related by a diagonal gauge transformation.
Thus, there are two distinct models ($H_{\mathrm{S,CI}}^{(x,y,z)}$ and $H_{\mathrm{S,CI}}^{(I,y,y)}$) as shown schematically in Fig.~\ref{figchiral2}. They are distinct in the sense that the unitary transformation relating them is off-diagonal.
Model $H_{\mathrm{S,CI}}^{(I,y,y)}$ has an additional version $H_{\mathrm{S,CI}}^{(z,y,x)}$ that is the one related by a trivial, diagonal transformation such that its A-B hopping parameters are imaginary rather than real, Table~\ref{tabless}. Table~\ref{tabless2} shows how the tight-binding parameters of the three models correspond to $x_n$, $y_n$ of the canonical form, Table~\ref{tablescf}. Since each model consists of nearest-neighbor only, $x_n = y_n = 0$ for $n \geq 2$, although it is straightforward to generalize to longer range hoppings, as described in Appendix~\ref{a:longrange}.

Although we only consider systems with two energy bands, these models fall into a Bogoliubov de Gennes symmetry class (CI) which nominally has four components. In Ref.~\cite{altland97}, the generic $4 \times 4$ Hamiltonian representing class CI is actually block diagonal with a single $2 \times 2$ block representing spin-up electrons and spin-down holes (say). Comparison with the form of $H_{\mathrm{S,CI}}^{(I,y,y)}$ shows that we may interpret the chain of A atoms (with onsite energies $-u$ and A-A hopping $t_{AA}$) as mimicking the spin-up electrons and the chain of B atoms (with onsite energies $u$ and B-B hopping $-t_{AA}$) as mimicking the spin-down holes.

\begin{table*}[t]
\begin{center}
\caption{\label{tablesns}Symmorphic models without chiral symmetry showing the values of the tight-binding parameters of the generalized Rice-Mele model Eq.~(\ref{cgrm2}), for each model. The same models with long-range hoppings are shown in Table~\ref{tableS}.}
\begin{tabular}{ L{1.2cm} || C{1.0cm} | C{1.0cm} | C{1.0cm}  | C{1.0cm} | C{1.0cm}| C{1.0cm} | C{1.0cm} | C{1.0cm} | C{1.0cm} | C{1.0cm} | C{1.0cm} | C{1.0cm}}
\hline
name & $\epsilon_0$ & $u$ & $t_{AA}$ & $\phi_{AA}$ &  $t_{BB}$ & $\phi_{BB}$ & $v$ & $\phi_v$ & $w$ & $\phi_w$ & $t_3$ & $\phi_3$  \\ [1pt]
\hline \hline
$H_{\mathrm{S,AI}}^{(I,0,0)}$ & arb. & arb. & arb. & $0$ & arb. & $0$ & arb. & $0$ & arb. & $0$ & arb. & $0$ \\ [1pt]
\hline
$H_{\mathrm{S,AI}}^{(x,0,0)}$ & arb. & $0$ & arb. & arb. & $t_{AA}$ & $-\phi_{AA}$ & arb. & arb. & arb. & arb. & $w$ & $-\phi_w$ \\ [1pt]
\hline
$H_{\mathrm{S,AI}}^{(z,0,0)}$ & arb. & arb. & arb. & $0$ & arb. & $0$ & arb. & $\pi/2$ & arb. & $\pi/2$ & arb. & $\pi/2$ \\ [1pt]
\hline \hline
$H_{\mathrm{S,AII}}^{(y,0,0)}$ & arb. & $0$ & arb. & arb. & $t_{AA}$ & $-\phi_{AA}$ & $0$ & n/a & arb. & arb. & $-w$ & $-\phi_w$ \\ [1pt]
\hline \hline
$H_{\mathrm{S,D}}^{(0,x,0)}$ & $0$ & arb. & arb. & arb. & $-t_{AA}$ & $-\phi_{AA}$ & $0$ & n/a & arb. & arb. & $-w$ & $-\phi_w$ \\ [1pt]
\hline
$H_{\mathrm{S,D}}^{(0,z,0)}$ & $0$ & $0$ & arb. & $\pi/2$ & arb. & $\pi/2$ & arb. & $0$ & arb. & $0$ & arb. & $0$ \\ [1pt]
\hline 
$H_{\mathrm{S,D}}^{(0,I,0)}$ & $0$ & $0$ & arb. & $\pi/2$ & arb. & $\pi/2$ & arb. & $\pi/2$ & arb. & $\pi/2$ & arb. & $\pi/2$ \\ [1pt]
\hline \hline
$H_{\mathrm{S,C}}^{(0,y,0)}$ & $0$ & arb. & arb. & arb. & $-t_{AA}$ & $-\phi_{AA}$ & arb. & arb. & arb. & arb. & $w$ & $-\phi_w$ \\ [1pt]
\hline
\end{tabular}
\end{center}
\end{table*}

\renewcommand{\arraystretch}{1.3}
\begin{table*}[t]
\begin{center}
\caption{\label{tablesns3}Canonical forms of the Bloch Hamiltonian $H(k)$ for noninteracting, one-dimensional tight-binding models with two bands and no chiral symmetry. The first column indicates the Cartan label of the symmetry class where `NS' indicates a nonsymmorphic subclass. Column $U_i$ indicates the form of either time-reversal ($i=T$) or charge-conjugation ($i=C$) symmetry in $k$ space, where $I$ is the identity matrix. 
Columns `$\tilde{d}_0$', `$\tilde{d}_x$', `$\tilde{d}_y$', and `$\tilde{d}_z$' indicate components of the canonical form $\tilde{H} (k,s) = \mbox{\boldmath$\sigma$} \cdot {\bf \tilde{d}}$ where intracell spacing $s=0$ for symmorphic systems, $s=a/2$ for nonsymmorphic systems.}
\begin{tabular}{ L{1.1cm} | C{1.4cm} | C{3.6cm} | C{3.6cm} | C{3.6cm}  | C{3.6cm}}
\hline
Class & $U_i$ & $\tilde{d}_0$ & $\tilde{d}_x$ & $\tilde{d}_y$ & $\tilde{d}_z$ \\
\hline \hline
AI & $U_T=I$ & $\frac{a_0}{2} + \sum_{n=1}^{\infty} a_n \cos (kna)$ & $\frac{x_0}{2} + \sum_{n=1}^{\infty} x_n \cos (kna)$ & $\sum_{n=1}^{\infty} \tilde{y}_n \sin (kna)$ & $\frac{z_0}{2} + \sum_{n=1}^{\infty} z_n \cos (kna)$ \\
\hline
AII & $U_T=\sigma_y$ & $\frac{a_0}{2} + \sum_{n=1}^{\infty} a_n \cos (kna)$ & $\sum_{n=1}^{\infty} \tilde{x}_n \sin (kna)$ & $\sum_{n=1}^{\infty} \tilde{y}_n \sin (kna)$ & $\sum_{n=1}^{\infty} \tilde{z}_n \sin (kna)$ \\
\hline
D & $U_C=\sigma_x$ & $\sum_{n=1}^{\infty} \tilde{a}_{n} \sin (kna)$ & $\sum_{n=1}^{\infty} \tilde{x}_{n} \sin (kna)$ & $\sum_{n=1}^{\infty} \tilde{y}_{n} \sin (kna)$ & $\frac{z_{0}}{2} + \sum_{n=1}^{\infty} z_{n} \cos (kna)$ \\
\hline
C & $U_C=\sigma_y$ & $\sum_{n=1}^{\infty} \tilde{a}_{n} \sin (kna)$ & $\frac{x_{0}}{2} + \sum_{n=1}^{\infty} x_{n} \cos (kna)$ & $\frac{y_{0}}{2} + \sum_{n=1}^{\infty} y_{n} \cos (kna)$ & $\frac{z_{0}}{2} + \sum_{n=1}^{\infty} z_{n} \cos (kna)$ \\
\hline
A (NS) & $U_T=\sigma_x$ & $\frac{a_0}{2} + \sum_{n=1}^{\infty} a_n \cos (kna)$ & $\sum_{n=0}^{\infty} \alpha_n \cos \left[ ka \left(n+\tfrac{1}{2} \right) \right]$ & $\sum_{n=0}^{\infty} \beta_n \cos \left[ ka \left(n+\tfrac{1}{2} \right) \right]$ & $\sum_{n=1}^{\infty} {\tilde z}_n \sin (kna)$ \\
\hline
A (NS) & $U_C=\sigma_y$ & $\sum_{n=1}^{\infty} \tilde{a}_n \sin (kna)$ & $\sum_{n=0}^{\infty} \alpha_n \cos \left[ ka \left(n+\tfrac{1}{2} \right) \right]$ & $\sum_{n=0}^{\infty} \beta_n \cos \left[ ka \left(n+\tfrac{1}{2} \right) \right]$ & $\frac{z_0}{2} + \sum_{n=1}^{\infty} z_n \cos (kna)$ \\
\hline
\end{tabular}
\end{center}
\end{table*}
\renewcommand{\arraystretch}{1.2}

\begin{table*}[t]
\begin{center}
\caption{\label{tablesns2}Symmorphic models without chiral symmetry showing how their tight-binding parameters correspond to the canonical form, Eqs.~(\ref{ad0})-(\ref{adz}) and Table~\ref{tablesns3}.}
\begin{tabular}{ L{0.4cm} || C{1.4cm} | C{2.1cm} | C{1.4cm} || C{2.1cm} || C{2.1cm} | C{1.6cm} | C{1.6cm} || C{2.1cm}}
\hline
& $H_{\mathrm{S,AI}}^{(I,0,0)}$ & $H_{\mathrm{S,AI}}^{(x,0,0)}$ &  $H_{\mathrm{S,AI}}^{(z,0,0)}$ & $H_{\mathrm{S,AII}}^{(y,0,0)}$ & $H_{\mathrm{S,D}}^{(0,x,0)}$ & $H_{\mathrm{S,D}}^{(0,z,0)}$ & $H_{\mathrm{S,D}}^{(0,I,0)}$ & $H_{\mathrm{S,C}}^{(0,y,0)}$ \\
\hline \hline
$a_0$ &  $2\epsilon_0$ & $2\epsilon_0$ &  $2\epsilon_0$ & $2\epsilon_0$ & $0$ & $0$ & $0$ & $0$ \\
\hline
$a_1$ &  $t_{AA}+t_{BB}$ & $2t_{AA} \cos \phi_{AA}$ &  $t_{AA}+t_{BB}$ & $2t_{AA} \cos \phi_{AA}$ & $0$ & $0$ & $0$ & $0$ \\
\hline
$\tilde{a}_1$ &  $0$ & $0$ & $0$ & $0$ & $-2t_{AA} \sin \phi_{AA}$ & $-t_{AA}-t_{BB}$ & $-t_{AA}-t_{BB}$ & $-2t_{AA} \sin \phi_{AA}$ \\
\hline
$x_0$ &  $2v$ & $-2v\sin\phi_v$ & $2u$ & $0$ & $0$ & $0$ & $0$ & $2v\cos \phi_v$ \\
\hline
$x_1$ &  $w+t_3$ & $2w\sin\phi_w$ &  $t_{AA}-t_{BB}$ & $0$ & $0$ & $0$ & $0$ & $2w\cos \phi_w$ \\
\hline
$\tilde{x}_1$ &  $0$ & $0$ & $0$  & $-2w \sin \phi_w$& $-2w\sin \phi_w$ & $-t_{AA}+t_{BB}$ & $-t_{AA}+t_{BB}$ & $0$ \\
\hline
$y_0$ &  $0$ & $0$ & $0$ & $0$ & $0$ & $0$ & $0$ & $-2v\sin \phi_v$ \\
\hline
$y_1$ &  $0$ & $0$ & $0$ & $0$ & $0$ & $0$ & $0$ & $2w\sin \phi_w$ \\
\hline
$\tilde{y}_1$ & $w-t_3$ & $-2t_{AA} \sin \phi_{AA}$ & $-w-t_3$ & $2w\cos\phi_w$ & $2w\cos \phi_w$ & $-w+t_3$ & $-w-t_3$ & $0$ \\
\hline
$z_0$ &  $2u$ & $2v\cos\phi_v$ & $-2v$ & $0$ & $2u$ & $2v$ & $-2v$ & $2u$ \\
\hline
$z_1$ &  $t_{AA}-t_{BB}$ & $2w\cos\phi_w$ & $w-t_3$ & $0$ & $2t_{AA} \cos \phi_{AA}$ & $w+t_3$ & $w-t_3$ & $2t_{AA}\cos \phi_{AA}$ \\
\hline
$\tilde{z}_1$ &  $0$ & $0$ & $0$ & $-2t_{AA} \sin \phi_{AA}$ & $0$ & $0$ & $0$ & $0$ \\
\hline
\end{tabular}
\end{center}
\end{table*}

\begin{figure}[t]
\includegraphics[scale=0.44]{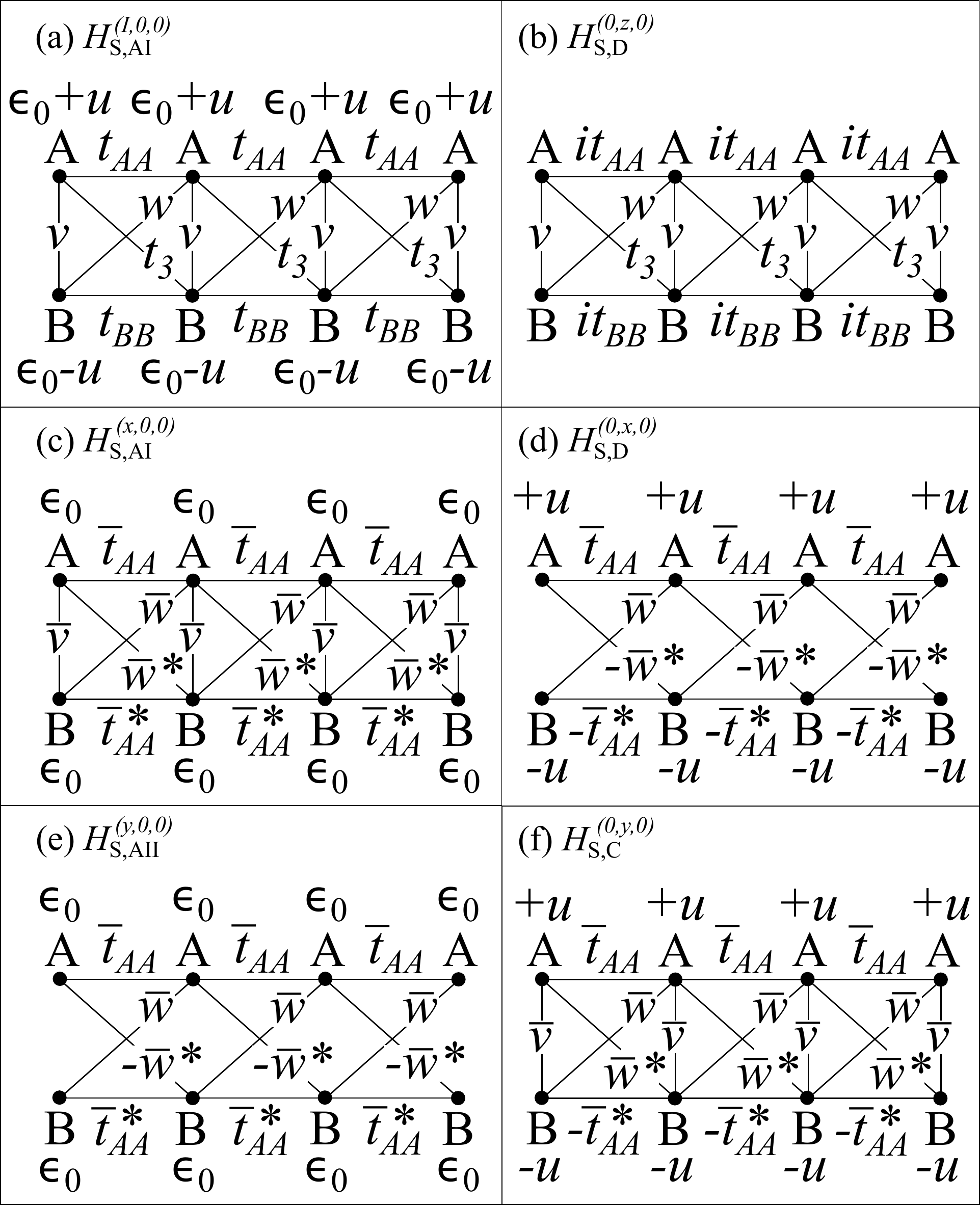}
\caption{Symmorphic models with either time-reversal symmetry (left side) or charge-conjugation (right side) symmetry only.
(a) Shows $H_{\mathrm{S,AI}}^{(I,0,0)}$ ($T^2=1$) with arbitrary, real hopping parameters.
(b) Shows $H_{\mathrm{S,D}}^{(0,z,0)}$ ($C^2=1$) with imaginary A-A and B-B hopping, but real A-B hopping parameters.
(c) $H_{\mathrm{S,AI}}^{(x,0,0)}$ ($T^2=1$) with complex intra- and intercell hopping parameters.
(d) $H_{\mathrm{S,D}}^{(0,x,0)}$ ($C^2=1$) with alternating on-site energies, complex intra- and intercell hopping parameters.
(e) $H_{\mathrm{S,AII}}^{(y,0,0)}$ ($T^2=-1$) with complex intercell hopping parameters.
(f) $H_{\mathrm{S,C}}^{(0,y,0)}$ ($C^2=-1$) with alternating on-site energies, complex intra- and intercell hopping parameters.
Bars over parameters indicate that they are complex numbers, in general.
A finite lateral spacing between A and B orbitals is shown for clarity, but this separation can be zero without loss of generality.
}\label{figchiral3}
\end{figure}

\section{Symmorphic models without chiral symmetry}\label{s:sns}

\subsection{Symmetry class A with $T^2 = 0$, $C^2 = 0$, $S^2 = 0$}\label{s:a}

We now consider symmorphic models without chiral symmetry. Symmetry class A is described by the generalized Rice-Mele model Eqs.~(\ref{cgrm1},\ref{cgrm2}), and the corresponding expression with tight-binding parameters of all ranges is given in Eqs.(\ref{d0all})-(\ref{dzall}). For intracell spacing $s=0$, Eq.~(\ref{cgrm2}) shows that, without any symmetry constraints, each component of the ${\bf d}$ vector at $s=0$ may be written as an arbitrary real Fourier series of a $2\pi$-periodic function.
Energies eigenvalues are $E_{\pm}(k) = {\tilde d}_0 \pm \sqrt{{\tilde d}_x^2 + {\tilde d}_y^2 + {\tilde d}_z^2}$, and the system is generally an insulator.
Since time-reversal and spatial-inversion symmetries are absent, the spectrum generally isn't an even function of $k$.

We write the canonical form for this class as $H_{\mathrm{sym,A}} (k,s) = \mbox{\boldmath$\sigma$} \cdot {\bf \tilde{d}} (k,s)$ where
\begin{eqnarray}
{\tilde d}_0 (k,0) \!&=&\! \frac{a_0}{2} + \sum_{n=1}^{\infty} \left[ a_n \cos (kna) + \tilde{a}_{n} \sin (kna) \right] \! , \label{ad0} \\
{\tilde d}_x (k,0) \!&=&\! \frac{x_{0}}{2} + \sum_{n=1}^{\infty} \left[ x_{n} \cos (kna) + \tilde{x}_{n} \sin (kna) \right] \! , \label{adx} \\
{\tilde d}_y (k,0) \!&=&\! \frac{y_{0}}{2} + \sum_{n=1}^{\infty} \left[ y_{n} \cos (kna) + \tilde{y}_{n} \sin (kna) \right] \! , \label{ady} \\
{\tilde d}_z (k,0) \!&=&\! \frac{z_{0}}{2} + \sum_{n=1}^{\infty} \left[  z_{n} \cos (kna) + \tilde{z}_{n} \sin (kna)  \right] \! , \label{adz}
\end{eqnarray}
where all the parameters are real. This canonical form expresses the fact that, without any symmetry constraints, each component of the ${\bf d}$ vector may be written as an arbitrary real Fourier series of a $2\pi$-periodic function of $ka$.
With nearest-neighbor parameters only, the parameters from the canonical form~(\ref{ad0})-(\ref{adz}) may be read from Eqs.~(\ref{cgrm1},\ref{cgrm2}), e.g. $a_0 = 2 \epsilon_0$, $z_0 = 2u$, wherein parameters of second order or above are zero due to the nearest-neighbor approximation.

\subsection{Symmetry class AI with $T^2 = 1$, $C^2 = 0$, $S^2 = 0$}\label{s:a1}

Models with time-reversal symmetry $T^2 = 1$, but no charge-conjugation or chiral symmetry are in the AI (orthogonal) class.
There are three possibilities, ${\cal T}= I$, $S_x$, or $S_z$, giving three models, Table~\ref{tablesns}.
However, multiplying ${\cal T}$ by $S_z$ shows that one of them is trivially related to another by a diagonal gauge transformation.
Thus, there are two distinct models ($H_{\mathrm{S,AI}}^{(I,0,0)}$ and $H_{\mathrm{S,AI}}^{(x,0,0)}$) as shown schematically in Fig.~\ref{figchiral3}. They are distinct in the sense that the unitary transformation relating them is off-diagonal.
Model $H_{\mathrm{S,AI}}^{(I,0,0)}$ has an additional version $H_{\mathrm{S,AI}}^{(z,0,0)}$ such that its A-B hopping parameters are imaginary rather than real, Table~\ref{tablesns}.
The Rice-Mele model~\cite{ricemele82} is $H_{\mathrm{S,AI}}^{(I,0,0)}$ with $\epsilon_0 = t_3 = t_{AA} = t_{BB} = 0$, i.e., the SSH model with alternating onsite energies $\pm u$.
The SSH model with additional real next-nearest neighbor coupling discussed in~\cite{li14} is the same as $H_{\mathrm{S,AI}}^{(I,0,0)}$ with $\epsilon_0 = u = t_3 = 0$ and the replacements  $v = t_1$, $w= t_2$, $t_{AA} = t_A$, and $t_{BB} = t_B$, where our parameters are listed first.
The tight-binding model on a triangular lattice considered in~\cite{schulze13} is the same as $H_{\mathrm{S,AI}}^{(I,0,0)}$ with $t_{BB} = t_3 = 0$ and the replacements $\epsilon_0 = (\epsilon_A + \epsilon_B)/2$, $u = (\epsilon_A - \epsilon_B)/2$, $t_{AA} = t_3$, $v = t_1$, and $w = t_2$, where our parameters are listed first.
The extended Creutz ladder discussed in~\cite{sun17} is the same as $H_{\mathrm{S,AI}}^{(x,0,0)}$ with $\epsilon_0 = \phi_w = 0$ and the replacements  $v = -J_Y$, $\phi_v = - \phi$, $w= -J_D$, $t_{AA} = -J_X$, and $\phi_{AA} = - \theta$, where our parameters are listed first.

The canonical form for this class, Table~\ref{tablesns3}, may be written as for class A with ${\tilde d}_0 (k,0)$, ${\tilde d}_x (k,0)$, and ${\tilde d}_z (k,0)$ as even functions of $k$, and ${\tilde d}_y (k,0)$ odd, i.e., set $\tilde{a}_n = \tilde{x}_n = y_n = \tilde{z}_n = 0$ in Eqs.~(\ref{ad0})-(\ref{adz}).
Then, time reversal symmetry is given by ${\cal T}= I$.
Relations of the tight-binding parameters for each model to the canonical form are given in Table~\ref{tablesns2}, where $H_{\mathrm{S,AI}}^{(x,0,0)}$ is rotated by $R_y^{-1}$ and $H_{\mathrm{S,AI}}^{(z,0,0)}$ is rotated by $R_y$.
Energies eigenvalues are $E_{\pm}(k) = {\tilde d}_0 \pm \sqrt{{\tilde d}_x^2 + {\tilde d}_y^2 + {\tilde d}_z^2}$, and the system is generally an insulator.
With time-reversal symmetry, the spectrum is an even function of $k$.

Although this class does not satisfy chiral symmetry, it is known that the Rice-Mele model~\cite{ricemele82} satisfies a combination of chirality and spatial inversion~\cite{chiu16,arkinstall17,allen22} as $\sigma_y H(k) \sigma_y = -H(-k)$. This is satisfied by $H_{\mathrm{S,AI}}^{(I,0,0)}$ with $d_0 = 0$, i.e., with $\epsilon_0 = 0$ and $t_{BB} = - t_{AA}$. Note that this additional symmetry doesn't affect the topological indices~\cite{shiozaki14} of these one-dimensional models as predicted by the ten-fold way based solely on nonspatial symmetries~\cite{schnyder08,kitaev09,ryu10,chiu16} .

\subsection{Symmetry class AII with $T^2 = -1$, $C^2 = 0$, $S^2 = 0$}\label{s:a2}

Models with time-reversal symmetry $T^2 = -1$, but no charge-conjugation or chiral symmetry are in the AII (symplectic) class.
There is only one possibility, ${\cal T}= S_y$, giving $H_{\mathrm{S,AII}}^{(y,0,0)}$, Table~\ref{tablesns} and Fig.~\ref{figchiral3}(e).
The canonical form for this class, Table~\ref{tablesns3}, may be written as for class A with ${\tilde d}_0 (k,0)$ as an even function of $k$, and ${\tilde d}_x (k,0)$, ${\tilde d}_y (k,0)$, ${\tilde d}_z (k,0)$ as odd functions, i.e., set $\tilde{a}_n = x_n = y_n = z_n = 0$ in Eqs.~(\ref{ad0})-(\ref{adz}).
Relations of the tight-binding parameters for $H_{\mathrm{S,AII}}^{(y,0,0)}$ to the canonical form are given in Table~\ref{tablesns2}.
Energy eigenvalues are $E_{\pm}(k) = {\tilde d}_0 \pm \sqrt{{\tilde d}_x^2 + {\tilde d}_y^2 + {\tilde d}_z^2}$. Owing to Kramer's degeneracy (with $T^2=-1$), states at $k=0$ and $k= \pm \pi/a$ are degenerate because these $k$ values are time-reversal invariant. As there are only two bands and the bands are degenerate, the system is gapless.

\begin{figure}[t]
\includegraphics[scale=0.39]{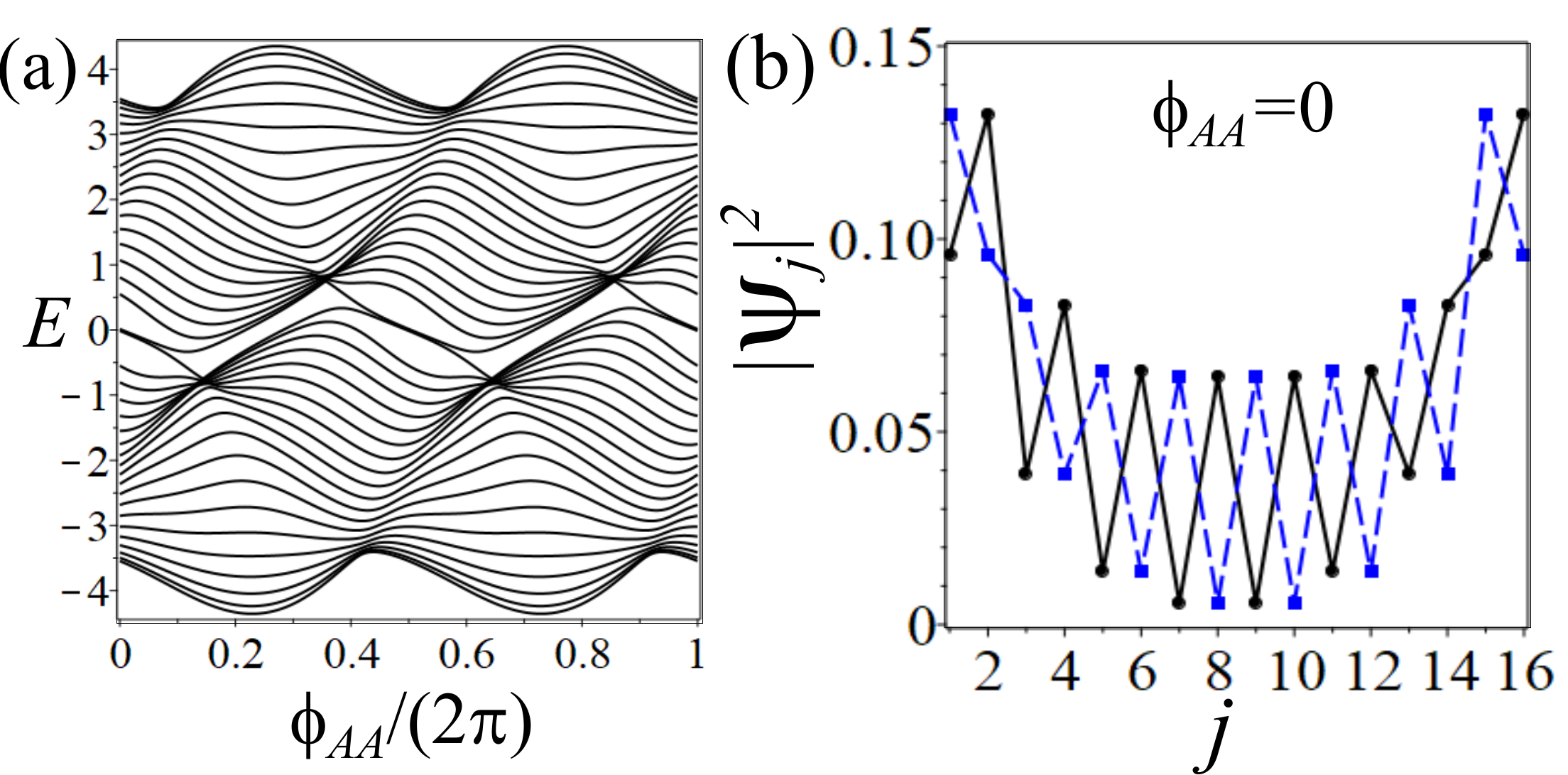}
\caption{(a) Energy levels $E$ in position space as a function of phase $\phi_{AA}$ of model $H_{\mathrm{S,D}}^{(0,x,0)}$ in the D symmorphic symmetry class ($T^2=0$, $C^2=1$, $S^2= 0$) with $J=40$ atoms.
(b) The probability density $|\psi_j|^2$ per site $j=1,2,\ldots,16$ for the two zero energy states for $\phi_{AA} = 0$ and $J=16$ atoms.
Other parameter values are $u = \sqrt{2}$, $t_{AA} = 1$, $w=1$, and $\phi_w = \pi/4$.
}\label{figedge1}
\end{figure}

\subsection{Symmetry class D with $T^2 = 0$, $C^2 = 1$, $S^2 = 0$}\label{s:d}

Models with charge-conjugation symmetry $C^2 = 1$, but no time-reversal or chiral symmetry are in the D class.
There are three possibilities, ${\cal C}= I$, $S_x$, or $S_z$, giving three models, Table~\ref{tablesns}.
However, multiplying ${\cal C}$ by $S_z$ shows that one of them is trivially related to another by a diagonal gauge transformation.
Thus, there are two distinct models ($H_{\mathrm{S,D}}^{(0,x,0)}$ and $H_{\mathrm{S,D}}^{(0,z,0)}$) as shown schematically in Fig.~\ref{figchiral3}. They are distinct in the sense that the unitary transformation relating them is off-diagonal.
Model $H_{\mathrm{S,D}}^{(0,z,0)}$ has an additional version $H_{\mathrm{S,D}}^{(0,I,0)}$ such that its A-B hopping parameters are imaginary rather than real, Table~\ref{tablesns}.
The SSH model with additional imaginary next-nearest neighbor coupling discussed in~\cite{hetenyi21} is the same as $H_{\mathrm{S,D}}^{(0,z,0)}$ with $t_3 = 0$ and $t_{BB} = - t_{AA}$ (hence $d_0=0$) and the replacements  $v = - J$, $w= - J^{\prime}$, $t_{AA} = -K$, and $t_{BB} = K$, where our parameters are listed first.

The canonical form for this class, Table~\ref{tablesns3}, may be written as for class A with ${\tilde d}_z (k,0)$ as an even function of $k$, and ${\tilde d}_0 (k,0)$,  ${\tilde d}_x (k,0)$, and ${\tilde d}_y (k,0)$ as odd functions, i.e., set $a_n = x_n = y_n = \tilde{z}_n = 0$ in Eqs.~(\ref{ad0})-(\ref{adz}).
Then, charge-conjugation symmetry is given by ${\cal C}= S_x$.
Relations of the tight-binding parameters for each model to the canonical form are given in Table~\ref{tablesns2}, where $H_{\mathrm{S,D}}^{(0,z,0)}$ is rotated by $R_x$ and $H_{\mathrm{S,D}}^{(0,I,0)}$ is rotated by $R_y$.
Energy eigenvalues are $E_{\pm}(k) = {\tilde d}_0 \pm \sqrt{{\tilde d}_x^2 + {\tilde d}_y^2 + {\tilde d}_z^2}$ and the system is generally an insulator.
With charge-conjugation symmetry, $E_{\pm}(-k) = - E_{\mp}(k)$.

Despite the absence of chiral symmetry, the constraint of charge-conjugation symmetry protects a $\mathbb{Z}_2$ topological index~\cite{li16} in class D in accord with predictions of the tenfold way classification~\cite{schnyder08,kitaev09,ryu10,chiu16}.
It may be interpreted in terms of the three-dimensional path of $(\tilde{d}_x, \tilde{d}_y, \tilde{d}_z )$ for $-\pi < ka \leq \pi$~\cite{li16}. With the canonical form, Table~\ref{tablesns3}, components $\tilde{d}_x$ and $\tilde{d}_y$ are odd functions of a $2\pi$-periodic function. Thus, the closed path begins and ends on either the north or south pole of the Bloch sphere, $\tilde{d}_z (-\pi) = \tilde{d}_z (\pi) = \frac{z_{0}}{2} + \sum_{n=1}^{\infty} (-1)^n z_{n}$, and it is also at either the north or south pole at $k=0$, $\tilde{d}_z (0) = \frac{z_{0}}{2} + \sum_{n=1}^{\infty} z_{n}$.
The $\mathbb{Z}_2$ topological index, $\nu_2$, may be defined~\cite{li16} as 
\begin{eqnarray}
\nu_2 = \begin{cases}
0 &\mbox{if } \mathrm{sgn} \;\!\!\big[ \tilde{d}_z (\pi) \big] = \mathrm{sgn} \;\!\!\big[  \tilde{d}_z (0) \big] , \\
1 & \mbox{if } \mathrm{sgn} \;\!\!\big[ \tilde{d}_z (\pi) \big] = - \mathrm{sgn} \;\!\!\big[ \tilde{d}_z (0) \big] .
\end{cases}
\end{eqnarray}
This can be written in terms of the canonical form including all orders of hopping using the expression for $\tilde{d}_z$,
\begin{eqnarray}
\nu_2 = \begin{cases}
0 &\mbox{if } |\frac{z_{0}}{2} + \sum_{m=1}^{\infty} z_{2m}| > |\sum_{m=0}^{\infty} z_{2m+1}| , \\
1 & \mbox{if } |\frac{z_{0}}{2} + \sum_{m=1}^{\infty} z_{2m}| < |\sum_{m=0}^{\infty} z_{2m+1}| ,
\end{cases}
\end{eqnarray}
For nearest-neighbor terms only (first order), the expression for $\nu_2$ simplifies as
\begin{eqnarray}
\nu_2 = \begin{cases}
0 &\mbox{if } |z_0| > 2 |z_1| , \\
1 & \mbox{if } |z_0| < 2 |z_1| ,
\end{cases}
\end{eqnarray}
and the system is gapless if $|z_0| = 2 |z_1|$.
Expressions for $z_0$ and $z_1$ in terms of tight-binding parameters for different models may be read off Table~\ref{tablesns2}. For example, for $H_{\mathrm{S,D}}^{(0,x,0)}$, then $\nu_2 = 1$ if $|u| <  |2t_{AA} \cos \phi_{AA}|$.

As an illustration, Fig.~\ref{figedge1}(a) shows the energy levels in position space of model $H_{\mathrm{S,D}}^{(0,x,0)}$, as a function of phase $\phi_{AA}$ with $u = \sqrt{2}$, $t_{AA} = 1$, $w=1$, and $\phi_w = \pi/4$. With these choices, the index $\nu_2 = 1$ if $|\cos \phi_{AA} | > 1 / \sqrt{2}$, i.e., for $0 < \phi_{AA} < \pi/4$, $3\pi/4 < \phi_{AA} < 5\pi/4$ and $7\pi/4 < \phi_{AA} < 2\pi$. As shown in Fig.~\ref{figedge1}, this phase supports two zero energy edge states as opposed to none for $\nu_2 = 0$.
The edge states have weight on both the A and B sites, unlike the SSH model, Fig.~\ref{figwind1}(d).
Note that this model doesn't have chiral symmetry, so $d_0 \neq 0$, and the whole spectrum tends to oscillate with $\phi_{AA}$, Fig.~\ref{figedge1}(a).

\subsection{Symmetry class C with $T^2 = 0$, $C^2 = -1$, $S^2 = 0$}\label{s:c}

Models with charge-conjugation symmetry $C^2 = -1$, but no time-reversal or chiral symmetry are in the C class.
There is only one possibility, ${\cal C}= S_y$, giving $H_{\mathrm{S,C}}^{(0,y,0)}$, Table~\ref{tablesns} and Fig.\ref{figchiral3}(f).
The canonical form for this class, Table~\ref{tablesns3}, may be written as for class A with ${\tilde d}_0 (k,0)$ as an odd function of $k$, and ${\tilde d}_x (k,0)$, ${\tilde d}_y (k,0)$, ${\tilde d}_z (k,0)$ as even functions, i.e., set $a_n = \tilde{x}_n = \tilde{y}_n = \tilde{z}_n = 0$ in Eqs.~(\ref{ad0})-(\ref{adz}).
Relations of the tight-binding parameters for $H_{\mathrm{S,C}}^{(0,y,0)}$ to the canonical form are given in Table~\ref{tablesns2}.
Energy eigenvalues are $E_{\pm}(k) = {\tilde d}_0 \pm \sqrt{{\tilde d}_x^2 + {\tilde d}_y^2 + {\tilde d}_z^2}$.
The system is generally an insulator and, with charge-conjugation symmetry, $E_{\pm}(-k) = - E_{\mp}(k)$.

\begin{table*}[t]
\begin{center}
\caption{\label{tablenss}Nonsymmorphic models with chiral symmetry showing the values of the tight-binding parameters of the generalized Rice-Mele model with $d_0 = 0$, Eq.~(\ref{cgrm4}), $\epsilon_0 = 0$, $t_{BB} = - t_{AA}$ and $\phi_{BB} = \phi_{AA}$. We also set $t_3 = 0$. `arb.' indicates that the parameter can take any arbitrary real value. The second column `$\mathcal{P}$' shows the form of spatial-inversion symmetry in position space where `$e$' (`$o$') indicates an even (odd) number of orbitals, and `$0$' indicates the symmetry is absent altogether. The same models with long-range hoppings are shown in Table~\ref{tableAns}.}
\begin{tabular}{ L{1.2cm} | C{0.9cm} || C{1.0cm} | C{1.0cm} | C{1.0cm} | C{1.0cm} | C{1.0cm} | C{1.0cm} | C{1.0cm}}
\hline
name & $\mathcal{P}$ & $u$ & $t_{AA}$ & $\phi_{AA}$ & $v$ & $\phi_v$ & $w$ & $\phi_w$  \\ [1pt]
\hline \hline
$H_{\mathrm{NS,A}}^{(0,0,y)}$ & $0$ & arb. & arb. & arb. & arb. & arb. & $v$ & $\phi_v$ \\ [1pt]
\hline
$H_{\mathrm{NS,A}}^{(0,0,x)}$ & $0$  & arb. & arb. & arb. & arb. & arb. & $-v$ & $\phi_v$ \\ [1pt]
\hline \hline
$H_{\mathrm{NS,AI}}^{(I,y,y)}$ & $P_x (o)$  & arb. & arb. & $0$ & arb. & $0$ & $v$ & $0$ \\ [1pt]
\hline
$H_{\mathrm{NS,AI}}^{(z,x,y)}$ & $P_y (o)$  & arb. & arb. & $0$ & arb. & $\pi/2$ & $v$ & $\pi/2$ \\ [1pt]
\hline
$H_{\mathrm{NS,AI}}^{(I,x,x)}$ & $P_y (o)$  & arb. & arb. & $0$ & arb. & $0$ & $-v$ & $0$ \\ [1pt]
\hline
$H_{\mathrm{NS,AI}}^{(z,y,x)}$ & $P_x (o)$  & arb. & arb. & $0$ & arb. & $\pi/2$ & $-v$ & $\pi/2$ \\ [1pt]
\hline \hline
$H_{\mathrm{NS,D}}^{(x,z,y)}$ & $P_x (e)$  & $0$ & arb. & $\pi/2$ & arb. & $0$ & $v$ & $0$ \\ [1pt]
\hline
$H_{\mathrm{NS,D}}^{(y,I,y)}$ & $P_y (e)$  & $0$ & arb. & $\pi/2$ & arb. & $\pi/2$ & $v$ & $\pi/2$ \\ [1pt]
\hline
$H_{\mathrm{NS,D}}^{(y,z,x)}$ & $P_x (e)$  & $0$ & arb. & $\pi/2$ & arb. & $0$ & $-v$ & $0$ \\ [1pt]
\hline
$H_{\mathrm{NS,D}}^{(x,I,x)}$ & $P_y (e)$  & $0$ & arb. & $\pi/2$& arb. & $\pi/2$ & $-v$ & $\pi/2$ \\ [1pt]
\hline \hline
$H_{\mathrm{NS,AIII}}^{(x,y,z)}$ & $P_x (o)$  & $0$ & $0$ & n/a & arb. & arb. & $v$ & $-\phi_v$ \\ [1pt]
\hline
$H_{\mathrm{NS,AIII}}^{(y,x,z)}$ & $P_y (o)$  & $0$ & $0$ & n/a & arb. & arb. & $-v$ & $-\phi_v$ \\ [1pt]
\hline
\end{tabular}
\end{center}
\end{table*}

\begin{table*}[t]
\begin{center}
\caption{\label{tablenss2}Nonsymmorphic models with chiral symmetry showing how their parameters, Table~\ref{tablenss}, correspond to the canonical forms in Table~\ref{tablescf}.}
\begin{tabular}{ L{1.2cm} | C{0.5cm} | C{1.9cm} | C{2.1cm} | C{1.4cm}  | C{1.5cm} | C{1.5cm}}
\hline
model & $x_0$ & $x_1$ & $\tilde{x}_1$ & $\alpha_0$ & $\beta_0$ & $\tilde{\beta}_0$ \\
\hline \hline
$H_{\mathrm{NS,A}}^{(0,0,y)}$ & $2u$ & $2t_{AA} \cos \phi_{AA}$ & $-2t_{AA} \sin \phi_{AA}$ & $0$ & $2v \cos \phi_v$ & $-2v \sin \phi_v$ \\ [1pt]
\hline
$H_{\mathrm{NS,A}}^{(0,0,x)}$ & $2u$ & $2t_{AA} \cos \phi_{AA}$ & $-2t_{AA} \sin \phi_{AA}$ & $0$ & $2v \sin \phi_v$ & $2v \cos \phi_v$ \\ [1pt]
\hline \hline
$H_{\mathrm{NS,AI}}^{(I,y,y)}$ & $2u$ & $2t_{AA}$ & $0$ & $0$ & $2v$ & $0$ \\ [1pt]
\hline
$H_{\mathrm{NS,AI}}^{(z,x,y)}$ & $2u$ & $-2t_{AA}$ & $0$ & $0$ & $2v$ & $0$ \\ [1pt]
\hline
$H_{\mathrm{NS,AI}}^{(I,x,x)}$ & $2u$ & $-2t_{AA}$ & $0$ & $0$ & $2v$ & $0$ \\ [1pt]
\hline
$H_{\mathrm{NS,AI}}^{(z,y,x)}$ & $2u$ & $2t_{AA}$ & $0$ & $0$ & $2v$ & $0$ \\ [1pt]
\hline \hline
$H_{\mathrm{NS,D}}^{(x,z,y)}$ & $0$ & $0$ & $-2t_{AA}$ & $0$ & $2v$ & $0$ \\ [1pt]
\hline
$H_{\mathrm{NS,D}}^{(y,I,y)}$ & $0$ & $0$ & $2t_{AA}$ & $0$ & $2v$ & $0$ \\ [1pt]
\hline
$H_{\mathrm{NS,D}}^{(y,z,x)}$ & $0$ & $0$ & $2t_{AA}$ & $0$ & $2v$ & $0$ \\ [1pt]
\hline
$H_{\mathrm{NS,D}}^{(x,I,x)}$ & $0$ & $0$ & $-2t_{AA}$ & $0$ & $2v$ & $0$ \\ [1pt]
\hline \hline
$H_{\mathrm{NS,AIII}}^{(x,y,z)}$ & $0$ & $0$ & $0$ & $2v \cos \phi_v$ & $-2v \sin \phi_v$ & $0$ \\ [1pt]
\hline
$H_{\mathrm{NS,AIII}}^{(y,x,z)}$ & $0$ & $0$ & $0$ & $2v \cos \phi_v$ & $-2v \sin \phi_v$ & $0$ \\ [1pt]
\hline
\end{tabular}
\end{center}
\end{table*}

\begin{figure}[t]
\includegraphics[scale=0.34]{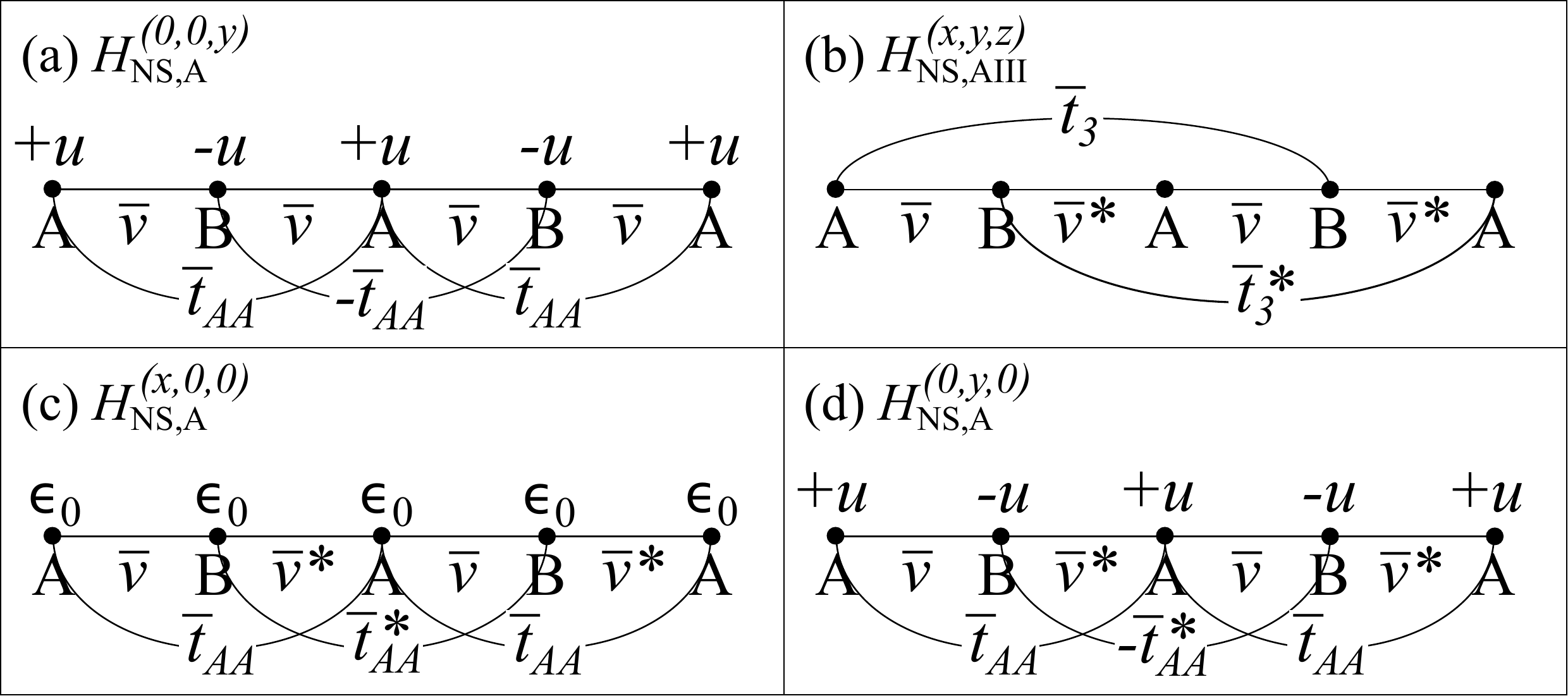}
\caption{Nonsymmorphic models.
(a) Shows $H_{\mathrm{NS,A}}^{(0,0,y)}$ ($S^2=\mathrm{NS}$) with alternating onsite energies $\pm u$, constant nearest-neighbor hopping ${\bar v}$, and alternating
second-nearest neighbor hopping $\pm {\bar t}_{AA}$.
(b) Shows $H_{\mathrm{NS,AIII}}^{(x,y,z)}$ ($T^2=\mathrm{NS}$, $C^2=\mathrm{NS}$, $S^2=1$) with alternating, complex nearest-neighbor coupling $ve^{\pm i \phi_v}$ and staggered, complex third nearest-neighbour couplings $t_3 e^{\pm i \phi_3}$ (we include $t_3$, otherwise the model is just a trivial linear chain).
(c) $H_{\mathrm{NS,A}}^{(x,0,0)}$ ($T^2=\mathrm{NS}$) with constant onsite energies $\epsilon_0$, alternating, complex nearest-neighbor coupling $ve^{\pm i \phi_v}$, and alternating, complex A-A coupling $t_{AA} e^{\pm i \phi_{AA}}$.
(d) $H_{\mathrm{NS,A}}^{(0,y,0)}$ ($C^2=\mathrm{NS}$) with alternating onsite energies $\pm u$, alternating, complex nearest-neighbor coupling $ve^{\pm i \phi_v}$, and alternating, complex A-A coupling $\pm t_{AA} e^{\pm i \phi_{AA}}$.
Bars over parameters indicate that they are complex numbers, in general.
Other nonsymmorphic models are shown in Figs.~\ref{figNSt1} and~\ref{figNSc1}.
}\label{figchiral5}
\end{figure}

\section{Nonsymmorphic models with chiral symmetry}\label{s:ns}

\subsection{Nonsymmorphic symmetry class A with $T^2 = 0$, $C^2 = 0$, $S^2 = \mathrm{NS}$}\label{s:gns}

The two possibilities for a nonsymmorphic nonspatial symmetry, $T_{a/2}$  and  $T_{a/2} S_z$, suggest that there are two possible models with nonsymmorphic chiral symmetry (only), Table~\ref{tablenss}. However, they are related by a trivial, diagonal gauge transformation because $S_z$ is diagonal; thus, there is only one distinct model in subclass A with $S^2 = \mathrm{NS}$.

The models written in the atomic basis may be identified explicitly using the generalized Rice-Mele Hamiltonian~(\ref{cgrm4}) with $d_0 = 0$.
For the Bloch Hamiltonian $H(k,s)$, nonsymmorphic chiral symmetry corresponds to cases for which $H(k,a/2)$ satisfies chiral symmetry~(\ref{sk}) with $U_S (k,a/2)$ being independent of $k$. In principle, there are three such models, corresponding to $U_S (k,a/2) = \sigma_z$, $\sigma_y$, or $\sigma_x$. 
However, chiral symmetry $U_S (k,s) = \sigma_z$ is independent of the intracell spacing $s$ because it and $U (k,s)$, Eq.(\ref{uks2}), are diagonal. Thus, $U_S (k,s) = \sigma_z$ is included as symmorphic chiral symmetry yielding $H_{\mathrm{S,AIII}}^{(0,0,z)} (k,s)$, the generalized SSH model.
Thus, there are two general nonsymmorphic models:
$H_{\mathrm{NS,A}}^{(0,0,y)}$ with $U_S (k,a/2) = \sigma_y$ in $k$ space and $T_{a/2} S_z$ in position space, $H_{\mathrm{NS,A}}^{(0,0,x)}$ with $U_S (k,a/2) = \sigma_x$ and $T_{a/2}$.
The values of the tight-binding parameters for each model are summarized in Table~\ref{tablenss}.

Considering one of these models in position space with an even number of atoms and periodic boundary conditions (thus, satisfying chiral symmetry), then, when we change to the chiral basis, the resulting Hamiltonian is block diagonal, with $2 \times 2$ blocks, but it is no longer a periodic lattice (with a period of $a$); this is why analysis of the generalized Rice-Mele model with $U_S (k,a/2) = \sigma_z$ is unable to realise a nonsymmorphic model.

Although the symmetry operators for intracell spacing $s=a/2$ are independent of $k$, they acquire $k$ dependences when written for zero intracell spacing $s=0$ (or, equivalently, when written in the type I or `periodic' representation~\cite{bena09,cayssol21}) according to Eqs.~(\ref{usks})-(\ref{upks}).
For example, for $H_{\mathrm{NS,A}}^{(0,0,y)}$, $U_S (k,a/2) = \sigma_y$ translates as
\begin{eqnarray}
U_S (k,0) = \begin{pmatrix}
0 & -i e^{-ika/2} \\
i e^{ika/2} & 
\end{pmatrix} .
\end{eqnarray}

The canonical form $H_{\mathrm{NS},A} (k,a/2) = \mbox{\boldmath$\sigma$} \cdot {\bf \tilde{d}}$ of the Hamiltonian is given in Table~\ref{tablescf} with ${\tilde d}_z (k,a/2) = 0$. Component ${\tilde d}_x$ is an arbitrary, real $2\pi$ periodic function of $ka$ whereas ${\tilde d}_y$ has a period of $4\pi$~\cite{shiozaki15}.
In the canonical form, chiral symmetry is given by $\sigma_z$.
Model $H_{\mathrm{NS,A}}^{(0,0,y)} (k,a/2)$ can be transformed to the canonical form, Table~\ref{tablescf}, using $R=R_y$, Eq.~(\ref{ry}), and $H_{\mathrm{NS,A}}^{(0,0,x)} (k,a/2)$ can be transformed to the canonical form using $R=R_x$, Eq.~(\ref{rx}).
The canonical form in Table~\ref{tablescf} is written for intracell spacing $s=a/2$, but it may be written for $s=0$ using the unitary transformation~Eqs.~(\ref{uks1},\ref{uks2}).
In terms of tight-binding parameters, nonsymmorphic models (Tables~\ref{tablenss} and~\ref{tablensns}) have definite relationships between nearest-neighbor hopping parameters $w$ and $v$ (generalizing to $t_{2n+1}^{\prime}$ and $t_{2n+1}$ for hoppings of arbitrary order, Appendix~\ref{a:longrange}). 
Symmorphic models (Tables~\ref{tabless} and~\ref{tablesns}) either have no particular relation between different hopping parameters or third-nearest neighbor hopping $t_3$ is related to next-nearest hopping $w$ (generalizing to $t_{2n+1}$ and $t_{2n-1}^{\prime}$ for hoppings of arbitrary order, Appendix~\ref{a:longrange}). 

In this symmetry class, energy eigenvalues are given by $E_{\pm}(k) = \pm \sqrt{{\tilde d}_x^2 + {\tilde d}_y^2}$, and the system is generally an insulator.
Since time-reversal and spatial-inversion symmetries are absent, the spectrum generally isn't an even function of $k$, but chiral symmetry imposes $E_{\pm}(k) = - E_{\mp}(k)$.
As ${\tilde d}_y$ at $s=a/2$ has a period of $4\pi$, the path of the ${\bf \tilde{d}}$ vector in the ${\tilde d}_x$-${\tilde d}_y$ plane is not closed and it is not possible to define a winding number. However, it is possible to identify a $\mathbb{Z}_2$ topological index $\mu_2$~\cite{shiozaki15,shiozaki16} by counting whether the number of times the trajectory crosses the negative ${\tilde d}_x$ axis is even ($\mu_2=0$) or odd ($\mu_2=1$) for $0 \leq ka < 2 \pi$.
The end point of the trajectory must have the same ${\tilde d}_x$ value as the start point, and a negated value of ${\tilde d}_y$. Thus, it is impossible to change the $\mathbb{Z}_2$ topological index by adiabatically changing parameters in order to move the start and end points or to adjust the trajectory, as long as the origin is avoided.

For nearest-neighbor terms only ($x_0$, $x_1$, $\tilde{x}_1$, $\beta_0$, and $\tilde{\beta}_0$ only), an expression for the $\mathbb{Z}_2$ topological index, $\mu_2$, may be found by examining the sign of $\tilde{d}_x$ at the point when 
$\tilde{d}_y = 0$ (i.e., when $\beta_0 \cos (ka/2) + \tilde{\beta}_0 \sin (ka/2) = 0$). This gives
\begin{eqnarray}
\mu_2 &=& \begin{cases}
0 &\mbox{if } f  > 0 , \\
1 & \mbox{if } f < 0 ,
\end{cases} \label{mu2A} \\
f &=& \frac{x_0}{2} (\beta_0^2 + \tilde{\beta}_0^2) - x_1 (\beta_0^2 - \tilde{\beta}_0^2) - 2 \tilde{x}_1 \beta_0 \tilde{\beta}_0 , \label{deff}
\end{eqnarray}
and the system is gapless if $f = 0$.
Expressions for the parameters $x_0$, $x_1$, $\tilde{x}_1$, $\beta_0$, and $\tilde{\beta}_0$ in terms of tight-binding parameters for the two models may be read off Table~\ref{tablenss2}.
For example, for $H_{\mathrm{NS,A}}^{(0,0,y)}$, then $\mu_2 = 1$ if $u < 2t_{AA}\cos (\phi_{AA} - 2 \phi_v)$.
The $\mathbb{Z}_2$ topological index, $\mu_2$, is discussed in more detail in Section~\ref{s:nst}.

As stated above, there is only one distinct model because $H_{\mathrm{NS,A}}^{(0,0,y)}$ and $H_{\mathrm{NS,A}}^{(0,0,x)}$ are related by a diagonal transformation in the atomic basis in position space,
$H_{\mathrm{NS,A}}^{(0,0,x)} = {\mathcal R}_d H_{\mathrm{NS,A}}^{(0,0,y)} {\mathcal R}_d$, 
where the transformation has period $2a$,
\begin{eqnarray}
{\mathcal R}_d \!=\! \begin{pmatrix}
1 & 0 & 0 & 0 & 0 & \cdots \\
0 & 1 & 0  & 0 & 0 & \cdots \\
0 & 0  & -1 & 0 & 0 & \cdots \\
0 & 0 & 0 & -1 & 0 & \cdots \\
0 & 0 & 0 & 0 & 1 & \cdots \\
\vdots & \vdots & \vdots & \vdots & \vdots & \ddots
\end{pmatrix} \!\! , \label{rd}
\end{eqnarray}
and this may require reversing the sign of some parameters (e.g. $t_{AA}$).
Thus, in terms of models which are distinguished by off-diagonal transformations in the atomic basis, there is actually only one, $H_{\mathrm{NS,A}}^{(0,0,y)}$, which is sketched in Fig.~\ref{figchiral5}(a).

This model, $H_{\mathrm{NS,A}}^{(0,0,y)}$, can be viewed as a generalized version of the CDW model~\cite{shiozaki15,brzezicki20,fuchs21,allen22} with alternating onsite energies and constant nearest-neighbor hopping ${\bar v} = v \exp(i \phi_v)$. Second-nearest neighbor hopping (if included) also alternates as ${\bar t}_{AA} = t_{AA} \exp(i \phi_{AA})$ for A-A hopping and $-{\bar t}_{AA} = - t_{AA} \exp(i \phi_{AA})$ for B-B hopping. It may be generalized to include arbitrary-range hoppings, Table~\ref{tableAns}.

\begin{figure}[t]
\includegraphics[scale=0.31]{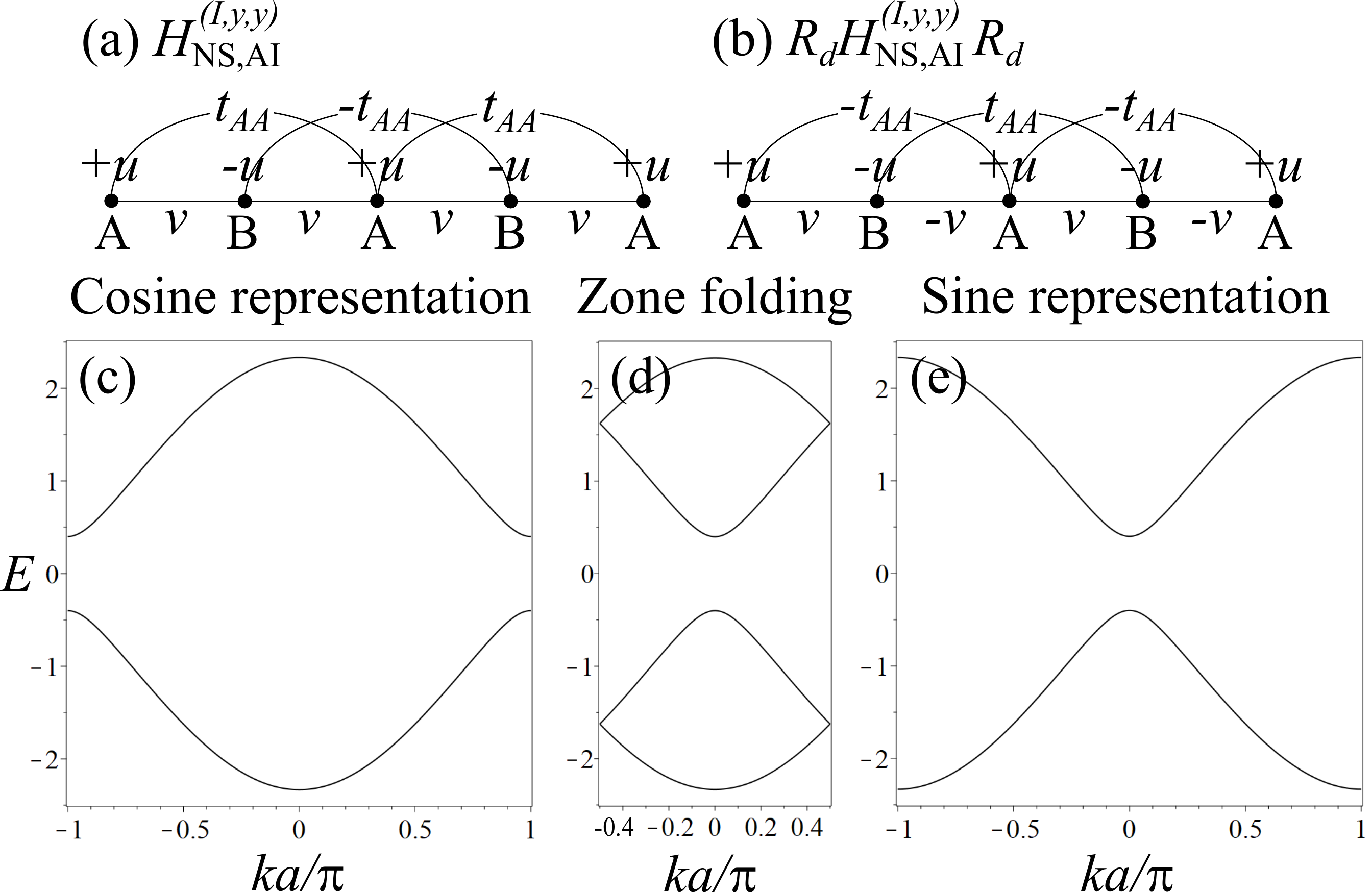}
\caption{Model $H_{\mathrm{NS,AI}}^{(I,y,y)}$ (CDW model) in the AI nonsymmorphic symmetry class ($T^2=1$, $C^2=\mathrm{NS}$, $S^2= \mathrm{NS}$).
(a) Shows the model with intracell spacing $s=a/2$, real and constant nearest-neighbor hopping $v$, alternating onsite energies $\pm u$, and alternating second-neighbor hopping $\pm t_{AA}$.
(b) Shows the model after a diagonal gauge transformation in position space ${\cal R}_d  H_{\mathrm{NS,AI}}^{(I,y,y)} {\cal R}_d$ yielding alternating nearest-neighbor hopping $\pm v$.
(c) The band structure $E(k)$ corresponding to the cosine representation (a).
(d) The band structure $E(k)$ with zone folding at $ka = \pm \pi/2$.
(e) The band structure $E(k)$ corresponding to the sine representation (b).
Zone folding (d) illustrates that (c) and (e) describe the same physical states.
Parameter values are $u = 0.8$, $v=1.0$, $t_{AA} = 0.2$,
equivalent to $x_0 = 1.6$, $x_1=0.4$, $\beta_0 = 2.0$.
}\label{figNSt1}
\end{figure}

\subsection{Nonsymmorphic symmetry class AI with $T^2 = 1$, $C^2 = \mathrm{NS}$, $S^2 = \mathrm{NS}$}\label{s:nst}

We now consider nonsymmorphic chiral symmetry in systems with charge-conjugation $C$ and time-reversal $T$ symmetries, too.
If chiral symmetry $S$ is nonsymmorphic, then one (and only one) of $C$ and $T$ must be, too.
If we consider that $S$ and $C$ are nonsymmorphic, each described by $T_{a/2}$  or  $T_{a/2} S_z$, then there are four possibilities, Table~\ref{tablenss}, with $T$ being symmorphic ($T^2 = 1$) for each of them.
However, only one of these is distinct, Fig.~\ref{figNSt1}, as can be seen by multiplying the symmetry operators by $S_z$.

To identify models in the atomic basis corresponding to each of these possibilities, we apply time-reversal symmetry~(\ref{tk}) to the models with chiral symmetry described in Section~\ref{s:gns}.
For the Bloch Hamiltonian, we consider $U_T (k,a/2)$ to be independent of $k$, i.e. $U_T (k,a/2) = I$, $\sigma_x$,  $\sigma_y$ or $\sigma_z$.
Then, with definite chiral symmetry and time-reversal symmetry, we determine the form of charge-conjugation symmetry $U_C$ using $U_S (k,a/2) =U_C^{\ast} (k,a/2) U_T (-k,a/2)$.
Thus, the four models with $T=1$, $C= \mathrm{NS}$, $S= \mathrm{NS}$ (label `NS' indicates nonsymmorphic) are listed in Table~\ref{tablenss}.
In this class, the canonical form of the Bloch Hamiltonian $\tilde{H}_{\mathrm{NS,AI}} (k,a/2) = \mbox{\boldmath$\sigma$} \cdot {\bf \tilde{d}}$ can be written as in Table~\ref{tablescf} with ${\tilde d}_z (k,a/2) = 0$.
Values of the tight-binding parameters of each model in terms of the parameters of the canonical form are given in Table~\ref{tablenss2}.

\begin{figure}[t]
\includegraphics[scale=0.42]{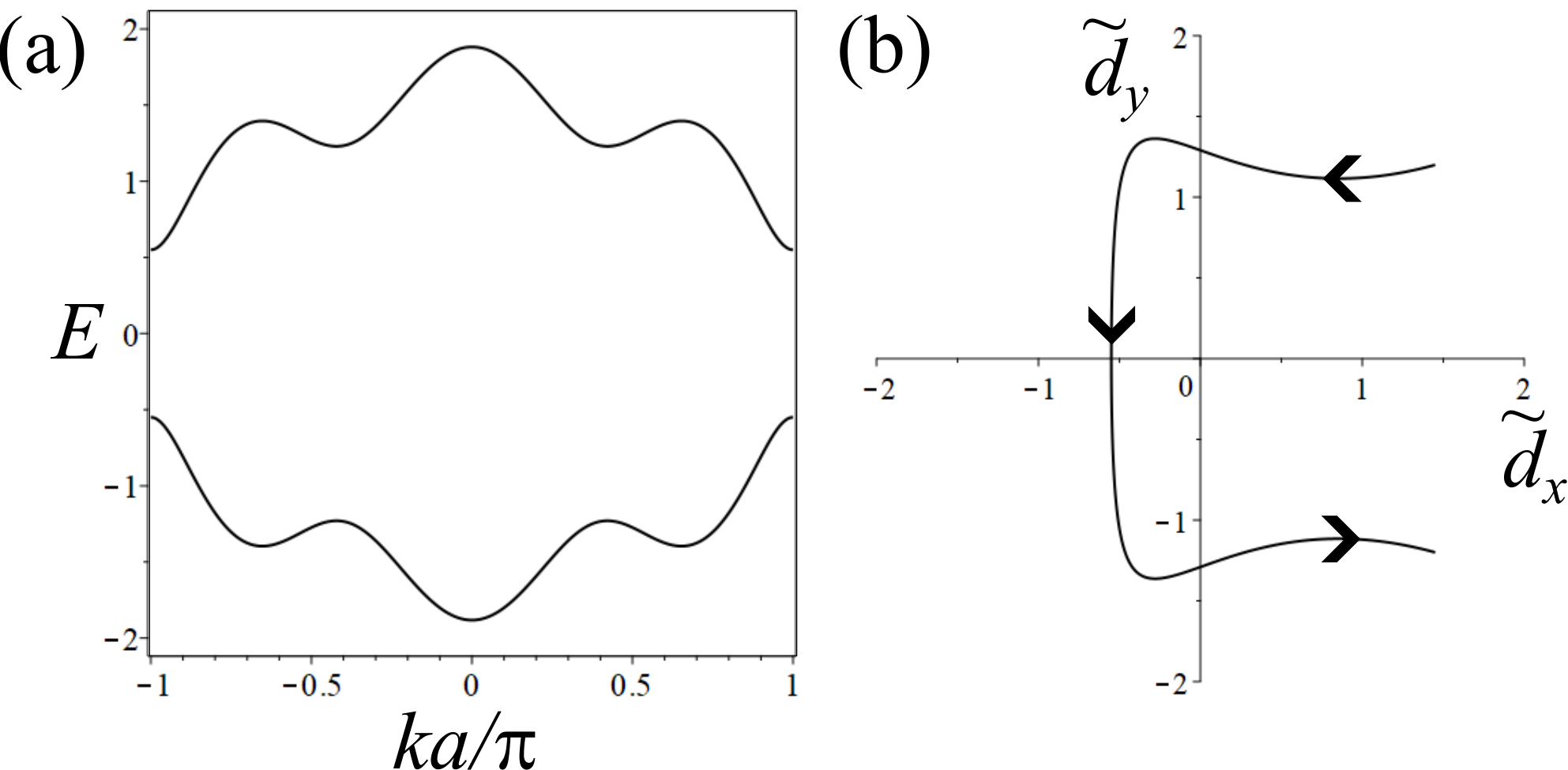}
\caption{(a) Band structure $E(k)$ in the AI nonsymmorphic symmetry class ($T^2 = 1$, $C^2 = \mathrm{NS}$, $S^2 = \mathrm{NS}$) using the canonical form,~Table~\ref{tablescf}. Parameter values are $x_0 = 0.5$, $x_1 = 1.0$, $x_2 = 0.2$, $\beta_0 = 1.5$, $\beta_1 = - 0.5$, $\beta_2 = 0.2$, with $x_n = \beta_n = 0$ for $n \geq 3$.
(b) The corresponding trajectory of the ${\bf \tilde{d}}$ vector in the ${\tilde d}_x$-${\tilde d}_y$ plane. For $0 \leq ka < 2 \pi$, it traces an open path which crosses the negative ${\tilde d}_x$ axis once, indicating a $\mathbb{Z}_2$ topological index~\cite{shiozaki15} of value $\mu_2 = 1$.
}\label{figwind3}
\end{figure}

Energy eigenvalues are $E_{\pm} (k) = \pm \sqrt{{\tilde d}_x^2 + {\tilde d}_y^2}$, and the system is generally an insulator, Fig.~\ref{figbulk1}(b).
As described in Section~\ref{s:gns}, ${\tilde d}_y (k,a/2)$ has a period of $4\pi$ and it is not possible to define a winding number, but there is a $\mathbb{Z}_2$ topological index $\mu_2$~\cite{shiozaki15,brzezicki20}. This is defined by counting whether the number of times the trajectory crosses the negative ${\tilde d}_x$ axis is even ($\mu_2=0$) or odd ($\mu_2=1$) for $0 \leq ka < 2 \pi$. An example is shown with band structure in Fig.~\ref{figwind3}(a), where we note that the higher-order terms in the canonical form, Table~\ref{tablescf}, produce strong features in the band structure at $ka \approx \pm \pi/2$ which would result in van Hove singularities in the density of states.
The path of the corresponding ${\bf \tilde{d}}$ vector is in Fig.~\ref{figwind3}(b) showing the trajectory crossing the negative ${\tilde d}_x$ axis once, indicating a $\mathbb{Z}_2$ topological index of value $\mu_2 = 1$ for this choice of parameters. The canonical form, Table~\ref{tablescf}, shows that ${\tilde d}_y (\pi/a,a/2) = 0$ so the path must cross the ${\tilde d}_x$ axis at $k = \pi/a$. The end point of the trajectory must have the same ${\tilde d}_x$ value as the start point, and a negated value of ${\tilde d}_y$. With these constraints, it is impossible to change the $\mathbb{Z}_2$ topological index by adiabatically changing parameters in order to move the start and end points or to adjust the trajectory, as long as the origin is avoided.
For nearest-neighbor terms only ($x_0$, $x_1$ and $\beta_0$ only), the $\mathbb{Z}_2$ topological index, $\mu_2$, may be simplified as
\begin{eqnarray}
\mu_2 = \begin{cases}
0 &\mbox{if } x_0 > 2 x_1 , \\
1 & \mbox{if } x_0 < 2 x_1 ,
\end{cases}
\end{eqnarray}
and the system is gapless if $x_0 = 2 x_1$.
Expressions for $x_0$, $x_1$ in terms of tight-binding parameters for different models may be read off Table~\ref{tablenss2}. For example, for $H_{\mathrm{NS,AI}}^{(I,y,y)}$, then $\mu_2 = 1$ if $u < 2t_{AA}$.

The one distinct model, $H_{\mathrm{NS,AI}}^{(I,y,y)}$, is the CDW model~\cite{kivelson83,shiozaki15,brzezicki20,cayssol21,fuchs21,allen22} with alternating onsite energies and constant nearest-neighbor hopping $v$, Fig.~\ref{figNSt1}(a).
Second-nearest neighbor hopping (if included) also alternates as $t_{AA}$ for A-A hopping and $- t_{AA}$ for B-B hopping.
For the nonsymmorphic cases, the Bloch Hamiltonian is not $2\pi$ periodic in $ka$, and there is ambiguiity in the choice of the canonical form, Table~\ref{tablescf}: here, we have chosen a `cosine' representation ${\tilde d}_y (k,a/2) = \sum_{n=0}^{\infty} p_n \cos \left[ ka \left(n+1/2 \right) \right]$, but it is equally valid to choose a `sine' representation ${\tilde d}_y (k,a/2) = \sum_{n=0}^{\infty} \tilde{p}_n \sin \left[ ka \left(n+1/2 \right) \right]$ as the two are related by the diagonal, gauge transformation ${\cal R}_d$, Eq.~(\ref{rd}), in position space.
Fig.~\ref{figNSt1}(b) shows the model schematically in position space after the transformation ${\cal R}_d  H_{\mathrm{NS,AI}}^{(I,y,y)} {\cal R}_d$, and the transformation introduces alternating nearest-neighbor hopping $\pm v$ (this is equivalent to model $H_{\mathrm{NS,AI}}^{(I,x,x)}$).
On Fourier transforming to $k$ space, the original model, Fig.~\ref{figNSt1}(a), yields the cosine representation,
$E_{\mathrm{cos}} (k) = \pm \sqrt{[u + 2t_{AA}\cos(ka)]^2 + [2v\cos(ka/2)]^2}$, as plotted in Fig.~\ref{figNSt1}(c).
However, the transformed model, Fig.~\ref{figNSt1}(b), gives the sine representation,
$E_{\mathrm{sine}} (k) = \pm \sqrt{[u - 2t_{AA}\cos(ka)]^2 - [2v\sin(ka/2)]^2}$, as plotted in Fig.~\ref{figNSt1}(e).
The two representations describe the same physical states
$E_{\mathrm{sine}} (k + \pi /a) = E_{\mathrm{cos}} (k)$
as can be seen by the fact they both give the same spectrum upon zone folding at $ka = \pm \pi/2$, Fig.~\ref{figNSt1}(d).

In this symmetry class, both chiral symmetry $S$ and charge-conjugation $C$ are nonsymmorphic, Table~\ref{tablesummary}, and one may wonder whether this symmetry class has $C^2 = 1$ or $C^2 = -1$. There are models ($H_{\mathrm{NS,AI}}^{(I,x,x)}$ and $H_{\mathrm{NS,AI}}^{(z,x,y)}$) where $C = \sigma_x$ in $k$ space and $C = T_{a/2}$ in position space. Hence $C^2 = T_{a/2}^2 = T_{a}$ is not the identity matrix as such but describes translation by a unit cell (of length $a$) which is a symmetry of the lattice. However, there are also models ($H_{\mathrm{NS,AI}}^{(I,y,y)}$ and $H_{\mathrm{NS,AI}}^{(z,y,x)}$) where $C = \sigma_y$ in $k$ space and $C = T_{a/2} S_z$ in position space. Since $T_{a/2}$ and $S_z$ anticommute, then $C^2 = T_{a/2} S_z T_{a/2} S_z = - T_{a/2}^2 = - T_{a}$. So, there appears to be both cases ($C^2 = 1$ or $C^2 = -1$), even though, as described above, all four models are equivalent (related by diagonal gauge transformations).

Thus, we choose to identify this symmetry (sub)class by $C^2 = \mathrm{NS}$ rather than $C^2 = 1$ or $C^2 = -1$.
Note that the apparent ambiguity ($C^2 = 1$ or $C^2 = -1$) is recognised as ambiguity in the sign, $\epsilon_{\sigma}$, of a product of symmetry operators in the classification of Ref.~\cite{shiozaki16}, and the classification does not depend on such ambiguity.
In the next section, we describe a similar situation with nonsymmorphic time-reversal symmetry, and show how the two apparently contradicting cases (of $T^2 = 1$ and $T^2 = -1$) give a consistent implementation of Kramer's degeneracy.

\begin{figure}[t]
\includegraphics[scale=0.32]{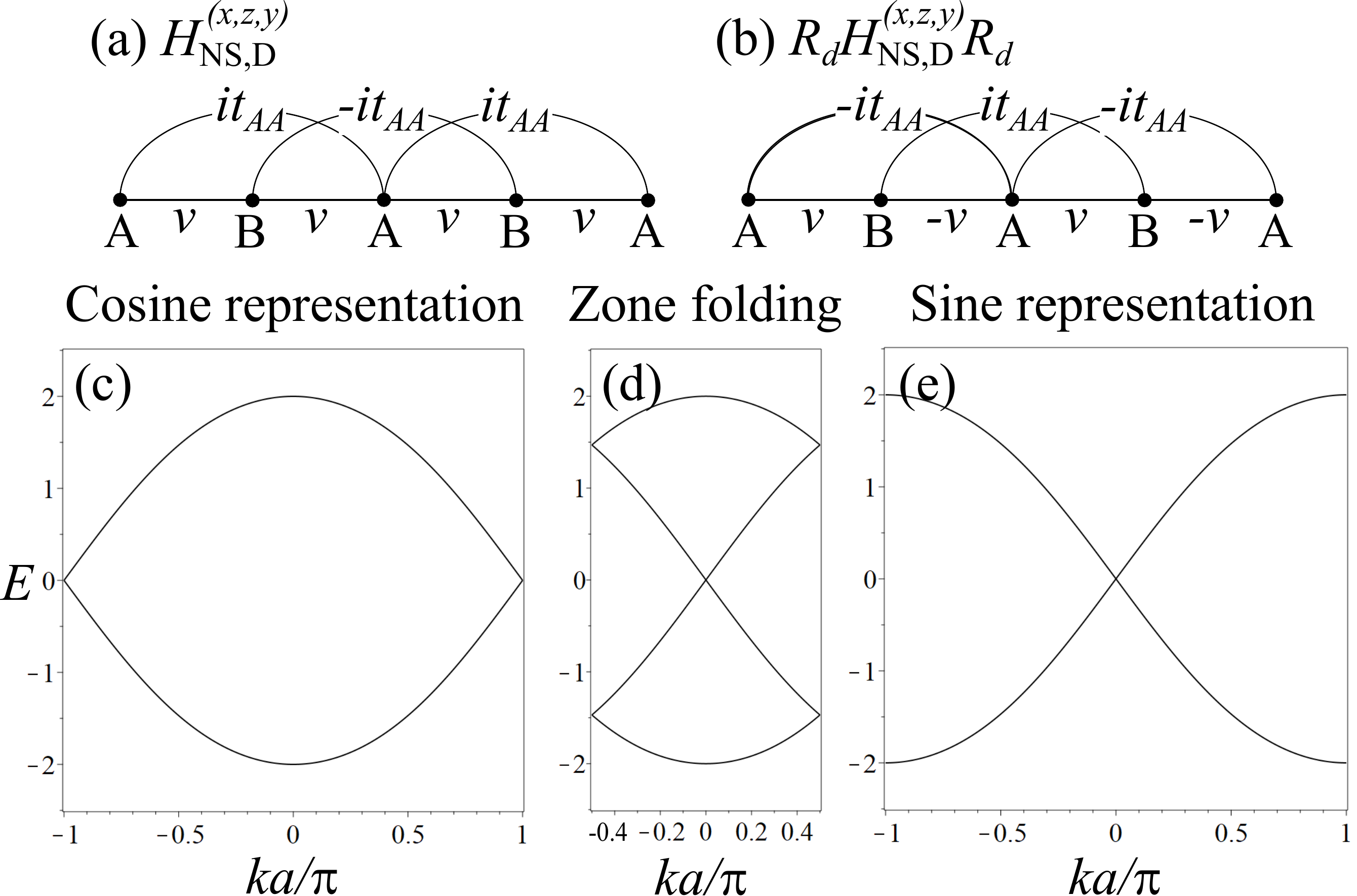}
\caption{Model $H_{\mathrm{NS,D}}^{(x,z,y)}$ in the D nonsymmorphic symmetry class ($T^2=\mathrm{NS}$, $C^2=1$, $S^2= \mathrm{NS}$).
(a) Shows the model with intracell spacing $s=a/2$, real and constant nearest-neighbor hopping $v$, and alternating, imaginary second-neighbor hopping $\pm i t_{AA}$.
(b) Shows the model after a diagonal gauge transformation in position space ${\cal R}_d  H_{\mathrm{NS,D}}^{(x,z,y)} {\cal R}_d$ yielding alternating nearest-neighbor hopping $\pm v$.
(c) The band structure $E(k)$ corresponding to the cosine representation (a).
(d) The band structure $E(k)$ with zone folding at $ka = \pm \pi/2$.
(e) The band structure $E(k)$ corresponding to the sine representation (b).
Zone folding (d) illustrates that (c) and (e) describe the same physical states.
Parameter values are $v=1.0$ and $t_{AA} = 0.2$,
equivalent to $\tilde{x}_1=-0.4$, $\beta_0 = 2.0$.
}\label{figNSc1}
\end{figure}

\subsection{Nonsymmorphic symmetry class D with $T^2=\mathrm{NS}$, $C^2 = 1$, $S^2 = \mathrm{NS}$}\label{s:nsd}

If we now choose $S$ and $T$ to be nonsymmorphic, each described by $T_{a/2}$  or  $T_{a/2} S_z$, then there are also four possibilities, Table~\ref{tablenss}, with $C$ being symmorphic ($C^2 = 1$) for each of them.
Again, one of these (e.g. $H_{\mathrm{NS,D}}^{(x,z,y)}$) is distinct as can be seen by multiplying the symmetry operators by $S_z$.
Model $H_{\mathrm{NS,D}}^{(x,z,y)}$, Fig.~\ref{figNSc1}(a), has real nearest-neighbor hopping $v$, but imaginary and alternating second-neighbor hopping $\pm i t_{AA}$.

The canonical form of the Bloch Hamiltonian $\tilde{H}_{\mathrm{NS,D}} (k,a/2) = \mbox{\boldmath$\sigma$} \cdot {\bf \tilde{d}}$ is given in Table~\ref{tablescf} with ${\tilde d}_z (k,a/2) = 0$.
Values of the tight-binding parameters of each model in terms of the parameters of the canonical form are given in Table~\ref{tablenss2}.
Energy eigenvalues are $E_{\pm} (k) = \pm \sqrt{{\tilde d}_x^2 + {\tilde d}_y^2}$, and the system is generally gapless, Fig.~\ref{figbulk1}(d).

Here, we have chosen the `cosine' representation ${\tilde d}_y (k,a/2) = \sum_{n=0}^{\infty} p_n \cos \left[ ka \left(n+1/2 \right) \right]$, but it is equally valid to choose a `sine' representation ${\tilde d}_y (k,a/2) = \sum_{n=0}^{\infty} \tilde{p}_n \sin \left[ ka \left(n+1/2 \right) \right]$ as the two are related by the diagonal, gauge transformation ${\cal R}_d$, Eq.~(\ref{rd}), in position space.
Fig.~\ref{figNSc1}(b) shows the model schematically in position space after the transformation ${\cal R}_d  H_{\mathrm{NS,D}}^{(x,z,y)} {\cal R}_d$, and the transformation introduces alternating nearest-neighbor hopping $\pm v$ (this is equivalent to model $H_{\mathrm{NS,D}}^{(y,z,x)}$).
On Fourier transforming to $k$ space, the original model, Fig.~\ref{figNSc1}(a), yields the cosine representation,
$E_{\mathrm{cos}} (k) = \pm \sqrt{4t_{AA}^2\sin^2(ka) + 4v^2\cos^2(ka/2)}$, as plotted in Fig.~\ref{figNSc1}(c).
However, the transformed model, Fig.~\ref{figNSc1}(b), gives the sine representation,
$E_{\mathrm{sine}} (k) = \pm \sqrt{4t_{AA}^2\sin^2(ka) + 4v^2\sin^2(ka/2)}$, as plotted in Fig.~\ref{figNSc1}(e).
The two representations describe the same physical states
$E_{\mathrm{sine}} (k + \pi /a) = E_{\mathrm{cos}} (k)$
as can be seen by the fact they both give the same spectrum upon zone folding at $ka = \pm \pi/2$, Fig.~\ref{figNSc1}(d). In each of the band plots, Fig.~\ref{figNSc1}(c), (d), (e), there is a single degeneracy point in the spectrum~\cite{zhao16}, and the system is gapless. We find that Kramer's degeneracy does not hold in general (most states are not degenerate with their time-reversed pair), but there is a single point which is doubly degenerate.

\begin{table*}[t]
\begin{center}
\caption{\label{tablensns}Nonsymmorphic models without chiral symmetry showing the values of the tight-binding parameters of the generalized Rice-Mele model, Eqs.~(\ref{cgrm1},\ref{cgrm2}), for each model with $t_3=0$. The same models with all long-range hoppings are shown in Table~\ref{tableAns2}.}
\begin{tabular}{ L{1.2cm} || C{1.0cm} | C{1.0cm} | C{1.0cm} | C{1.0cm} | C{1.0cm} | C{1.0cm} | C{1.0cm} | C{1.0cm} | C{1.0cm} | C{1.0cm}}
\hline
name & $\epsilon_0$  & $u$ & $t_{AA}$ & $\phi_{AA}$ & $t_{BB}$ & $\phi_{BB}$ & $v$ & $\phi_v$ & $w$ & $\phi_w$  \\ [1pt]
\hline \hline
$H_{\mathrm{NS,A}}^{(x,0,0)}$ & arb. & $0$ & arb. & arb. & $t_{AA}$ & $-\phi_{AA}$ & arb. & arb. & $v$ & $-\phi_v$ \\ [1pt]
\hline
$H_{\mathrm{NS,A}}^{(y,0,0)}$ & arb. & $0$ & arb. & arb. & $t_{AA}$ & $-\phi_{AA}$ & arb. & arb. & $-v$ & $-\phi_v$ \\ [1pt]
\hline \hline
$H_{\mathrm{NS,A}}^{(0,y,0)}$ & $0$ & arb. & arb. & arb. & $-t_{AA}$ & $-\phi_{AA}$ & arb. & arb. & $v$ & $-\phi_v$ \\ [1pt]
\hline
$H_{\mathrm{NS,A}}^{(0,x,0)}$ & $0$ & arb. & arb. & arb. & $-t_{AA}$ & $-\phi_{AA}$ & arb. & arb. & $-v$ & $-\phi_v$ \\ [1pt]
\hline \hline
\end{tabular}
\end{center}
\end{table*}

To understand this, we briefly review the Kramer's degeneracy argument.
In general, time-reversal symmetry ensures that the energy $E_{\pm} (k)$ is an even function with respect to $k$, Eq.~(\ref{et}). Particularly, for time-reversal invariant wave vectors $k_{T} = - k_{T}$ (i.e., $k_T = 0$ or $\pm \pi/a$ in the first Brillouin zone), then time-reversal symmetry creates a degeneracy, but only if the state $| \psi (k_{T}) \rangle$ and its time-reversed partner $T (k_{T}) | \psi (k_{T}) \rangle$ are orthogonal. 
We denote the time-reversal operator as $T(k) = U_T (k) K$ where $U_T (k) \equiv U_T (k,s=0)$ is a unitary operator and $K$ is the complex conjugation operator. We explicitly include the possiblity of $k$ dependence and consider zero intracell separation $s=0$.
Then, for two states $| \psi (k) \rangle$ and $| \phi (k) \rangle$,
\begin{eqnarray}
\langle T(k) \psi (k) | T(k) \phi (k) \rangle &=& 
\langle U_T (k) \psi^{\ast} (k) | U_T (k) \phi^{\ast} (k) \rangle , \nonumber \\
&=& 
\langle \psi^{\ast} (k) | U_T^{\dagger} (k) U_T (k) \phi^{\ast} (k) \rangle , \nonumber \\
&=& \langle \phi (k) | \psi (k) \rangle , \label{kramer1}
\end{eqnarray}
using $U_T^{\dagger} (k) U_T (k) = I$ and $\langle \psi | \phi \rangle^{\ast} = \langle \phi | \psi \rangle$.
Substituting $\phi (k) \equiv T(k) \psi (k)$ into Eq.~(\ref{kramer1}),
\begin{eqnarray}
\langle T(k) \psi (k) | \psi (k) \rangle = 
\langle T(k) \psi (k) | T^2(k) \psi (k) \rangle . \label{kramer2}
\end{eqnarray}
The right side may be written as
\begin{eqnarray*}
\langle T(k) \psi (k) |  T^2(k) \psi (k) \rangle
&=& \langle T(k) \psi (k) | U_T (k) U_T^{\ast} (k) \psi (k) \rangle .
\end{eqnarray*}
If $U_T (k) U_T^{\ast} (k) = - I$,
then
\begin{eqnarray*}
\langle T(k) \psi (k) |  T^2(k) \psi (k) \rangle
= - \langle T(k) \psi (k) | \psi (k) \rangle .
\end{eqnarray*}
Comparison with Eq.~(\ref{kramer2}) requires that
\begin{eqnarray}
U_T (k) U_T^{\ast} (k) = - I \quad \Rightarrow \quad \langle T(k) \psi (k) | \psi (k) \rangle = 0 . \label{kramer3}
\end{eqnarray}
This means that every state has an orthogonal time-reversed partner which is at the same energy, and that $k = k_{T}$ values that are time-reversal invariant are points of degeneracy.

Note that there is a subtle difference between the requirement~(\ref{tone}) on the operator to represent time reversal, $U_T (k) U_T^{\ast} (-k)  = \pm I$, and the requirement~(\ref{kramer3}) for Kramer's degeneracy, $U_T (k) U_T^{\ast} (k) = - I$.
In the symmorphic symmetry classes AII and DIII, $U_T (k) = \sigma_y$ is independent of $k$, and $U_T U_T^{\ast} = - I$. Thus, every state has a degenerate partner, and there are degeneracy points at time-reversal invariant $k_{T}$ values ($k_{T}=0$ and $k= \pm \pi/a$).

When time is nonsymmorphic, $U_T (k)$ is dependent on $k$; while $U_T (k) U_T^{\ast} (-k)  = \pm I$ for all $k$, Kramer's degeneracy $U_T (k) U_T^{\ast} (k) = - I$ is only satisfied for certain $k$ values.
The form of $U_T (k)$ depends on the representation of the nonsymmorphic terms in the ${\bf \tilde{d}}$ vector. For the cosine representation, $U_T (k,a/2) = \sigma_x$ and Eq.~(\ref{utks}) shows that
\begin{eqnarray*}
U_T (k) &=& \begin{pmatrix}
0 & e^{ika/2} \\
e^{-ika/2} & 0
\end{pmatrix} , \\
\Rightarrow
\qquad
U_T (k) U_T^{\ast} (k) &=&
\begin{pmatrix}
e^{ika} & 0  \\
0 & e^{-ika}
\end{pmatrix} ,
\end{eqnarray*}
so this is equal to $-1$ for $ka=\pm \pi$ only; this is the location of the degeneracy point in Fig.~\ref{figNSc1}(c).

For the sine representation, $U_T (k,a/2) = \sigma_y$ and
\begin{eqnarray*}
U_T (k) &=& \begin{pmatrix}
0 & -i e^{ika/2} \\
i e^{-ika/2} & 0
\end{pmatrix} , \\
\Rightarrow
\qquad
U_T (k) U_T^{\ast} (k) &=&
\begin{pmatrix}
- e^{ika} & 0  \\
0 & - e^{-ika}
\end{pmatrix} ,
\end{eqnarray*}
so this is equal to $-1$ for $k = 0$ only, Fig.~\ref{figNSc1}(e).
Regardless of the representation (e.g. see Fig.~\ref{figNSc1}(d) for the zone-folded band structure), Kramer's degeneracy of the nonsymmorphic time-reversal symmetry dictates that there is one degeneracy point as opposed to none in the symmorphic case with $T^2 = 1$ and two in the symmorphic case with $T^2 = - 1$.

\begin{table*}[t]
\begin{center}
\caption{\label{tablensns2}Nonymmorphic models without chiral symmetry showing how their parameters, Table~\ref{tablensns}, correspond to the canonical form Table~\ref{tablesns3}.}
\begin{tabular}{ L{1.2cm} | C{1.0cm} | C{2.1cm} | C{2.1cm} | C{1.5cm} | C{1.5cm} | C{1.0cm} | C{2.1cm} | C{2.1cm} | }
\hline
name & $a_0$ & $a_1$ & $\tilde{a}_1$ & $\alpha_0$ & $\beta_0$ & $z_0$ & $z_1 $ & $\tilde{z}_1$ \\
\hline
\hline
$H_{\mathrm{NS,A}}^{(x,0,0)}$ & $2\epsilon_0$ & $2t_{AA} \cos \phi_{AA}$ & $0$ & $2v \cos \phi_v$ & $-2v \sin \phi_v$ & $0$ & $0$ & $-2t_{AA} \sin \phi_{AA}$ \\
\hline
$H_{\mathrm{NS,A}}^{(y,0,0)}$ & $2\epsilon_0$ & $-2t_{AA} \cos \phi_{AA}$ & $0$ & $2v \cos \phi_v$ & $-2v \sin \phi_v$ & $0$ & $0$ & $2t_{AA} \sin \phi_{AA}$ \\
\hline \hline
$H_{\mathrm{NS,A}}^{(0,y,0)}$ & $0$ & $0$ & $-2t_{AA} \sin \phi_{AA}$ & $2v \cos \phi_v$ & $-2v \sin \phi_v$ & $2u$  & $2t_{AA} \cos \phi_{AA}$ & $0$ \\
\hline
$H_{\mathrm{NS,A}}^{(0,x,0)}$ & $0$ & $0$ & $2t_{AA} \sin \phi_{AA}$ & $2v \cos \phi_v$ & $-2v \sin \phi_v$ & $2u$  & $-2t_{AA} \cos \phi_{AA}$ & $0$ \\
\hline
\end{tabular}
\end{center}
\end{table*}

\subsection{Nonsymmorphic symmetry class AIII with $T^2=\mathrm{NS}$, $C^2=\mathrm{NS}$, $S^2=1$}

Finally, we can choose $C$ and $T$ to be nonsymmorphic, each described by $T_{a/2}$  or  $T_{a/2} S_z$. In this case there are only two possibilities, Table~\ref{tablenss}, because $\mathcal{S} = I$ is not possible; for both possibilities $\mathcal{S} = S_z$ ($S^2=1$).
Again, only one model is distinct, e.g. $H_{\mathrm{NS,AIII}}^{(x,y,z)}$, Fig.~\ref{figchiral5}(b).
To find the form of the models in position space, with $\mathcal{S} = S_z$, we begin by considering the generalized SSH Bloch Hamiltonian $H_{\mathrm{S,AIII}}^{(0,0,z)} (k,s)$ and apply time-reversal symmetry, considering $U_T (k,a/2)$ to be independent of $k$.

The canonical form of the Bloch Hamiltonian $\tilde{H}_{\mathrm{NS,AIII}} (k,a/2) = \mbox{\boldmath$\sigma$} \cdot {\bf \tilde{d}}$ with ${\tilde d}_z (k,a/2) = 0$ is given in Table~\ref{tablescf}, which is the cosine represenation. Note that the sine representation would have
${\tilde d}_x (k,a/2) = \sum_{n=0}^{\infty} {\tilde \alpha}_n \sin \left[ ka \left(n+1/2 \right) \right]$ and ${\tilde d}_y (k,a/2) = \sum_{n=0}^{\infty} {\tilde \beta}_n \sin \left[ ka \left(n+1/2 \right) \right]$.
Energy eigenvalues are $E_{\pm} (k) = \pm \sqrt{{\tilde d}_x^2 + {\tilde d}_y^2}$, and the system is generally gapless, Fig.~\ref{figbulk1}(f), due to nonsymmorphic time-reversal symmetry and Kramer's degeneracy, as described in Section~\ref{s:nsd}.

\section{Nonsymmorphic models without chiral symmetry}\label{s:nsns}

\subsection{Nonsymmorphic symmetry class A with $T^2=\mathrm{NS}$, $C^2 =0$, $S^2 = 0$}

To explicitly confirm that the presence of a single degeneracy point is due to
Kramer's degeneracy with nonsymmorphic time-reversal symmetry, we construct a model with nonsymmorphic time-reversal symmetry only, i.e. no chiral, charge-conjugation or spatial inversion symmetry.
We begin from the full model with $d_0 \neq 0$, Eqs.~(\ref{cgrm1},\ref{cgrm2}) and set the intracell spacing $s=a/2$.

We can choose $U_T (k,a/2) = \sigma_x$ (${\mathcal T} = T_{a/2}$ in position space) yielding the cosine representation and model
$H_{\mathrm{NS,A}}^{(x,0,0)}$, Table~\ref{tablensns},
or $U_T (k,a/2) = \sigma_y$ (${\mathcal T} = T_{a/2} S_z$) giving the sine representation and model $H_{\mathrm{NS,A}}^{(y,0,0)}$.
Only one of these is distinct, e.g. $H_{\mathrm{NS,A}}^{(x,0,0)}$, because $H_{\mathrm{NS,A}}^{(y,0,0)}$ is related to it by the diagonal transformation $\mathcal{R}_i$, Eq.~(\ref{ri}). Model $H_{\mathrm{NS,A}}^{(x,0,0)}$ is shown schematically in Fig.~\ref{figchiral5}(c).

For the canonical form, we use the cosine representation, Table~\ref{tablesns3}.
Table~\ref{tablensns2} shows values of the tight-binding parameters for the two models and the corresponding values of the parameters $a_n$, $\alpha_n$, $\beta_n$, and ${\tilde z}_n$ of the canonical form, Table~\ref{tablesns3}.
Band energies are $E_{\pm} (k) = {\tilde d}_0 \pm \sqrt{{\tilde d}_x^2 + {\tilde d}_y^2 + {\tilde d}_z^2}$.
Owing to the nonsymmorphic time-reversal symmetry and Kramer's degeneracy, there is a single degeneracy point
(at $ka = \pi$ in the canonical form, Table~\ref{tablesns3}).

\subsection{Nonsymmorphic symmetry class A with $T^2=0$, $C^2 =\mathrm{NS}$, $S^2 = 0$}

To complete the description of nonsymmorphic models, we conclude with models having nonsymmorphic charge-conjugation symmetry (only).
Again, we begin from the full model with $d_0 \neq 0$, Eqs.~(\ref{cgrm1},\ref{cgrm2}) and set the intracell spacing $s=a/2$.
We can choose $U_C (k,a/2) = \sigma_y$ (${\mathcal C} = T_{a/2} S_z$ in position space) yielding the cosine representation and model
$H_{\mathrm{NS,A}}^{(0,y,0)}$, Table~\ref{tablensns}, or $U_C (k,a/2) = \sigma_x$ (${\mathcal C} = T_{a/2}$) giving the sine representation and model $H_{\mathrm{NS,A}}^{(0,x,0)}$.
Only one of these is distinct, e.g. $H_{\mathrm{NS,A}}^{(0,y,0)}$, because $H_{\mathrm{NS,A}}^{(0,x,0)}$ is related to it by the diagonal transformation $\mathcal{R}_i$, Eq.~(\ref{ri}). Model $H_{\mathrm{NS,A}}^{(0,y,0)}$ is shown schematically in Fig.~\ref{figchiral5}(d).

For the canonical form, we use the cosine representation, Table~\ref{tablesns3}.
Table~\ref{tablensns2} shows values of the tight-binding parameters for the two models and the corresponding values of the parameters $\tilde{a}_n$, $\alpha_n$, $\beta_n$, and $z_n$ of the canonical form, Table~\ref{tablesns3}.
Band energies are $E_{\pm} (k) = {\tilde d}_0 \pm \sqrt{{\tilde d}_x^2 + {\tilde d}_y^2 + {\tilde d}_z^2}$, and the form of ${\tilde d}_z$ shows that this is generally insulating.

$K$ theory predicts~\cite{shiozaki16} that this class has a $\mathbb{Z}_2$ topological index even in the absence of time-reversal or chiral symmetry.
The description of the $\mathbb{Z}_2$ index in the presence of nonsymmorphic chiral symmetry~\cite{shiozaki15,brzezicki20}, discussed in Sections~\ref{s:gns} and~\ref{s:nst}, may be generalized to the canonical form, Table~\ref{tablesns3}. Now, we consider the trajectory for $0 \leq ka < 2 \pi$ in three-dimensional space with Cartesian axes ${\tilde d}_x$, ${\tilde d}_y$, ${\tilde d}_z$, and count whether the number of times the trajectory crosses the negative ${\tilde d}_z$ axis is even ($\mu_2=0$) or odd ($\mu_2=1$).
The $\mathbb{Z}_2$ index may be defined because of contraints on the trajectory as described by the canonical form, Table~\ref{tablesns3}: 
${\tilde d}_x (\pi/a,a/2) = {\tilde d}_y (\pi/a,a/2) = 0$ so the path must cross the ${\tilde d}_z$ axis at $k = \pi/a$; the end point of the trajectory must have the same ${\tilde d}_z$ value as the start point, with negated values of ${\tilde d}_x$ and ${\tilde d}_y$. With these constraints, it is impossible to change the $\mathbb{Z}_2$ index by adiabatically changing parameters in order to move the start and end points or to adjust the trajectory, as long as the origin is avoided.
For nearest-neighbor terms only ($z_0$, $z_1$, $\alpha_0$ and $\beta_0$ only), the $\mathbb{Z}_2$ topological index, $\mu_2$, may be simplified as
\begin{eqnarray}
\mu_2 = \begin{cases}
0 &\mbox{if } z_0 > 2 z_1 , \\
1 & \mbox{if } z_0 < 2 z_1 ,
\end{cases}
\end{eqnarray}
and the system is gapless if $z_0 = 2 z_1$. Expressions for $z_0$, $z_1$ in terms of tight-binding parameters for the two models may be read off Table~\ref{tablensns2}. For example, for $H_{\mathrm{NS,A}}^{(0,y,0)}$, then $\mu_2 = 1$ if $u < 2 t_{AA} \cos \phi_{AA}$.

\begin{table*}[t]
\begin{center}
\caption{\label{tableS}Symmorphic models without chiral symmetry including all long-range hoppings, showing the values of the tight-binding parameters of the generalized Rice-Mele model, Eqs.~(\ref{d0all},\ref{dxall},\ref{dyall},\ref{dzall}), for each model where $n \geq 1$.
The last four rows show models with spatial-inversion symmetry only as described by unitary matrix $U_P$~(\ref{pk}) and discussed in Appendix~\ref{a:spatial}.
}
\begin{tabular}{ L{1.5cm} || C{1.0cm} | C{1.0cm} | C{1.0cm}  | C{1.0cm} | C{1.0cm}| C{1.0cm} | C{1.0cm} | C{1.0cm} | C{1.0cm} | C{1.0cm} | C{1.0cm} | C{1.0cm}}
\hline
name & $\epsilon_0$ & $u$ & $t_{2n}$ & $\phi_{2n}$ &  $t_{2n}^{\prime}$ & $\phi_{2n}^{\prime}$ & $v$ & $\phi_v$ & $t_{2n-1}^{\prime}$ & $\phi_{2n-1}^{\prime}$ & $t_{2n+1}$ & $\phi_{2n+1}$  \\ [1pt]
\hline \hline
$H_{\mathrm{S,AI}}^{(I,0,0)}$ & arb. & arb. & arb. & $0$ & arb. & $0$ & arb. & $0$ & arb. & $0$ & arb. & $0$ \\ [1pt]
\hline
$H_{\mathrm{S,AI}}^{(x,0,0)}$ & arb. & $0$ & arb. & arb. & $t_{2n}$ & $-\phi_{2n}$ & arb. & arb. & arb. & arb. & $t_{2n-1}^{\prime}$ & $-\phi_{2n-1}^{\prime}$ \\ [1pt]
\hline
$H_{\mathrm{S,AI}}^{(z,0,0)}$ & arb. & arb. & arb. & $0$ & arb. & $0$ & arb. & $\pi/2$ & arb. & $\pi/2$ & arb. & $\pi/2$ \\ [1pt]
\hline \hline
$H_{\mathrm{S,AII}}^{(y,0,0)}$ & arb. & $0$ & arb. & arb. & $t_{2n}$ & $-\phi_{2n}$ & $0$ & n/a & arb. & arb. & $-t_{2n-1}^{\prime}$ & $-\phi_{2n-1}^{\prime}$ \\ [1pt]
\hline \hline
$H_{\mathrm{S,D}}^{(0,x,0)}$ & $0$ & arb. & arb. & arb. & $-t_{2n}$ & $-\phi_{2n}$ & $0$ & n/a & arb. & arb. & $-t_{2n-1}^{\prime}$ & $-\phi_{2n-1}^{\prime}$ \\ [1pt]
\hline
$H_{\mathrm{S,D}}^{(0,z,0)}$ & $0$ & $0$ & arb. & $\pi/2$ & arb. & $\pi/2$ & arb. & $0$ & arb. & $0$ & arb. & $0$ \\ [1pt]
\hline
$H_{\mathrm{S,D}}^{(0,I,0)}$ & $0$ & $0$ & arb. & $\pi/2$ & arb. & $\pi/2$ & arb. & $\pi/2$ & arb. & $\pi/2$ & arb. & $\pi/2$ \\ [1pt]
\hline \hline
$H_{\mathrm{S,C}}^{(0,y,0)}$ & $0$ & arb. & arb. & arb. & $-t_{2n}$ & $-\phi_{2n}$ & arb. & arb. & arb. & arb. & $t_{2n-1}^{\prime}$ & $-\phi_{2n-1}^{\prime}$ \\ [1pt]
\hline \hline
$H_{\mathrm{S,A}}^{(U_P=I)}$ & arb. & arb. & arb. & $0$ & arb. & $0$ & arb. & arb. & arb. & arb. & $t_{2n-1}^{\prime}$ & $-\phi_{2n-1}^{\prime}$ \\ [1pt]
\hline
$H_{\mathrm{S,A}}^{((U_P=\sigma_x)}$ & arb. & $0$ & arb. & arb. & $t_{2n}$ & $-\phi_{2n}$ & arb. & $0$ & arb. & $0$ & arb. & $0$ \\ [1pt]
\hline
$H_{\mathrm{S,A}}^{((U_P=\sigma_y)}$ & arb. & $0$ & arb. & arb. & $t_{2n}$ & $-\phi_{2n}$ & arb. &  $\pi/2$ & arb. &  $\pi/2$ & arb. &  $\pi/2$ \\ [1pt]
\hline
$H_{\mathrm{S,A}}^{(U_P=\sigma_z)}$ & arb. & arb. & arb. & $0$ & arb. & $0$ & $0$ & n/a & arb. & arb. & $-t_{2n-1}^{\prime}$ & $-\phi_{2n-1}^{\prime}$ \\ [1pt]
\hline
\end{tabular}
\end{center}
\end{table*}

\begin{table*}[t]
\begin{center}
\caption{\label{tableAns2}Nonsymmorphic models without chiral symmetry including all long-range hoppings, showing the values of the tight-binding parameters of the generalized Rice-Mele model, Eqs.~(\ref{d0all},\ref{dxall},\ref{dyall},\ref{dzall}), for each model.
The last two rows show models with spatial-inversion symmetry only as described by unitary matrix $U_P$~(\ref{pk}) and discussed in Appendix~\ref{a:spatial}.}
\begin{tabular}{ L{1.5cm} || C{1.0cm} | C{1.0cm} | C{1.0cm} | C{1.0cm} | C{1.0cm} | C{1.0cm} | C{1.0cm} | C{1.0cm} | C{1.0cm} | C{1.0cm}}
\hline
name & $\epsilon_0$  & $u$ & $t_{2n}$ & $\phi_{2n}$ & $t_{2n}^{\prime}$ & $\phi_{2n}^{\prime}$ & $t_{2n+1}$ & $\phi_{2n+1}$ & $t_{2n+1}^{\prime}$ & $\phi_{2n+1}^{\prime}$  \\ [1pt]
\hline \hline
$H_{\mathrm{NS,A}}^{(x,0,0)}$ & arb. & $0$ & arb. & arb. & $t_{2n}$ & $-\phi_{2n}$ & arb. & arb. & $t_{2n+1}$ & $-\phi_{2n+1}$ \\ [1pt]
\hline
$H_{\mathrm{NS,A}}^{(y,0,0)}$ & arb. & $0$ & arb. & arb. & $t_{2n}$ & $-\phi_{2n}$ & arb. & arb. & $-t_{2n+1}$ & $-\phi_{2n+1}$ \\ [1pt]
\hline \hline
$H_{\mathrm{NS,A}}^{(0,y,0)}$ & $0$ & arb. & arb. & arb. & $-t_{2n}$ & $-\phi_{2n}$ & arb. & arb. & $t_{2n+1}$ & $-\phi_{2n+1}$ \\ [1pt]
\hline
$H_{\mathrm{NS,A}}^{(0,x,0)}$ & $0$ & arb. & arb. & arb. & $-t_{2n}$ & $-\phi_{2n}$ & arb. & arb. & $-t_{2n+1}$ & $-\phi_{2n+1}$ \\ [1pt]
\hline \hline
$H_{\mathrm{NS,A}}^{(U_P=I)}$ & arb. & arb. & arb. & $0$  & arb. & $0$ & arb. & arb. & $t_{2n+1}$ & $-\phi_{2n+1}$ \\ [1pt]
\hline
$H_{\mathrm{NS,A}}^{(U_P=\sigma_z)}$ & arb. & arb. & arb. & $0$ & arb. & $0$ & arb. & arb. & $-t_{2n+1}$ & $-\phi_{2n+1}$ \\ [1pt]
\hline
\end{tabular}
\end{center}
\end{table*}

\begin{table*}[t]
\begin{center}
\caption{\label{tableAs}Symmorphic models with chiral symmetry including all long-range hoppings, showing the values of the tight-binding parameters of the generalized Rice-Mele model, Eqs.~(\ref{dxsall},\ref{dysall},\ref{dzsall}), for each model where $n \geq 1$, $\epsilon_0 = 0$, $t_{2n}^{\prime} = - t_{2n}$ and $\phi_{2n}^{\prime} = \phi_{2n}$.}
\begin{tabular}{ L{1.2cm} || C{1.0cm} | C{1.0cm} | C{1.0cm} | C{1.0cm} | C{1.0cm} | C{1.0cm} | C{1.0cm} | C{1.0cm} | C{1.0cm}}
\hline
name & $u$ & $t_{2n}$ & $\phi_{2n}$ & $v$ & $\phi_v$ & $t_{2n-1}^{\prime}$ & $\phi_{2n-1}^{\prime}$ & $t_{2n+1}$ & $\phi_{2n+1}$  \\ [1pt]
\hline \hline
$H_{\mathrm{S,AIII}}^{(0,0,z)}$ & $0$ & $0$ & n/a & arb. & arb. & arb. & arb. & arb. & arb. \\ [1pt]
\hline
$H_{\mathrm{S,AIII}}^{(0,0,y)}$ & arb. & arb. & arb. & arb. & $0$ & arb. & arb. & $t_{2n-1}^{\prime}$ & $\phi_{2n-1}^{\prime}$ \\ [1pt]
\hline
$H_{\mathrm{S,AIII}}^{(0,0,x)}$ & arb. & arb. & arb. & arb. & $\pi/2$ & arb. & arb. & $-t_{2n-1}^{\prime}$ & $\phi_{2n-1}^{\prime}$ \\ [1pt]
\hline \hline
$H_{\mathrm{S,BDI}}^{(I,z,z)}$ & $0$ & $0$ & n/a & arb. & $0$ & arb. & $0$ & arb. & $0$ \\ [1pt]
\hline
$H_{\mathrm{S,BDI}}^{(x,z,y)}$ & $0$ & arb. & $\pi/2$ & arb. & $0$ & arb. & $0$ & $t_{2n-1}^{\prime}$ & $0$ \\ [1pt]
\hline
$H_{\mathrm{S,BDI}}^{(I,x,x)}$ & arb. & arb. & $0$ & $0$ & n/a & arb. & $0$ & $-t_{2n-1}^{\prime}$ & $0$ \\ [1pt]
\hline
$H_{\mathrm{S,BDI}}^{(z,I,z)}$ & $0$ & $0$ & n/a & arb. & $\pi/2$ & arb. & $\pi/2$ & arb. & $\pi/2$ \\ [1pt]
\hline
$H_{\mathrm{S,BDI}}^{(x,I,x)}$ & $0$ & arb. & $\pi/2$ & arb. & $\pi/2$ & arb. & $\pi/2$ & $-t_{2n-1}^{\prime}$ & $\pi/2$ \\ [1pt]
\hline
$H_{\mathrm{S,BDI}}^{(z,x,y)}$ & arb. & arb. & $0$ & $0$ & n/a & arb. & $\pi/2$ & $t_{2n-1}^{\prime}$ & $\pi/2$ \\ [1pt]
\hline \hline
$H_{\mathrm{S,DIII}}^{(y,x,z)}$ & $0$ & $0$ & n/a & $0$ & n/a & arb. & arb. & $-t_{2n-1}^{\prime}$ & $-\phi_{2n-1}^{\prime}$ \\ [1pt]
\hline
$H_{\mathrm{S,DIII}}^{(y,z,x)}$ & $0$ & arb. & $\pi/2$ & $0$ & n/a & arb. & $0$ & $-t_{2n-1}^{\prime}$ & $0$ \\ [1pt]
\hline
$H_{\mathrm{S,DIII}}^{(y,I,y)}$ & $0$ & arb. & $\pi/2$ & $0$ & n/a & arb. & $\pi/2$ & $t_{2n-1}^{\prime}$ & $\pi/2$ \\ [1pt]
\hline \hline
$H_{\mathrm{S,CI}}^{(x,y,z)}$ & $0$ & $0$ & n/a & arb. & arb. & arb. & arb. & $t_{2n-1}^{\prime}$ & $-\phi_{2n-1}^{\prime}$ \\ [1pt]
\hline
$H_{\mathrm{S,CI}}^{(I,y,y)}$ & arb. & arb. & $0$ & arb. & $0$ & arb. & $0$ & $t_{2n-1}^{\prime}$ & $0$  \\ [1pt]
\hline
$H_{\mathrm{S,CI}}^{(z,y,x)}$ & arb. & arb. & $0$ & arb. & $\pi/2$ & arb. & $\pi/2$ & $-t_{2n-1}^{\prime}$ & $\pi/2$ \\ [1pt]
\hline
\end{tabular}
\end{center}
\end{table*}

\begin{table*}[t]
\begin{center}
\caption{\label{tableAns}Nonsymmorphic models with chiral symmetry including all long-range hoppings, showing the values of the tight-binding parameters of the generalized Rice-Mele model, Eqs.~(\ref{dxnsall},\ref{dynsall},\ref{dznsall}), for each model with $\epsilon_0 = 0$, $t_{2n}^{\prime} = - t_{2n}$ and $\phi_{2n}^{\prime} = \phi_{2n}$.}
\begin{tabular}{ L{1.2cm} || C{1.0cm} | C{1.0cm} | C{1.0cm} | C{1.0cm} | C{1.0cm} | C{1.0cm} | C{1.0cm}}
\hline
name & $u$ & $t_{2n}$ & $\phi_{2n}$ & $t_{2n+1}$ & $\phi_{2n+1}$ & $t_{2n+1}^{\prime}$ & $\phi_{2n+1}^{\prime}$  \\ [1pt]
\hline \hline
$H_{\mathrm{NS,A}}^{(0,0,y)}$ & arb. & arb. & arb. & arb. & arb. & $t_{2n+1}$ & $\phi_{2n+1}$ \\ [1pt]
\hline
$H_{\mathrm{NS,A}}^{(0,0,x)}$ & arb. & arb. & arb. & arb. & arb. & $-t_{2n+1}$ & $\phi_{2n+1}$ \\ [1pt]
\hline \hline
$H_{\mathrm{NS,AI}}^{(I,y,y)}$ & arb. & arb. & $0$ & arb. & $0$ & $t_{2n+1}$ & $0$ \\ [1pt]
\hline
$H_{\mathrm{NS,AI}}^{(z,x,y)}$ & arb. & arb. & $0$ & arb. & $\pi/2$ & $t_{2n+1}$ & $\pi/2$ \\ [1pt]
\hline
$H_{\mathrm{NS,AI}}^{(I,x,x)}$ & arb. & arb. & $0$ & arb. & $0$ & $-t_{2n+1}$ & $0$ \\ [1pt]
\hline
$H_{\mathrm{NS,AI}}^{(z,y,x)}$ & arb. & arb. & $0$ & arb. & $\pi/2$ & $-t_{2n+1}$ & $\pi/2$ \\ [1pt]
\hline \hline
$H_{\mathrm{NS,D}}^{(x,z,y)}$ & $0$ & arb. & $\pi/2$ & arb. & $0$ & $t_{2n+1}$ & $0$ \\ [1pt]
\hline
$H_{\mathrm{NS,D}}^{(y,I,y)}$ & $0$ & arb. & $\pi/2$ & arb. & $\pi/2$ & $t_{2n+1}$ & $\pi/2$ \\ [1pt]
\hline
$H_{\mathrm{NS,D}}^{(y,z,x)}$ & $0$ & arb. & $\pi/2$ & arb. & $0$ & $-t_{2n+1}$ & $0$ \\ [1pt]
\hline
$H_{\mathrm{NS,D}}^{(x,I,x)}$ & $0$ & arb. & $\pi/2$& arb. & $\pi/2$ & $-t_{2n+1}$ & $\pi/2$ \\ [1pt]
\hline \hline
$H_{\mathrm{NS,AIII}}^{(x,y,z)}$ & $0$ & $0$ & n/a & arb. & arb. & $t_{2n+1}$ & $-\phi_{2n+1}$ \\ [1pt]
\hline
$H_{\mathrm{NS,AIII}}^{(y,x,z)}$ & $0$ & $0$ & n/a & arb. & arb. & $-t_{2n+1}$ & $-\phi_{2n+1}$ \\ [1pt]
\hline
\end{tabular}
\end{center}
\end{table*}

\section{Conclusion}\label{s:conc}

We considered noninteracting, one-dimensional tight-binding models on a periodic lattice with two energy bands.
For each symmetry class, Table~\ref{tablesummary}, we identified distinct models in the atomic basis in position space and showed that they may be written in a unified canonical form.
Generally, states of matter were found to be in agreement with predictions of the tenfold way classification for the symmorphic models~\cite{schnyder08,kitaev09,ryu10,chiu16} and with the classification of Ref.~\cite{shiozaki16} for nonsymmorphic models.
Exceptions arise because we only consider systems with two bands: they are the symmorphic DIII and the nonsymmorphic D class which we find are both gapless due to the presence of time-reversal symmetry with Kramer's degeneracy. With two bands, there are also not enough degrees of freedom to realize the CII (chiral symplectic) symmetry class~\cite{matveeva22}.
In principle, however, our approach could be extended to include more than two orbitals per unit cell or to incorporate superconducting pairing, yielding a four-orbital Bogoliubov de Gennes representation.

We described models with nonsymmorphic nonspatial symmetries and found that they separate into only two types of state of matter: an insulator with a $\mathbb{Z}_2$ topological index~\cite{shiozaki15,shiozaki16} in the absence of nonsymmorphic time-reversal symmetry or, in the presence of nonsymmorphic time-reversal symmetry, a metallic state~\cite{zhao16}.
The models identified in the atomic basis in position space provide a recipe for experimental realization of examples of each symmetry group in a relatively simple way with only two orbitals per cell, and there are a variety of different platforms for physical synthesis of one-dimensional tight-binding models including engineered atomic lattices~\cite{kim12,cheon15,shim15,kim17,drost17,lee19,huda20,kiczynski22} and cold atoms in optical lattices~\cite{atala13,przysiezna15,meier16,meier18,cooper19,kang2020}.

All relevant data presented in this paper can be accessed~\cite{datanote}.

\begin{acknowledgments}
The author thanks S. T. Carr, C. Y. Leung, A. Romito and H. Schomerus for helpful discussions.
\end{acknowledgments}

\appendix

\section{Long-range hoppings}\label{a:longrange}

We generalize the Rice-Mele model~(\ref{cgrm2}) to include all possible long-range couplings,
\begin{eqnarray}
d_0 (k,s) &=& \epsilon_0 + \sum_{n=1}^{\infty} \big[ t_{2n} \cos (nka + \phi_{2n} ) \nonumber \\
&& + t_{2n}^{\prime} \cos (nka + \phi_{2n}^{\prime} ) \big] , \label{d0all} \\
d_x (k,s) &=& \sum_{n=0}^{\infty} \big[ t_{2n+1} \cos [k (na + s) + \phi_{2n+1}] \nonumber \\
&& \!\!\!\!\!\!  + t_{2n+1}^{\prime} \cos [ k((n+1)a - s) + \phi_{2n+1}^{\prime}] \big] , \label{dxall} \\
d_y (k,s) &=& \sum_{n=0}^{\infty} \big[ - t_{2n+1} \sin [k (na + s) + \phi_{2n+1}] \nonumber \\
&& \!\!\!\!\!\!  + t_{2n+1}^{\prime} \sin [ k((n+1)a - s) + \phi_{2n+1}^{\prime}] \big] , \label{dyall} \\
d_z (k,s) &=& u + \sum_{n=1}^{\infty} \big[ t_{2n} \cos (nka + \phi_{2n} )
\nonumber \\
&& - t_{2n}^{\prime} \cos (nka + \phi_{2n}^{\prime} ) \big] , \label{dzall}
\end{eqnarray}
where $t_{2n}$ and $t_{2n}^{\prime}$ are A-A and B-B hoppings with phases $\phi_{2n}$ and $\phi_{2n}^{\prime}$, $t_{2n+1}$ and $t_{2n+1}^{\prime}$ are A-B hoppings with phases $\phi_{2n+1}$ and $\phi_{2n+1}^{\prime}$, for integer $n$.
In terms of the nearest-neighbor parameters in the main text, $t_{2} = t_{AA}$, $t_{2}^{\prime} = t_{BB}$, $\phi_{2} = \phi_{AA}$, $\phi_{2}^{\prime} = \phi_{BB}$, $t_1 = v$, $t_1^{\prime} = w$, $\phi_1 = \phi_v$, $\phi_1^{\prime} = \phi_w$.
For symmorphic models without chiral symmetry, application of time-reversal or charge-conjugation symmetry requires that each component $d_i$ is an even or odd function of $k$ at $s=0$, resulting in the same models as described in the main text with all hoppings as listed in Table~\ref{tableS}. For nonsymmorphic models without chiral symmetry, we consider intracell spacing $s=a/2$, and they are listed in Table~\ref{tableAns2}.

Chiral symmetry imposes $d_0=0$ and one of the other components must be zero, too.
With $d_0 = 0$, then $\epsilon_0 = 0$, $t_{2n}^{\prime} = - t_{2n}$ and $\phi_{2n}^{\prime} = \phi_{2n}$.
Then, for symmorphic models, we consider $s=0$ and simplify as
\begin{eqnarray}
d_x (k,0) &=& v \cos \phi_v + \sum_{n=1}^{\infty} \big[ t_{2n+1} \cos (kna  + \phi_{2n+1}) \nonumber \\
&& \qquad + t_{2n-1}^{\prime} \cos ( kna + \phi_{2n-1}^{\prime}) \big] , \label{dxsall} \\
d_y (k,0) &=& - v \sin \phi_v + \sum_{n=1}^{\infty} \big[ - t_{2n+1} \sin (kna  + \phi_{2n+1}) \nonumber \\
&& \qquad + t_{2n-1}^{\prime} \sin ( kna + \phi_{2n-1}^{\prime}) \big] , \label{dysall} \\
d_z (k,0) &=& u + 2 \sum_{n=1}^{\infty} t_{2n} \cos (nka + \phi_{2n} ) , \label{dzsall} 
\end{eqnarray}
using $t_1 = v$ and $\phi_1 = \phi_v$. Applying chiral and time-reversal symmetries introduces constraints on the parameters giving the models discussed in the main text for nearest-neighbor hopping and listed in Table~\ref{tableAs} including all long-range hoppings. 
For nonsymmorphic models, we consider $s=a/2$ and simplify as
\begin{eqnarray}
d_x (k,a/2) &=& \sum_{n=0}^{\infty} \big[ t_{2n+1} \cos (ka[n+\tfrac{1}{2}]  + \phi_{2n+1}) \nonumber \\
&& \quad + t_{2n+1}^{\prime} \cos ( ka[n+\tfrac{1}{2}] + \phi_{2n+1}^{\prime}) \big] , \label{dxnsall} \\
d_y (k,a/2) &=& \sum_{n=0}^{\infty} \big[ - t_{2n+1} \sin (ka[n+\tfrac{1}{2}]  + \phi_{2n+1}) \nonumber \\
&& \quad + t_{2n+1}^{\prime} \sin ( ka[n+\tfrac{1}{2}] + \phi_{2n+1}^{\prime}) \big] , \label{dynsall} \\
d_z (k,a/2) &=& u + 2 \sum_{n=1}^{\infty} t_{2n} \cos (nka + \phi_{2n} ) . \label{dznsall} 
\end{eqnarray}
Again, application of chiral and time-reversal symmetries introduces constraints on the parameters giving the models discussed in the main text for nearest-neighbor hopping and listed in Table~\ref{tableAns} including all long-range hoppings. 

\section{Spatial-inversion symmetry and quantized Zak phase}\label{a:spatial}

The catalog of tight-binding models in the main text, Table~\ref{tablesummary}, is based on the ten-fold way classification of nonspatial symmetries (time-reversal, charge-conjugation and chiral)~\cite{wigner51,wigner58,dyson62a,gade93,verbaarschot93,verbaarschot94,altland97,schnyder08,kitaev09,ryu10,chiu16}.
An alternative approach is to classify models based on spatial symmetries~\cite{teo08,chiu13,morimoto13,shiozaki14,chiu16} such as
spatial-inversion symmetry~\cite{hughes11,chen11,fuchs21} as described by Eq.~(\ref{pk}).
We consider the generalized Rice-Mele model, Eqs.~(\ref{d0all},\ref{dxall},\ref{dyall},\ref{dzall}), and apply spatial-inversion symmetry in the form of $U_P = I$, $\sigma_x$,  $\sigma_y$,  or $\sigma_z$. This forces the individual components of the ${\bf d}$ vector to be either even or odd with respect to $k$.

With intracell spacing $s=0$, there are four models, denoted $H_{\mathrm{S,A}}^{(U_P=i)}$ for $i = I$, $\sigma_x$,  $\sigma_y$,  or $\sigma_z$ with parameters listed in Table~\ref{tableS}. In position space, these models satisfy spatial inversion~(\ref{ppos}) with ${\mathcal P} = P_I$, $P_x$, $P_y$, or $P_z$, respectively, for an even number of atoms, $J$.
For model $H_{\mathrm{S,A}}^{(U_P=I)}$ with $U_P=I$, all four components of the ${\bf d}$ vector are even with respect to $k$. Apart from component $d_0$, this is the same as class C with $U_C^2 = -1$ described in the main text, Table~\ref{tablesns3}. Hence, as with class C, $H_{\mathrm{S,A}}^{(U_P=I)}$ will be an insulator with a single phase only.

Of the three models $H_{\mathrm{S,A}}^{(U_P=i)}$ for $i = \sigma_x$,  $\sigma_y$, or $\sigma_z$, only two of them ($H_{\mathrm{S,A}}^{(U_P=\sigma_x)}$ and $H_{\mathrm{S,A}}^{(U_P=\sigma_z)}$) are distinct in the atomic basis because $H_{\mathrm{S,A}}^{(U_P=\sigma_y)}$ is a version of $H_{\mathrm{S,A}}^{(U_P=\sigma_x)}$ with imaginary A-B coupling parameters and is thus related to it by a diagonal gauge transformation.
With these models, component $d_0$ is an even function of $k$, one of the other components is also even and the other two are odd. They may be rotated into the same canonical form using rotations such as $R_x$ or $R_y$. Apart from component $d_0$, they have the same form as class D with $U_C^2 = 1$ described in the main text, Table~\ref{tablesns3}. Like class D, they are insulators with a $\mathbb{Z}_2$ topological index. In the context of one-dimensional systems with spatial inversion symmetry, the $\mathbb{Z}_2$ index is Zak phase quantized to either $0$ or $\pi$~\cite{zak89,hughes11,chen11,cayssol21,fuchs21}.

With intracell spacing $s=a/2$, application of $U_P = \sigma_x$ or $\sigma_y$ just reproduces models $H_{\mathrm{S,A}}^{(U_P=\sigma_x)}$ and $H_{\mathrm{S,A}}^{(U_P=\sigma_y)}$ discussed above.
Application of $U_P = I$ or $\sigma_z$ produces two models denoted $H_{\mathrm{NS,A}}^{(U_P=I)}$ and $H_{\mathrm{NS,A}}^{(U_P=\sigma_z)}$, with tight-binding parameters listed in Table~\ref{tableAns2}.
In position space, these models satisfy spatial inversion~(\ref{ppos}) with ${\mathcal P} = P_x$ (for $U_T = I$) or $P_y$ (for $U_T = \sigma_z$), for an odd number of atoms, $J$. We label these models as `NS' even though the symmetry is not nonsymmorphic, because the form of the Hamiltonian is similar to that of the nonsymmorphic ones, i.e., they also have definite relationships between hoppings $t_{2n+1}^{\prime}$ and $t_{2n+1}$, Table~\ref{tableAns2}.
Only one of these models is distinct, $H_{\mathrm{NS,A}}^{(U_P=I)}$, because $H_{\mathrm{NS,A}}^{(U_P=\sigma_z)}$ is related to it by the diagonal transformation ${\mathcal R}_d$, Eq.~(\ref{rd}).
Apart from component $d_0$ (which is even), $H_{\mathrm{NS,A}}^{(U_P=I)}$ has the same form as class A with $U_C^2 = \mathrm{NS}$ described in the main text, Table~\ref{tablesns3}. Like this class, $H_{\mathrm{NS,A}}^{(U_P=I)}$ describes an insulator with a $\mathbb{Z}_2$ topological index (the quantized Zak phase).

One can subsequently refine the classification by applying time-reversal or charge-conjugation symmetry which will also place requirements on components of the ${\bf d}$ vector to be even or odd functions of $k$. In situations where these constraints clash with those of spatial-inversion symmetry, the component must be set equal to zero.
With time-reversal symmetry, the resulting models would be like those with chiral symmetry but with $d_0 \neq 0$.
For example, applying $U_T = I$ to model $H_{\mathrm{S,A}}^{(U_P=\sigma_x)}$ requires $d_z =0$ with $d_x$ being even and $d_y$ odd, similar to class BDI, Table~\ref{tablescf} (with $d_0 \neq 0$).
Finally, we note that it is possible to consider other spatial symmetries, for example in a four band topological insulator~\cite{hetenyi18} with charge-conjugation and reflection symmetries~\cite{teo08,chiu13,morimoto13,shiozaki14,chiu16}.


\begin{thebibliography}{99}

\bibitem{wigner51}
E. P. Wigner,
On the statistical distribution of the widths and spacings of nuclear resonance levels,
Proc.\ Camb.\ Philos.\ Soc.\ {\bf 47}, 790 (1951).

\bibitem{wigner58}
E. P. Wigner,
On the distribution of the roots of certain symmetric matrices,
Ann.\ Math.\ {\bf 67}, 325 (1958).

\bibitem{dyson62a}
F. J. Dyson,
Statistical theory of the energy levels of complex systems. I,
J.\ Math.\ Phys.\ {\bf 3}, 140 (1962).

\bibitem{gade93}
R. Gade,
Anderson localization for sublattice models,
Nucl.\ Phys.\ B {\bf 398}, 499 (1993).

\bibitem{verbaarschot93}
J. J. M. Verbaarschot and I. Zahed,
Spectral density of the QCD Dirac operator near zero virtuality,
Phys.\ Rev.\ Lett.\ {\bf 70}, 3852 (1993).

\bibitem{verbaarschot94}
J. Verbaarschot,
Spectrum of the QCD Dirac operator and chiral random matrix theory,
Phys.\ Rev.\ Lett.\ {\bf 72}, 2531 (1994).

\bibitem{altland97}
A. Altland and M. R. Zirnbauer,
Nonstandard symmetry classes in mesoscopic normal-superconducting hybrid structures,
Phys.\ Rev.\ B {\bf 55}, 1142 (1997).

\bibitem{schnyder08}
A. P. Schnyder, S. Ryu, A. Furusaki, and A. W. W. Ludwig,
Classification of topological insulators and superconductors in three spatial dimensions,
Phys.\ Rev.\ B {\bf 78}, 195125 (2008).

\bibitem{kitaev09}
A. Kitaev,
Periodic table for topological insulators and superconductors,
AIP Conf.\ Proc.\ {\bf 1134}, 22 (2009).

\bibitem{ryu10}
S. Ryu, A. P. Schnyder, A. Furusaki, and A. W. W. Ludwig,
Topological insulators and superconductors: tenfold way and dimensional hierarchy,
New J.\ Phys.\ {\bf 12}, 065010 (2010).

\bibitem{chiu16}
C.-K.Chiu, J. C. Y. Teo, A. P. Schnyder, and S. Ryu,
Classification of topological quantum matter with symmetries,
Rev.\ Mod.\ Phys.\ {\bf 88}, 035005 (2016).

\bibitem{liu14}
C.-X. Liu, R.-X. Zhang, and B. K. VanLeeuwen,
Topological nonsymmorphic crystalline insulators,
Phys.\ Rev.\ B {\bf 90}, 085304 (2014).

\bibitem{shiozaki14}
K. Shiozaki and M. Sato,
Topology of crystalline insulators and superconductors,
Phys.\ Rev.\ B {\bf 90}, 165114 (2014).

\bibitem{young15}
S. M. Young and C. L. Kane,
Dirac semimetals in two dimensions,
Phys.\ Rev.\ Lett.\ {\bf 115}, 126803 (2015).

\bibitem{wang16}
Z. Wang, A. Alexandradinata, R. J. Cava, and B. A. Bernevig,
Hourglass fermions,
Nature {\bf 532}, 189 (2016).

\bibitem{shiozaki16}
K. Shiozaki, M. Sato, and K. Gomi,
Topology of nonsymmorphic crystalline insulators and superconductors,
Phys.\ Rev.\ B {\bf 93}, 195413 (2016).

\bibitem{varjas17}
D. Varjas, F. de Juan, and Y.-M. Lu,
Space group constraints on weak indices in topological insulators,
Phys.\ Rev.\ B {\bf 96}, 035115 (2017).

\bibitem{kruthoff17}
J. Kruthoff, J. de Boer, J. van Wezel, C. L. Kane, and R.-J. Slager,
Topological classification of crystalline insulators through band structure combinatorics,
Phys.\ Rev.\ X {\bf 7}, 041069 (2017).

\bibitem{herrera22}
M. A. J. Herrera and D. Bercioux,
Tunable Dirac points in a two-dimensional nonsymmorphic wallpaper group lattice,
Commun.\ Phys.\ {\bf 6}, 42 (2023).

\bibitem{lax74}
M. Lax,
{\it Symmetry principles in solid state and molecular physics}
(Wiley-Interscience, 1974).

\bibitem{mong10}
R. S. K. Mong, A. M. Essin, and J. E. Moore,
Antiferromagnetic topological insulators,
Phys. Rev. B {\bf 81}, 245209 (2010).

\bibitem{otrokov19}
M. M. Otrokov {\it et al.},
Prediction and observation of an antiferromagnetic topological insulator,
Nature {\bf 576}, 416 (2019).

\bibitem{gong19}
Y. Gong {\it et al.},
Experimental realization of an intrinsic magnetic topological insulator,
Chin.\ Phys.\ Lett.\ {\bf 36}, 076801 (2019).

\bibitem{zhang19}
D. Zhang, M. Shi, T. Zhu, D. Xing, H. Zhang, and J. Wang,
Topological axion states in the magnetic insulator MnBi$_2$Te$_4$ with the quantized magnetoelectric effect,
Phys.\ Rev.\ Lett.\ {\bf 122}, 206401 (2019).

\bibitem{niu20}
C. Niu, H. Wang, N. Mao, B. Huang, Y. Mokrousov, and Y. Dai,
Antiferromagnetic topological insulator with nonsymmorphic protection in two dimensions,
Phys.\ Rev.\ Lett.\ {\bf 124}, 066401 (2020).

\bibitem{fang15}
C. Fang and L. Fu,
New classes of three-dimensional topological crystalline insulators: Nonsymmorphic and magnetic,
Phys.\ Rev.\ B {\bf 91}, 161105(R) (2015).

\bibitem{shiozaki15}
K. Shiozaki, M. Sato, and K. Gomi,
$\mathbb{Z}_2$ topology in nonsymmorphic crystalline insulators: M\"obius twist in surface states,
Phys.\ Rev.\ B {\bf 91}, 155120 (2015).

\bibitem{zhao16}
Y. X. Zhao and A. P. Schnyder,
Nonsymmorphic symmetry-required band crossings in topological semimetals,
Phys.\ Rev.\ B {\bf 94}, 195109 (2016).

\bibitem{arkinstall17}
J. Arkinstall, M. H. Teimourpour, L. Feng, R. El-Ganainy, and H. Schomerus,
Topological tight-binding models from nontrivial square roots,
Phys.\ Rev.\ B {\bf 95}, 165109 (2017).

\bibitem{marques19}
A. M. Marques and R. G. Dias,
One-dimensional topological insulators with noncentered inversion symmetry axis,
Phys.\ Rev.\ B {\bf 100}, 041104(R) (2019).

\bibitem{brzezicki20}
W. Brzezicki and T. Hyart,
Topological domain wall states in a nonsymmorphic chiral chain,
Phys.\ Rev.\ B {\bf 101}, 235113 (2020).

\bibitem{allen22}
R. E. J. Allen, H. V. Gibbons, A. M. Sherlock, H. R. M. Stanfield, and E. McCann,
Nonsymmorphic chiral symmetry and solitons in the Rice-Mele model,
Phys.\ Rev.\ B {\bf 106}, 165409 (2022).

\bibitem{yang22}
Y. Yang, H. C. Po, V. Liu, J. D. Joannopoulos, L. Fu, and M. Solja\v{c}i\'{c},
Non-Abelian nonsymmorphic chiral symmetries,
Phys.\ Rev.\ B {\bf 106}, L161108 (2022).

\bibitem{ssh79}
W. P. Su, J. R. Schrieffer, and A. J. Heeger,
Solitons in polyacetylene,
Phys.\ Rev.\ Lett.\ {\bf 42}, 1698 (1979).

\bibitem{ssh80}
W. P. Su, J. R. Schrieffer, and A. J. Heeger,
Soliton excitations in polyacetylene,
Phys.\ Rev.\ B {\bf 22}, 2099 (1980).

\bibitem{hasan10}
M. Z. Hasan and C. L. Kane,
Colloquium: Topological insulators,
Rev.\ Mod.\ Phys.\ {\bf 82}, 3045 (2010).

\bibitem{asboth16}
J. K. Asb\'oth, L. Oroszl\'any, and A. P\'alyi,
{\it A Short Course on Topological Insulators}
(Springer, Switzerland, 2016).

\bibitem{cayssol21}
J. Cayssol and J.-N. Fuchs,
Topological and geometrical aspects of band theory,
J.\ Phys.\ Mater.\ {\bf 4}, 034007 (2021).

\bibitem{takayama80}
H. Takayama, Y. R. Lin-Liu, and K. Maki,
Continuum model for solitons in polyacetylene,
Phys.\ Rev.\ B {\bf 21}, 2388 (1980).

\bibitem{kivelson83}
S. Kivelson,
Solitons with adjustable charge in a commensurate Peierls insulator,
Phys.\ Rev.\ B {\bf 28}, 2653 (1983).

\bibitem{fulga11}
I. C. Fulga, F. Hassler, A. R. Akhmerov, and C. W. J. Beenakker,
Scattering formula for the topological quantum number of a disordered multimode wire,
Phys.\ Rev.\ B {\bf 83}, 155429 (2011).

\bibitem{chen11}
K. T. Chen and P. A. Lee,
Static electric field in one-dimensional insulators without boundaries,
Phys.\ Rev.\ B {\bf 84}, 113111 (2011).

\bibitem{gangadharaiah12}
S. Gangadharaiah, L. Trifunovic, and D. Loss,
Localized end states in density modulated quantum wires and rings,
Phys.\ Rev.\ Lett.\ {\bf 108}, 136803 (2012).

\bibitem{pershoguba12}
S. S. Pershoguba and V. M. Yakovenko,
Shockley model description of surface states in topological insulators,
Phys.\ Rev.\ B {\bf 86}, 075304 (2012).

\bibitem{li14}
L. Li, Z. Xu, and S. Chen,
Topological phases of generalized Su-Schrieffer-Heeger models,
Phys.\ Rev.\ B {\bf 89}, 085111 (2014).

\bibitem{li15}
L. Li and S. Chen, 
Characterization of topological phase transitions via topological properties of transition points,
Phys.\ Rev.\ B {\bf 92}, 085118 (2015).

\bibitem{guo16}
H.-M. Guo,
A brief review on one-dimensional topological insulators and superconductors,
Sci.\ China Phys.\ Mech.\ {\bf 59}, 637401 (2016).

\bibitem{velasco17}
C. G. Velasco and B. Paredes,
Realizing and detecting a topological insulator in the AIII symmetry class,
Phys.\ Rev.\ Lett.\ {\bf 119}, 115301 (2017).

\bibitem{bercioux17}
D. Bercioux, O. Dutta, and E. Rico,
Solitons in one-dimensional lattices with a flat band,
Ann.\ Phys.\ {\bf 529}, 1600262 (2017).

\bibitem{rhim17}
J.-W. Rhim, J. Behrends, and J. H. Bardarson,
Bulk-boundary correspondence from the intercellular Zak phase,
Phys.\ Rev.\ B {\bf 95}, 035421 (2017).

\bibitem{rhim18}
J.-W. Rhim, J. H. Bardarson, and R.-J. Slager,
Unified bulk-boundary correspondence for band insulators,
Phys.\ Rev.\ B {\bf 97}, 115143 (2018).

\bibitem{liu18}
T. Liu and H. Guo,
Topological phase transition in the quasiperiodic disordered Su–Schriffer–Heeger chain,
Phys.\ Lett.\ A {\bf 382}, 3287 (2018).

\bibitem{munoz18}
F. Munoz, F. Pinilla, J. Mella, and M. I. Molina,
Topological properties of a bipartite lattice of domain wall states,
Sci.\ Rep.\ {\bf 8}, 17330 (2018).

\bibitem{perezgonzalez19}
B. P\'{e}rez-Gonz\'alez, M. Bello, \'{A}. G\'{o}mez-Le\'{o}n, and G. Platero,
Interplay between long-range hopping and disorder in topological systems,
Phys.\ Rev.\ B {\bf 99}, 035146 (2019).

\bibitem{chen20}
B.-H. Chen and D.-W. Chiou,
An elementary rigorous proof of bulk-boundary correspondence in the generalized Su-Schrieffer-Heeger model,
Phys. Lett. A {\bf 384}, 126168 (2020).

\bibitem{scollon20}
M. Scollon and M. P. Kennett,
Persistence of chirality in the Su-Schrieffer-Heeger model in the presence of on-site disorder,
Phys.\ Rev.\ B {\bf 101}, 144204 (2020).

\bibitem{pletyukhov20}
M. Pletyukhov, D. M. Kennes, K. Piasotski, J. Klinovaja, D. Loss, and H. Schoeller,
Rational boundary charge in one-dimensional systems with interaction and disorder,
Phys.\ Rev.\ Research {\bf 2}, 033345 (2020).

\bibitem{lin20}
Y.-T. Lin, D. M. Kennes, M. Pletyukhov, C. S. Weber, H. Schoeller, and V. Meden,
Interacting Rice-Mele model: Bulk and boundaries,
Phys.\ Rev.\ B {\bf 102}, 085122 (2020).

\bibitem{vanmiert20}
G. van Miert and C. Ortix,
On the topological immunity of corner states in two-dimensional crystalline insulators,
npj Quantum Mater.\ {\bf 5}, 63 (2020).

\bibitem{han20}
S.-H. Han, S.-G. Jeong, S.-W. Kim, T.-H. Kim, and S. Cheon,
Topological features of ground states and topological solitons in generalized Su-Schrieffer-Heeger models using generalized time-reversal, particle-hole, and chiral symmetries,
Phys.\ Rev.\ B {\bf 102}, 235411 (2020).

\bibitem{hetenyi21}
B. Het\'enyi, Y. Pulcu, and S. Do\u{g}an,
Calculating the polarization in bipartite lattice models: Application to an extended Su-Schrieffer-Heeger model,
Phys.\ Rev.\ B {\bf 103}, 075117 (2021).

\bibitem{chen21}
H.-T. Chen, C.-H. Chang, and H.-c. Kao,
Connection between the winding number and the Chern number,
Chin.\ J.\ Phys.\ {\bf 72}, 50 (2021).

\bibitem{fuchs21}
J.-N. Fuchs and F. Pi\'{e}chon,
Orbital embedding and topology of one-dimensional two-band insulators,
Phys.\ Rev.\ B {\bf 104}, 235428 (2021).

\bibitem{zurita21}
J. Zurita, C. Creffield, and G. Platero,
Tunable zero modes and quantum interferences in flat-band topological insulators,
Quantum {\bf 5}, 591 (2021).

\bibitem{kim12}
T.-H. Kim and H. W. Yeom,
Topological solitons versus nonsolitonic phase defects in a
quasi-one-dimensional charge-density wave,
Phys.\ Rev.\ Lett.\ {\bf 109}, 246802 (2012).

\bibitem{cheon15}
S. Cheon, T.-H. Kim, S.-H. Lee, and H. W. Yeom,
Chiral solitons in a coupled double Peierls chain,
Science {\bf 350}, 182 (2015).

\bibitem{shim15}
H. Shim, G. Lee, J.-M. Hyun, and H. Kim,
Cooperative interplay between impurities and charge density wave in the phase transition of atomic wires,
New J.\ Phys.\ {\bf 17}, 093026 (2015).

\bibitem{kim17}
T.-H. Kim, S. Cheon, and H. W. Yeom,
Switching chiral solitons for algebraic operation of topological quaternary digits,
Nat. Phys. {\bf 13}, 444 (2017).

\bibitem{drost17}
R. Drost, T. Ojanen, A. Harju, and P. Liljeroth,
Topological states in engineered atomic lattices,
Nat. Phys. {\bf 13}, 668 (2017).

\bibitem{lee19}
G. Lee, H. Shim, J.-M. Hyun, and H. Kim,
Intertwined solitons and impurities in a quasi-one-dimensional charge-density-wave system: In/Si(111),
Phys.\ Rev.\ Lett.\ {\bf 122}, 016102 (2019).

\bibitem{huda20}
M. N. Huda, S. Kezilebieke, T. Ojanen, R. Drost, and P. Liljeroth,
Tuneable topological domain wall states in engineered atomic chains,
npj Quantum Mater.\ {\bf 5}, 17 (2020).

\bibitem{kiczynski22}
M. Kiczynski, S. K. Gorman, H. Geng, M. B. Donnelly, Y. Chung, Y. He, J. G. Keizer, and M. Y. Simmons,
Engineering topological states in atom-based semiconductor quantum dots,
Nature {\bf 606}, 694 (2022).

\bibitem{atala13}
M. Atala, M. Aidelsburger, J. T. Barreiro, D. Abanin, T. Kitagawa, E. Demler, and I. Bloch,
Direct measurement of the Zak phase in topological Bloch bands,
Nat.\ Phys.\ {\bf 9}, 795 (2013).

\bibitem{przysiezna15}
A. Przysiezna, O. Dutta, and J. Zakrzewski,
Rice–Mele model with topological solitons in an optical lattice,
New J.\ Phys.\ {\bf 17}, 013018 (2015).

\bibitem{meier16}
E. J. Meier, F. A. An, and B. Gadway,
Observation of the topological soliton state in the Su–Schrieffer–Heeger model,
Nat.\ Commun.\ {\bf 7}, 13986 (2016).

\bibitem{meier18}
E. J. Meier, F. A. An, A. Dauphin, M. Maffei, P. Massignan, T. L. Hughes, and B. Gadway,
Observation of the topological Anderson insulator in disordered atomic wires,
Science {\bf 362}, 929 (2018).

\bibitem{cooper19}
N. R. Cooper, J. Dalibard, and I. B. Spielman,
Topological bands for ultracold atoms,
Rev.\ Mod.\ Phys.\ {\bf 91}, 015005 (2019).

\bibitem{kang2020}
J. H. Kang, J. H. Han, and Y Shin,
Creutz ladder in a resonantly shaken 1D optical lattice,
New J.\ Phys.\ {\bf 22}, 013023 (2020).

\bibitem{ricemele82}
M. J. Rice and E. J. Mele,
Elementary excitations of a linearly conjugated diatomic polymer,
Phys.\ Rev.\ Lett.\ {\bf 49}, 1455 (1982).

\bibitem{matveeva22}
P. Matveeva, T. Hewitt, D. Liu, K. Reddy, D. Gutman, and S. T. Carr,
One-dimensional noninteracting topological insulators with chiral symmetry,
Phys.\ Rev.\ B {\bf 107}, 075422 (2023).

\bibitem{creutz99}
M. Creutz,
End states, ladder compounds, and domain-wall fermions,
Phys.\ Rev.\ Lett.\ {\bf 83}, 2636 (1999).

\bibitem{shockley39}
W. Shockley,
On the surface states associated with a periodic potential,
Phys.\ Rev.\ {\bf 56}, 317 (1939).

\bibitem{vanderbilt93}
D. Vanderbilt and R. D. King-Smith,
Electric polarization as a bulk quantity and its relation to surface charge,
Phys.\ Rev.\ B {\bf 48}, 4442 (1993).

\bibitem{canonicalnote}
For each symmetry class, we can write all models in the same canonical form using unitary transformations. In some cases, the canonical form is not unique, e.g. for class BDI, Table~\ref{tablescf}, we have chosen ${\tilde d}_x$ even and ${\tilde d}_y$ odd, but one could also choose ${\tilde d}_x$ odd and ${\tilde d}_y$ even. For nonsymmorphic systems, there is a `cosine' or a `sine' representation, as described in the main text: this corresponds to the occasional ambiguity in the sign, $\epsilon_{\sigma}$, of a product of symmetry operators in Ref.~\cite{shiozaki16}.

\bibitem{budich13}
J. C. Budich and E. Ardonne,
Topological invariant for generic one-dimensional time-reversal-symmetric superconductors in class DIII,
Phys.\ Rev.\ B {\bf 88}, 134523 (2013).

\bibitem{li16}
L. Li, C. Yang, and S. Chen,
Topological invariants for phase transition points of one-dimensional $\mathbb{Z}_2$ topological systems,
Eur.\ Phys.\ J.\ B {\bf 89}, 195 (2016).

\bibitem{gholizadeh18}
S. Gholizadeh, M. Yahyavi and B. Het\'enyi,
Extended Creutz ladder with spin-orbit coupling: A one-dimensional analog of the Kane-Mele model,
EPL {\bf 122}, 27001 (2018).

\bibitem{teo08}
J. C. Y. Teo, L. Fu, and C. L. Kane,
Surface states and topological invariants in three-dimensional topological insulators: Application to Bi$_{1-x}$Sb$_x$,
Phys.\ Rev.\ B {\bf 78}, 045426 (2008).

\bibitem{hughes11}
T. L. Hughes, E. Prodan, and B. A. Bernevig,
Inversion-symmetric topological insulators,
Phys.\ Rev.\ B {\bf 83}, 245132 (2011).

\bibitem{chiu13}
C.-K. Chiu, H. Yao, and S. Ryu,
Classification of topological insulators and superconductors in the presence of reflection symmetry,
Phys.\ Rev.\ B {\bf 88}, 075142 (2013).

\bibitem{morimoto13}
T. Morimoto and A. Furusaki,
Topological classification with additional symmetries from Clifford algebras,
Phys.\ Rev.\ B {\bf 88}, 125129 (2013).

\bibitem{bena09}
C. Bena and G. Montambaux,
Remarks on the tight-binding model of graphene,
New J.\ Phys.\ {\bf 11}, 095003 (2009).

\bibitem{symmetrynotes}
The condition $U_S (k,s) =U_C^{\ast} (k,s) U_T (-k,s)$ may sometimes yield irrelevant additional factors (e.g. of `$i$') appearing as products with the Pauli spin matrices; we neglect such factors.

\bibitem{junemann17}
J. J\"unemann, A. Piga, S.-J. Ran, M. Lewenstein, M. Rizzi, and A. Bermudez,
Exploring interacting topological insulators with ultracold atoms: The synthetic Creutz-Hubbard model,
Phys.\ Rev.\ X {\bf 7}, 031057 (2017).

\bibitem{kitaev01}
A. Y. Kitaev,
Unpaired Majorana fermions in quantum wires,
Phys.-Usp.\ {\bf 44}, 131 (2001).

\bibitem{bychkovrashba84}
Y. A. Bychkov and E. I. Rashba,
Oscillatory effects and the magnetic susceptibility of carriers in inversion layers,
J.\ Phys.\ C {\bf 17}, 6039 (1984).

\bibitem{mii14}
T. Mii, N. Shima, K. Kano, and K. Makoshi,
Spin-orbit interaction in the tight-binding model - toward the comprehension of the Rashba effect at surfaces,
J.\ Phys.\ Soc.\ Jpn.\ {\bf 83}, 064706 (2014).

\bibitem{schulze13}
M. Schulze, D. Bercioux, and D. F. Urban,
Adiabatic pumping in the quasi-one-dimensional triangle lattice,
Phys.\ Rev.\ B {\bf 87}, 024301 (2013).

\bibitem{sun17}
N. Sun and L.-K. Lim,
Quantum charge pumps with topological phases in a Creutz ladder,
Phys.\ Rev.\ B {\bf 96}, 035139 (2017).

\bibitem{datanote}
\href{https://doi.org/10.17635/lancaster/researchdata/597}{https://doi.org/10.17635/lancaster/researchdata/597}

\bibitem{zak89}
J. Zak,
Berry’s phase for energy bands in solids,
Phys.\ Rev.\ Lett.\ {\bf 62}, 2747 (1989).

\bibitem{hetenyi18}
B. Het\'enyi and M. Yahyavi,
Topological insulation in a ladder model with particle-hole and reflection symmetries,
J.\ Phys.: Condens.\ Matter {\bf 30}, 10LT01 (2018).

\end{thebibliography}
\end{document}